\documentclass[a4paper,usenames,dvipsnames,11pt]{article}
\pdfoutput=1

\usepackage{jheppub}
\usepackage{slashed}
\usepackage{mathrsfs,booktabs,multirow,tabularx}
\usepackage{stmaryrd}
\usepackage{xspace}
\usepackage{fancyvrb}
\usepackage[makeroom]{cancel}
\usepackage{amsmath}    
\usepackage{amssymb}    %
\usepackage{graphicx}   
\usepackage{verbatim}   
\usepackage{lscape}
\usepackage{subfig}
\usepackage{listings}
\usepackage{mathtools}

\renewcommand\arraystretch{1.1}

\def\beq{\begin{equation}}
\def\eeq{\end{equation}}
\def\beqn{\begin{eqnarray}}
\def\eeqn{\end{eqnarray}}

\newcommand{\bqa}{\begin{eqnarray}}
\newcommand{\eqa}{\end{eqnarray}}
\chardef\MyArticleWithColor=\pdfcolorstackinit page direct{0 g}

\def\cCode#1{\begin{lstlisting}[mathescape,basicstyle=\small
\ttfamily,frame=leftline,aboveskip=4mm,belowskip=4mm,xleftmargin=20pt,framexleftmargin=10pt,
numbers=none,framerule=2pt,abovecaptionskip=0.0mm,belowcaptionskip=3.5mm #1]}

\newcommand\sss{\scriptscriptstyle}

\newcommand\as{\alpha_{\sss S}}

\newcommand{\tev}{\,\textrm{TeV}}
\newcommand{\gev}{\,\textrm{GeV}}

\newcommand\mf{{\sc\small MadFKS}}
\newcommand\ml{{\sc\small MadLoop}}
\newcommand\ct{{\sc\small CutTools}}
\newcommand\nin{{\sc\small Ninja}}
\newcommand\collier{{\sc\small Collier}}

\newcommand\sherpa{{\sc\small Sherpa}}
\newcommand\alpgen{{\sc\small AlpGen}}

\newcommand{\LO}{{\rm LO}}

\newcommand{\NLO}{{\rm NLO}}

\newcommand{\ord}{{\cal O}}

\def\beq{\begin{equation}}
\def\eeq{\end{equation}}
\def\beqar{\begin{eqnarray}}
\def\eeqar{\end{eqnarray}}
\def\barr#1{\begin{array}{#1}}
\def\earr{\end{array}}
\def\bfi{\begin{figure}}
\def\efi{\end{figure}}
\def\btab{\begin{table}}
\def\etab{\end{table}}
\def\bce{\begin{center}}
\def\ece{\end{center}}

\def\nl{\nonumber\\}

\def\de{\delta}
\def\la{\lambda}
\def\si{\sigma}

\def\refeq#1{\mbox{(\ref{#1})}}

\newcommand{\ri}{{\mathrm{i}}}

\newcommand{\rR}{{\mathrm{R}}}
\newcommand{\rT}{{\mathrm{T}}}
\newcommand{\rL}{{\mathrm{L}}}

\newcommand{\M}{{\cal{M}}}
\newcommand{\Mew}{{\tilde{\cal{M}}}}

\def\mathswitchr#1{#1}
\newcommand{\PW}{\mathswitchr W}

\newcommand{\PZ}{\mathswitchr Z}
\newcommand{\PA}{\mathswitchr A}
\newcommand{\PH}{\mathswitchr H}

\newcommand{\Pt}{\mathswitchr t}

\newcommand{\PWpm}{\mathswitchr {W^\pm}}

\def\mathswitch#1{\relax\ifmmode#1\else$#1$\fi}
\newcommand{\MW}{\mathswitch {M_\PW}}
\newcommand{\MZ}{\mathswitch {M_\PZ}}
\newcommand{\MH}{\mathswitch {M_\PH}}

\newcommand{\Mt}{\mathswitch {m_\Pt}}

\newcommand{\ntad}{n_{\rm tad}}

\newcommand{\cw}{\mathswitch {c_\mathrm{w}}}
\newcommand{\sw}{\mathswitch {s_\mathrm{w}}}

\newcommand{\NCf}{\mathswitch {N_{\mathrm{C}}^f}}
\newcommand{\NCt}{\mathswitch {N_{\mathrm{C}}^{\Pt}}}

\newcommand{\elm}{{\mathrm{em}}}
\newcommand{\ew}{{\mathrm{ew}}}

\newcommand{\SC}{{\mathrm{LSC}}}
\renewcommand{\SS}{{\mathrm{SSC}}}
\newcommand{\cc}{{\mathrm{C}}}

\newcommand{\pre}{{\mathrm{PR}}}

\newcommand{\bew}{b^{\ew}}

\newcommand{\cew}{C^{\ew}}

\newcommand{\NB}{N}
\newcommand{\GB}{V}

\newcommand{\ls}{l(s)}
\newcommand{\lu}{l(\mu^2)}

\newcommand{\lrMwithabs}{l(|r_{kl}|,M^2)}

\newcommand{\lsM}{l(s,M^2)}
\newcommand{\lsW}{l(s,\MW^2)}

\newcommand{\lWfnew}{l^{\rm reg}(\MW^2,m_f^2)}

\newcommand{\lWlanew}{l(\MW^2,Q^2)}

\newcommand{\lWZ}{l(\MW^2,\MZ^2)}
\newcommand{\lWM}{l(\MW^2,M^2)}
\newcommand{\ltW}{l(\Mt^2,\MW^2)}
\newcommand{\lHW}{l(\MH^2,\MW^2)}
\newcommand{\lemf}{l^\elm(m_f^2)}

\newcommand{\lemfsi}{l^\elm(m_{f_\si}^2)}

\newcommand{\lemW}{l^\elm(\MW^2)}
\newcommand{\Ls}{L(s)}
\newcommand{\LrM}{L(|r_{kl}|,M^2)}

\newcommand{\Lrs}{L(|r_{kl}|,s)}

\newcommand{\LsM}{L(s,M^2)}
\newcommand{\LsW}{L(s,\MW^2)}

\newcommand{\Lklanew}{L^{\rm{reg}}(m_k^2,Q^2)}

\newcommand{\LWlanew}{L(\MW^2,Q^2)}

\newcommand{\Lemknew}{L^\elm(s,Q^2,m_k^2)}

\newcommand{\lrs}{\log{\frac{|r_{kl}|}{s}}}
\newcommand{\lrsalpha}{l(|r_{kl}|,s)}

  \newcommand{\TO}{\rightarrow}
\newcommand{\mglong}{{\sc\small Mad\-Graph5\_aMC\-@NLO}}

\newcommand{\ALOHA}{{\sc\small Aloha}}

\newcommand{\denpoz}{{\sc\small DP}}

\newcommand{\deltaEW}{\delta^{\rm EW}_{\rm LA}}
\newcommand{\deltaQCD}{\delta^{\rm QCD}_{\rm LA}}
\newcommand{\Ltop}{L^t(s)}
\def\Las#1{l^{\as}(#1)}
\newcommand{\dmtQCD}{(\de\Mt)^{\rm QCD}}
 \newcommand{\MSbar}{{\rm \overline{MS}}}

 \usepackage{comment}

\title{One-loop electroweak Sudakov logarithms:\\ a revisitation and automation}

\author[a,b]{Davide Pagani,}
\author[c]{Marco Zaro}

\affiliation[a]{INFN, Sezione di Bologna, Via Irnerio 46, 40126 Bologna, Italy}
\affiliation[b]{Deutsches Elektronen-Synchrotron DESY, Notkestr. 85, 22607 Hamburg, Germany}
\affiliation[c]{TIFLab, Universit\`a degli Studi di Milano \& INFN, Sezione di Milano, Via Celoria 16, 20133 Milano, Italy}

\emailAdd{davide.pagani@bo.infn.it}
\emailAdd{marco.zaro@mi.infn.it}

\abstract{In this work we revisit the algorithm of Denner and Pozzorini for the calculation of one-loop electroweak Sudakov logarithms and we automate it in the {\mglong} framework. We adapt the formulas for modern calculations, keeping light-quarks and photons strictly massless and dealing with infrared divergences via dimensional regularisation. We improve the approximation by taking into account additional logarithms that are angular dependent. We prove that an imaginary term has been previously omitted and we show that it cannot be in general neglected for $2\TO n$ processes with $n> 2$. We extend the algorithm to NLO EW corrections to squared matrix-elements that involve also QCD corrections on top of subleading LO terms. Furthermore, we discuss the usage of this algorithm for approximating physical observables and cross sections. We propose a new approach in which the QED component is consistently removed and we show how it can be superior to the commonly used approaches.  The relevance of all the novelties introduced in this work is corroborated by numerical results obtained for several processes in a completely automated way. We thoroughly compare exact NLO EW corrections and their Sudakov approximations both at the amplitude level and for physical observables in high-energy hadronic collisions.}

\preprint{
\begin{flushright}
DESY-21-145\\
TIF-UNIMI-2021-13\\
\end{flushright}
}

\begin{document}
\maketitle
\flushbottom

\newpage

\section{Introduction\label{sec:intro}}

After more than ten years of operation of the Large Hadron Collider (LHC), we have tremendously improved our knowledge of the fundamental interactions of elementary particles. Above all, the long-sought Higgs boson has been observed~\cite{Aad:2012tfa,Chatrchyan:2012ufa} and especially its properties have been studied in detail: they have been found to be compatible with those predicted by the Standard Model (SM) \cite{Aad:2019mbh}. In general, no clear and unambiguous sign of beyond-the-SM (BSM) physics has been found at colliders. In parallel, at the LHC the SM itself has been stress-tested in all its sectors: {\it e.g.}, electroweak (EW) interactions, QCD dynamics and flavour physics.  
 However, the BSM search programme at colliders is only at its initial phase. At the LHC, $20$ times more data will be collected in the next years, large part of it during the High-Luminosity (HL) runs \cite{Azzi:2019yne,Cepeda:2019klc,CidVidal:2018eel,Cerri:2018ypt,Citron:2018lsq,Chapon:2020heu}. Moreover, several options  are possible for future colliders (see {\it e.g.}~Ref.~\cite{Gray:2021jij} for a recent review), involving collisions at higher energies between a pair of protons or leptons (both electrons/positrons and muons).
  
The success of this ambitious research programme is interconnected with the availability of precise and reliable SM predictions. A plethora of new calculations and techniques have already appeared in the literature for improving  both SM and BSM predictions. QCD radiative corrections at fixed order, going from Next-to-Leading-Order (NLO) to Next-to-NLO (NNLO) or even Next-to-NNLO ($\rm N^3LO$), have become available and techniques for the resummation of large logarithms appearing at fixed order have also been improved. On the other hand, an enormous effort has been done for the calculations of NLO QCD and also NLO EW corrections for processes with high-multiplicity final states. In particular, such corrections have been implemented in Monte Carlo generators and they have been even automated  \cite{Kallweit:2014xda, Frixione:2015zaa, Chiesa:2015mya, Biedermann:2017yoi, Chiesa:2017gqx,  Frederix:2018nkq, Pagani:2021iwa}, at different levels in the different frameworks, using various one-loop matrix-element providers \cite{Hirschi:2011pa, Cullen:2011ac, Cascioli:2011va, Actis:2012qn, Actis:2016mpe, Denner:2017wsf, Buccioni:2019sur}.  

A particular feature of EW corrections are the so-called ``Sudakov enhancements'' or ``Sudakov logarithms'' \cite{Sudakov:1954sw}, which enhance $\ord(\alpha^n)$ fixed-order corrections, the so-called ${\rm N}^n{\rm LO}$ EW corrections, with terms of order  $\sim - \alpha^n \log^k(s/\MW^2)$ with $k$ in the range $1\le k \le 2n$. In high-energy collisions, these logarithms involve two separate ranges of energy scales:  the $W$-boson mass $\MW$ and the centre-of-mass energy $\sqrt{s}$.  Most importantly, at variance with QCD, EW Sudakov logarithms do not cancel in IR-safe physical observables  \cite{Ciafaloni:1998xg, Ciafaloni:1999ub, Ciafaloni:2000df, Manohar:2014vxa}, which typically  are  not inclusive on the additional emission of neither $W$ nor $Z$ bosons. Therefore, they induce large and negative corrections, especially in the tails of the distributions.
With the future runs of the LHC, and especially with future colliders, higher energies will be probed at higher precision and therefore a reliable evaluation of such corrections and its automation in modern Monte Carlo generators  is necessary.

From a theoretical point of view, a lot of work has already been done in the past for what concerns EW Sudakov logarithms from virtual corrections~\cite{Beccaria:1998qe,Ciafaloni:1998xg,Fadin:1999bq,Melles:2000gw,Melles:2000ed,Melles:2000ia,Melles:2001mr,Melles:2001ye,Ciafaloni:1999ub,Ciafaloni:2000df,Ciafaloni:2000rp,Ciafaloni:2000gm,Denner:2000jv,Melles:2001dh,Ciafaloni:2001vt,Ciafaloni:2001vu,Denner:2001gw,Denner:2003wi,Denner:2004iz,Denner:2006jr,Chiu:2007yn,Chiu:2007dg,Chiu:2008vv,Denner:2008yn,Chiu:2009mg}. Especially, general algorithms for the calculation
at one and two-loop accuracy were
derived  in
Refs.~\cite{Denner:2000jv,Denner:2001gw}
and~\cite{Denner:2003wi,Denner:2004iz,Denner:2006jr,Denner:2008yn},
respectively. Moreover, their resummation has also been studied~\cite{Kuhn:1999nn,Fadin:1999bq,Ciafaloni:1999ub,Beccaria:2000jz,Hori:2000tm,Ciafaloni:2000df,Denner:2000jv,Denner:2001gw,Melles:2001ye,Beenakker:2001kf,Denner:2003wi,Pozzorini:2004rm,Feucht:2004rp,Jantzen:2005xi,Jantzen:2005az,Jantzen:2006jv,Chiu:2007yn,Chiu:2008vv,Manohar:2012rs, Bauer:2017bnh}
and in Refs.~\cite{Chiu:2007yn,Chiu:2008vv, Manohar:2018kfx} a general method to resum
such logarithms for an arbitrary process was developed, based on the framework of 
soft-collinear effective
theory (SCET)~\cite{Bauer:2000ew,Bauer:2000yr,Bauer:2001ct,Bauer:2001yt}.

On the other hand, the similar case of real weak-boson emission has been addressed
in Refs.~\cite{Ciafaloni:2001vt, Ciafaloni:2000rp,Ciafaloni:2000df,Ciafaloni:2001vu,Ciafaloni:2001mu,Ciafaloni:2003xf,Ciafaloni:2005fm,Ciafaloni:2006qu,Ciafaloni:2008cr,Bell:2010gi,Ciafaloni:2010ti,Stirling:2012ak,Bauer:2016kkv}
and for specific processes on a more phenomenological level in 
Refs.~\cite{Baur:2006sn,Bell:2010gi,Bern:2012vx,Chiesa:2013yma,Christiansen:2014kba,Krauss:2014yaa,Frixione:2014qaa,Frixione:2015zaa}.
The resummation of double-logarithmic corrections from the real
radiation has also been studied in Ref.~\cite{Bauer:2016kkv} and the simulation of multiple real weak boson radiation in a parton-shower approach has already been formulated and performed  in Refs.~\cite{Christiansen:2014kba, Kleiss:2020rcg, Brooks:2021kji, Masouminia:2021kne}. The SM LO  collinear splitting functions \cite{Chen:2016wkt} and  evolution of  parton  fragmentation  functions \cite{Bauer:2018xag} are also known and phenomenological results at high energy have been studied. Finally, studies on parton-distribution-functions (PDFs) in the EW SM exist in the literature \cite{Bauer:2018arx,Fornal:2018znf}.

It is an established fact (see {\it e.g.}~section 17 of Ref.~\cite{Mangano:2016jyj}) that  EW Sudakov logarithms from both virtual corrections and real emissions are sizeable at high energies (a few TeV's) for several processes and, especially for the former class, resummation is necessary in order to achieve precise predictions or even just sensible results. Indeed,  fixed-order corrections, precisely due to the EW Sudakov logarithms, can approach the size of $-100\%$ of the LO.  In other words, in order to reach the percent accuracy, not only the exact $\ord(\alpha)$ (NLO EW) corrections have to be calculated, but the Sudakov-enhanced component must be identified and resummed. Although the computation of exact NLO EW corrections is technically more involved than the case of its virtual Sudakov-logarithm subclass, only the former has been (fully) automated and implemented in Monte-Carlo's by different collaborations \cite{Kallweit:2014xda, Frixione:2015zaa, Chiesa:2015mya, Biedermann:2017yoi, Chiesa:2017gqx,  Frederix:2018nkq, Pagani:2021iwa}. Recently, the pioneering algorithm of Denner and Pozzorini \cite{Denner:2000jv,Denner:2001gw}, which allows to calculate both single and double  one-loop virtual EW Sudakov logarithms at $\ord(\alpha)$, has been  automated for the first time \cite{Bothmann:2020sxm} in  the {\sherpa} framework~\cite{Sherpa:2019gpd}. Previously, only specific (classes of) processes had been considered, as done {\it e.g.}~in Ref.~\cite{Chiesa:2013yma} within {\alpgen}~\cite{Mangano:2002ea}. 

In this work we automate the algorithm of Denner and Pozzorini \cite{Denner:2000jv,Denner:2001gw} in the {\mglong} framework  \cite{Alwall:2014hca, Frederix:2018nkq}, which already allows for the fully-automated calculation of NLO QCD and EW corrections and more in general Complete-NLO predictions \cite{Frederix:2018nkq, Pagani:2021iwa}. Thus, with {\mglong}, it is now possible not only to calculate in a completely automated approach NLO EW corrections, but also their subcomponent  that is typically dominant: the double and single virtual Sudakov logarithms. On the one hand, this work opens up the possibility of further building an automated framework for resumming Sudakov EW logarithms and match the result to NLO EW calculations in the {\mglong} framework. On the other hand, since virtual Sudakov logarithms are the dominant component of EW corrections and their evaluation is much faster and stable of the exact $\ord(\alpha)$ result,  this implementation leads also to a very good and fast approximation of NLO EW corrections at high energy.

Before implementing the algorithm of Denner and Pozzorini \cite{Denner:2000jv,Denner:2001gw}, however, we have revisited it. This work therefore does not only describe the technical steps underlying the implementation in {\mglong}, but it also provides an algorithm that is based on the one of Denner and Pozzorini and introduces relevant novelties w.r.t.~it. First, we have reframed the algorithm by setting the mass of the photon and light-fermion masses exactly to zero, regularising IR divergences by mean of Dimensional Regularisation (DR), as in modern NLO EW calculations and in general Monte Carlo implementations. Second, we have identified an imaginary term that was omitted in Ref.~\cite{Denner:2000jv}, which however cannot be in general neglected for $2\TO n$ processes with $n> 2$. Third, we have modified part of the expressions in order to take into account additional angular dependences, without assuming that all the invariants are of the same size of $s$.
Fourth, since (virtual) NLO EW corrections originate from both ``genuine'' EW corrections  on top of the dominant LO contributions and QCD corrections on top of the subdominant ones, we provide additional terms for taking into account the logarithmic dependence not only of the former, as done in the original work of Denner and Pozzorini, but also of the latter. All these modifications concern the approximation of the matrix elements or, more precisely, the interference of tree-level and renormalised one-loop amplitudes leading to the ultra-violet (UV) finite virtual corrections. The fourth point mentioned before was not considered in the pioneering work of Ref.~\cite{Denner:2000jv} precisely because therein the focus was on the one-loop {\it amplitudes} and not their {\it interferences} with tree-level amplitudes, which is what yields the virtual contribution to NLO EW corrections. The effects of all these modifications and their validation are then showcased presenting numerical results obtained via the implementation in {\mglong}. In a fully automated way, for several different processes, we compare exact results for virtual contributions obtained via  \ml~\cite{Hirschi:2011pa}, one of the modules of {\mglong}, and via the new implementation of the modified algorithm of Denner and Pozzorini for calculating one-loop virtual Sudakov logarithms.

Besides purely virtual contributions, which are unphysical and IR divergent, we also show comparisons between the Sudakov approximation and   the exact NLO EW corrections for production processes in proton--proton collisions, taking into account also the necessary additional terms to achieve IR finiteness: real emission of massless particles (photons, quarks and gluons) and PDF counter-terms. We show how, for a large class of processes and IR-finite observables, at variance with the case of virtual amplitudes, the exclusion of  the contribution of photons from the Denner and Pozzorini  algorithm \cite{Denner:2000jv,Denner:2001gw} leads to a better approximation of NLO EW corrections. We describe in detail how the algorithm has to be altered in order to exclude the QED component (photons) and keep the purely weak one ($W$ and $Z$ bosons).

The article is organised as follows. In Sec.~\ref{sec:revisitation} we revisit the work of Denner and Pozzorini \cite{Denner:2000jv,Denner:2001gw} at the pure amplitude level. We set the notation, using as much as possible the same one of Ref.~\cite{Denner:2000jv}, and we introduce three of the main novelties mentioned before: the formulation with strictly massless light-fermions and photons, the missing imaginary term, and additional terms that better take into account angular dependences and differences among the invariants. In Sec.~\ref{sec:NLOew} we move to the virtual NLO EW level, considering the interference of tree-level and one-loop amplitudes. Therein, we provide the additional term for taking into account logarithms of QCD origin. In Sec.~\ref{sec:SudMC} we present a modification of the algorithm such that the contribution of photons and gluons is excluded.  We discuss how this can be a superior approach for approximating physical (IR-safe) cross sections at NLO EW accuracy.
In Sec.~\ref{sec:implementation} we describe the technical steps for the implementation of the algorithm in the {\mglong} framework. We explain in detail the procedure for the generation of all the additional amplitudes that are necessary for the evaluation of the Sudakov logarithms and especially for the evaluation of the amplitudes themselves. This procedure requires new features of the code, such as the possibility to evaluate the interference between amplitudes 
with different external legs or numerical derivatives of the matrix elements.
In Sec.~\ref{sec:Res_Amp} we provide numerical results comparing NLO EW virtual contributions obtained via  {\ml} and via the new implementation of  the Sudakov approximation. We show the relevance of the novelties introduced w.r.t.~Ref.~\cite{Denner:2000jv} and at the same time we validate the new implementation. In Sec.~\ref{sec:Res_MC} we repeat a similar comparison for physical observables from selected hadronic production processes in high-energy collisions. We discuss the numerical results and show how the exclusion of the contribution of photons leads to better approximations of the NLO EW results. Both in Sec.~\ref{sec:Res_Amp} and Sec.~\ref{sec:Res_MC} results are obtained in a completely automated way via the new implementation in the {\mglong} framework. We give our conclusions and outlook in Sec.~\ref{sec:conclusions}.

\section{ The  Denner and Pozzorini algorithm revisited }
\label{sec:revisitation}

We start by revisiting the pioneering work of Denner and Pozzorini \cite{Denner:2000jv,Denner:2001gw}, which provides an algorithm for calculating one-loop EW double-logarithmic (DL) and single-logarithmic (SL) corrections of the form 
\beq
\frac{\alpha}{4\pi }\log^2{\frac{s}{\MW^2}} \qquad{\rm and}\qquad \frac{\alpha}{4\pi }\log{\frac{s}{\MW^2}}\,,  \label{eq:formallogs}
\eeq
for any individual helicity configuration of a generic SM partonic processes.
We introduce three novel features w.r.t.~the algorithm of Ref.~\cite{Denner:2000jv}, which we will denoted in the following as  {\denpoz}  algorithm:
\begin{enumerate}
   \item In all formulas light-fermions, photons and gluons are strictly massless, as in modern higher-order calculations. In other words, IR divergences are regularised via DR, introducing a IR-regularisation scale $Q$.
   \item We correct the expressions for the Subleading Soft-Collinear terms (see Secs.~\ref{sec:logsplit} and \ref{sec:SSC}), taking into account imaginary contributions that are  relevant for $2\TO n$ processes with $n>2$.   
\item We keep track of  terms that are proportional to the (squared) logarithm of the ratio between two invariants that can be built via two different pairs of momenta among those of the external particles.  Namely, using the notation that will be introduced later in this section, the terms proportional to $\log\frac{r_{k l}}{ r_{k' l'}}$ and $\log^2\frac{r_{k l}}{ r_{k' l'}}$.
  \end{enumerate} 
In this section, we will try to use as much as possible the notation of Ref.~\cite{Denner:2000jv}, where the {\denpoz} algorithm has been formulated. In this way the reader can easily detect the differences introduced in our work. Moreover, we will try to avoid unnecessary repetitions of the content of Ref.~\cite{Denner:2000jv}, but we will also define all the quantities that are entering the actual implementation in {\mglong}, which is then discussed in Sec.~\ref{sec:implementation}.  

\subsection{Range of validity and conventions}

The  {\denpoz} algorithm strictly relies on the assumption that  processes with on-shell external legs are considered and, especially, that all invariants are much larger than the gauge-boson masses.
In other words, if $k$ and $l$ are two generic external particles with momenta $p_k$ and $p_l$ respectively, then 
\beq \label{eq:Sudaklim} 
r_{kl}\equiv(p_k+p_l)^2 \simeq 2p_kp_l \gg \MW^2 \simeq \MH^2,\Mt^2,\MW^2,\MZ^2.
\eeq
It is interesting to notice that the condition \eqref{eq:Sudaklim} still allows for kinematic configurations with $r_{k l}\gg r_{k' l'}\gg \MW^2$, where the quantities $r_{k l}$ and $r_{k' l'}$ represent a generic pair of the many possible invariants that one can build with two external momenta. However,
since the required formal accuracy consists of the DL and SL in \eqref{eq:formallogs}, although logarithms of the form 
\beq
\frac{\alpha}{4\pi }\log^2{\frac{r_{k l}}{r_{k' l'}}} \qquad{\rm and}\qquad  \frac{\alpha}{4\pi }\log{\frac{r_{k l}}{r_{k' l'}}}\,,  
\eeq 
are present at $\ord(\alpha)$ and  can be non-negligible for configurations with  $r_{k l}\gg r_{k' l'}\gg \MW ^2$, they are not taken into account. In other words, the algorithm assumes the regime \eqref{eq:Sudaklim}, but large logarithms may be anyway not captured unless the condition
\beq \label{eq:rijnice} 
r_{k l}/ r_{k' l'}\simeq 1
\eeq
is satisfied for any possible pair of $r_{k l}$ and $ r_{k' l'}$ invariants.

In fact, condition \eqref{eq:rijnice} is quite unrealistic for actual calculations in collider physics, since  cross sections are dominated precisely by regions where one or more $r_{kl}$ invariants tend to be much smaller than $s\equiv r_{12} \gg \MW^2$. Indeed, the $r_{kl}$ invariants are related to those entering the propagators. Even if cuts are devised in order to maximise any possible value of $r_{kl}$ for a given $s$, 
the fulfilment of condition \eqref{eq:rijnice} is strictly impossible. For instance, if \eqref{eq:Sudaklim} is valid, one has that $\min(r_{kl}/s)<0.5$ for a $2\TO2$ process. This bound is even tighter and tighter for a generic $2\TO n$ process with $n$ growing.\footnote{\label{footnote1} The determination of the
    configuration where the smallest invariant is maximal in a $2\TO n$ process is related to the determination of the largest possible value for the minimum 
    angle between any two of the $n$ final-state momenta. This is the typical example of  a mathematical problem that it is easy to define and with a solution that is far from trivial. See for example \url{http://neilsloane.com/packings/index.html\#I}.}

It is worth to remind the reader an important limitation of the {\denpoz} algorithm. For a given process, at least one helicity configuration of the matrix element must not be mass suppressed, {\it i.e.}, it must not vanish in the limit $\MW^2/s\TO 0$.\footnote{
An equivalent formulation of this condition is that the scaling of the matrix element $\mathcal M$ with the centre-of-mass energy $\sqrt s$ must coincide with
what one expects from dimensional analysis: a non mass-suppressed helicity configuration of a matrix elements with $n$ external legs should scale as $\sqrt s ^{4-n}$. See footnote~\ref{foot:vbf} for a counterexample.} Indeed, such an assumption is one of the hypotheses under which the algorithm has been derived.
On the other hand, most of the processes do satisfy this hypothesis, having at least one helicity configuration that is {\it not} mass suppressed\footnote{\label{foot:vbf} Exceptions are possible, an important one is Higgs production via vector-boson fusion. Dimensional analysis for a $2\TO3$ matrix element
    requires $ [\mathcal M ] = \gev^{-1}$, and for this specific process the matrix element  scales with the energy as $ \mathcal M  \propto \frac{\MW}{s}$.}. Moreover, thanks to the condition \eqref{eq:Sudaklim},  helicity configurations that are {\it not} mass suppressed are by definition also dominant in the kinematic regime considered. The condition \eqref{eq:Sudaklim} also implies that processes including unstable particles and their decays cannot be treated in this approximation if physical observables are dominated by resonant configurations. Rather, the process without decays should be first considered and the decays should be then taken into account only after applying the {\denpoz} algorithm.

Being aware of all the possible limitations given by the conditions \eqref{eq:Sudaklim} and \eqref{eq:rijnice}, we describe the {\denpoz} algorithm and some modifications we have introduced in order to achieve the formal leading and subleading logarithmic accuracy (LA), {\it i.e.}, taking
into account only enhanced DL and SL terms of the form \eqref{eq:formallogs}, for one-loop EW virtual corrections to any SM amplitudes, in DR and therefore with possibly massless particles. The problems related to the validity of condition \eqref{eq:rijnice} will be also addressed, giving a pragmatic solution.

The starting point of the {\denpoz} algorithm is that since all the terms considered are logarithmic, they can be expressed via the quantities  
\beq
\LrM\equiv\frac{\alpha}{4\pi}\log^2{\frac{|r_{kl}|}{M^2}} \qquad{\rm and}\qquad 
\lrMwithabs\equiv\frac{\alpha}{4\pi}\log{\frac{|r_{kl}|}{M^2}}\,, \label{eq:generallogs}
\eeq
where $r_{kl}$ can be any of the invariants\footnote{As it will be also explained later (see eq.~\eqref{eq:process}), the {\denpoz} algorithm is derived for $n\TO 0$ processes with all the momenta incoming, but it can be easily adapted to the usual $2\TO n-2$ processes via crossing symmetry. Momentum conservation therefore  implies that some of the momenta must have, {\it e.g.}, negative energy and that  some of the $r_{jk}$ are negative. For instance, crossing a $4\TO 0$ process into a  $2\TO 2$ process $r_{13}=(p_1+(-p_3))^2=t$. } and $M$ any of the masses among $\MW, \MH,$  $\Mt$ and $\MZ$, depending on the associated Feynman diagrams. Moreover, in the case of massless particles, the regularisation of the divergences will lead to logarithms of the form \eqref{eq:generallogs} where $M\TO Q$ and $Q$ is the IR-regularisation scale. The most important point, in order to understand the novelties introduced in this section, is that the {\denpoz} algorithm splits twice the logarithms of the form in \eqref{eq:generallogs}; both splittings are connected to the modifications of the {\denpoz} algorithm that we present in this work.

 First, logarithms of the form in \eqref{eq:generallogs} are split into two classes: a symmetric and solely energy-dependent class, which is associated to the scales $\MW$ and $\sqrt{s}$ and parametrised by the quantities
\beq \label{dslogs}
\Ls\equiv\LsW \qquad{\rm and}\qquad 
\ls\equiv\lsW\,,
\eeq
and  a remaining class of logarithms involving mass ratios and ratios of invariants. This splitting involves the imaginary component that we are going to introduce in the formulas and that is not present in Ref.~\cite{Denner:2000jv}. It also involves the modifications that take care of the violation of condition \eqref{eq:rijnice}.

Second, while above the scale $\MW$ all one-loop EW contributions are treated in an unified approach, without separating purely QED from purely weak effects, below the $\MW$ scale only the QED component is present, involving logarithms between $\MW$ and the IR scale. In other words, for the contribution from QED loops $\MW$ works  as a technical separator. Above $\MW$ we have for example (see eq.~\eqref{eq:deSC}) quantities parametrised via the electroweak Casimir operator $\cew$, which involves the entire $\rm SU(2)\times U(1)$ group, while below $\MW$ we have only quantities that involve the charges $Q_k$ of the external particles. The latter class of contributions is denoted by the apex ``em'', standing for electromagnetic, and in Ref.~\cite{Denner:2000jv} it arises from the energy hierarchy  $\MH,\Mt,\MW,\MZ \gg m_{f\ne\Pt} \gg  \lambda$, where $\lambda$ is the mass of the photon. In this separation the logarithms $\lWZ$, $\ltW$, and $\lHW$, similarly to the quantities $\log{(|r_{kl}|/s)}$ and
$\log^2{(|r_{kl}|/s)}$, are neglected when they do not multiply the term $\ls$. 

 In Ref.~\cite{Denner:2000jv} all the quantities denoted as electromagnetic (``em'') depend on $m_{f\ne\Pt}$ and $\lambda$, which here are considered exactly equal to zero,
\beq
m_{f\ne\Pt} =  \lambda=0\,.\label{eq:massless}
\eeq
The consequence of \eqref{eq:massless} is that DR becomes necessary. In $D=4-2\epsilon$ dimensions, electromagnetic logarithms transform into $1/\epsilon$ poles plus finite terms and logarithms involving the IR-regularisation scale $Q$. In this context we are not interested in the structure of the IR poles, which for NLO EW corrections is discussed, {\it e.g.}, in Refs.~\cite{Frederix:2018nkq, Schonherr:2017qcj, Pagani:2021iwa, Buccioni:2019sur}; we are interested only in the logarithmic dependence of the finite part. This can be simply derived via the substitutions
\beq
 \log(\lambda^2)\rightarrow\log(Q^2)\,, \qquad \log(m^2_{f\ne\Pt})\rightarrow\log(Q^2)\, , \label{eq:subml}
\eeq
in the expressions of Ref.~\cite{Denner:2000jv}. 

We want to comment on the dependence on $Q$, the IR-regularisation scale, which is introduced here and it is not present in the original {\denpoz} algorithm.
The derivation of the formulas in  Ref.~\cite{Denner:2000jv} depends on the assumption that $\mu^2=s$, but therein $\mu$ is the UV-regularisation scale since all the IR divergences are regularised via $m_{f\ne\Pt}$ and $\lambda$.  However, similarly to this work, therein formulas have been derived assuming an on-shell renormalisation scheme, such as the $\alpha(\MZ)$ or $G_\mu$ ones. With such a renormalisation scheme, no {\it renormalisation}-scale dependence is present for one-loop renormalised amplitudes, both if exactly calculated or using the LA. Therefore, the {\denpoz} algorithm, although derived assuming a specific value for $\mu $ ($\mu^2=s$),  returns results that do not depend on $\mu$. The substitution in \eqref{eq:subml}, that we perform due to the condition \eqref{eq:massless}, does not depend on the condition \eqref{eq:Sudaklim}. Moreover, it affects only the regularisation of IR divergences and does not concern the UV ones. Therefore, this substitution introduces the correct dependence on $Q$ even if a common regulator for UV and IR divergences ($Q=\mu$) is used. Exceptions are discussed in Sec.~\ref{sec:NLOew}. 

Before providing the expressions necessary for automating one-loop EW Sudakov logarithms, we introduce further conventions and notations according to Ref.~\cite{Denner:2000jv}.
Amplitudes are assumed with $n$ arbitrary external particles and all momenta
$p_k$  incoming. Needless to say, any $2\TO n-2$ amplitude can be rewritten into a $n\TO 0$ amplitude via crossing symmetry. 
Processes are denoted as
\beq \label{eq:process}
\varphi_{i_1}(p_1)\dots \varphi_{i_n}(p_n)\rightarrow 0\,,
\eeq
where the (anti)particles  $\varphi_{i_k}$ are the
components of the various multiplets
$\varphi$ of the SM:
\begin{itemize}
\item $f^\kappa_\si$ and $\bar{f}^\kappa_\si$: chiral fermions and antifermions, with chiralities
$\kappa=\rR,\rL$ and the isospin indices $\si=\pm$,
\item $\GB_a=\PA,\PZ,\PWpm$: gauge bosons transversely (T) or
longitudinally (L) polarised. Neutral gauge bosons are also denoted as $\NB=\PA,\PZ$,
\item $\Phi_i$: the scalar doublet containing the Higgs particle $\PH$ and the
 neutral and charged Goldstone bosons $\chi,\phi^\pm$.
\end{itemize}

An important technical point of the {\denpoz} algorithm is that, since the high-energy limit is assumed, the Goldstone-boson equivalence theorem can be used. In fact, with this algorithm, contributions from longitudinal gauge-bosons  are always evaluated via the Goldstone-boson equivalence theorem. We will return to this point in Sec.~\ref{sec:generation}.

Following the same notation of Ref.~\cite{Denner:2000jv}, the couplings of each external field  $\varphi_{i_k}$  to the gauge bosons
$\GB_a$ is denoted by $\ri eI^{\GB_a}(\varphi)$, namely,
 $\ri eI^{\GB_a}_{\varphi_i\varphi_{i'}}(\varphi)$ is the
coupling corresponding to the $\GB_a\bar{\varphi}_i\varphi_{i'}$ vertex,
with all fields that are incoming. For simplicity, in the formulas 
 the components $\varphi_{i_k}$ are
replaced by their indices $i_k$, namely, $I^a_{i_ki'_k}(k)$.  All the values and formulas for the quantities $I^a_{i_ki'_k}(k)$, as many other terms appearing in the next sections are reported in detail in the appendices of Ref.~\cite{Denner:2000jv}. We do not repeat them here, but we want to warn the reader that the same exact conventions for Feynman rules have to be used in order obtain consistent results.

 For any process denoted as in \refeq{eq:process}, the Born matrix element reads
\begin{equation}
\M_0^{i_1 \ldots i_n}(p_1,\ldots, p_n). \label{eq:M0}
\end{equation}
The $\ord(\alpha)$ corrections to $\M_0$ in LA,  $\delta \M$, has the form  
\beq \label{LAfactorization} 
\delta \M^{i_1 \ldots i_n}(p_1,\ldots, p_n)= 
\M_0^{i'_1 \ldots i'_n}(p_1, \ldots, p_n)\delta_{i'_1i_1 \ldots i'_ni_n}.
\end{equation}
Equation \refeq{LAfactorization} means that the result can be written in a factorised form, but that involves Born amplitudes for different processes.
The contributions to $\delta \M$ have different origins:
\beq
\delta=\de^{\SC}+\de^{\SS}+\de^{\cc}+\de^\pre. \label{eq:deltatodeltas}
\eeq
The quantities  $\de^\SC$ and $\de^\SS$ are respectively the leading and subleading soft-collinear logarithms. They both emerge from the DL, which in turn originate from the eikonal approximation of one-loop diagrams  where  gauge bosons are exchanged between 
  external legs and are soft-collinear. The former represents the symmetric and solely energy-dependent class of logarithms, while the latter  involves mass ratios and ratios of invariants. 
 The quantity
$\de^\cc$  consists of the collinear logarithms, originating from virtual collinear gauge bosons from external
  lines and field renormalisation
constants. The logarithms resulting from parameter
renormalisation, which can be determined by the running of the
couplings, correspond to the term $\de^\pre$. 
In the case of longitudinally polarised bosons the equivalences
\beqar \label{eq:borneet} 
\M_0^{\ldots \PW^\pm_\rL \ldots} &=&\M_0^{\ldots \phi^\pm \ldots},\nl
\M_0^{\ldots \PZ_\rL \ldots} &=&\ri\M_0^{\ldots \chi \ldots},
\eeqar
are used and can be applied also for what concerns the different terms entering the definition of $\delta$.

In the following subsections we provide the formulas entering the implementation in {\mglong}, which is described in Sec.~\ref{sec:implementation}. We will discuss in detail only the aspects concerning the differences w.r.t.~Ref.~\cite{Denner:2000jv}.

\subsection{Logarithm splittings}
\label{sec:logsplit}

As already mentioned, the DL corrections come from loop diagrams with virtual gauge
bosons $\GB_a=\PA,\PZ,\PW^\pm$ 
connecting two external
legs. In particular, they originate from regions where the gauge boson is  soft and collinear to one of the external
legs. Their expressions can be derived by evaluating them in the eikonal
approximation. 

In Ref.~\cite{Denner:2000jv}, DL have been in general identified as
\beqar \label{eq:angsplit}
\LrM&=&\LsM+2\lsM\lrs+\Lrs \nonumber\\
         &=&\Ls+2\ls \log\frac{\MW^2}{M^2}+2\ls \lrs+\cdots
\eeqar
where the invariant $r_{kl}$ depends on the angle
between the momenta $p_k$ and $p_l$.
Equation \eqref{eq:angsplit} precisely represents the first of the logarithm splittings that has been mentioned before.  In the first line of $\eqref{eq:angsplit}$ the quantity $\LrM$ is split into $\LsM$, which is symmetric and energy dependent, and other two terms, of which the second can be neglected in the approximation \eqref{eq:rijnice}. Moving to the second line, the remaining terms are further rearranged such that if $M\neq \MW$, the mass-ratio logarithm $\log\frac{\MW^2}{M^2}$ is kept only when multiplying $\ls$. The dots at the end stand for the terms that are dropped in the splitting of the logarithms. In Ref.~\cite{Denner:2000jv}, the first two terms in the second line of eq.~\eqref{eq:angsplit} are identified as the leading soft-collinear ($\SC$)
contribution, which as already mentioned is angular-independent and involves only the $s/\MW^2$ ratio in the logarithms. The remaining term leads to the angular-dependent   subleading soft-collinear ($\SS$) contribution.

 When loop diagrams with virtual gauge
bosons $\GB_a=\PA,\PZ,\PW^\pm$ connecting two external legs are evaluated in the eikonal approximation, the logarithmic dependence can be derived by the expansion of the $C_0$ function in the high-energy limit, namely condition \eqref{eq:Sudaklim}. The expression can be found in Ref.~\cite{Roth:1996pd}. If the gauge boson $\GB$ with mass $M$ is exchanged between the external particles $\phi_k$ and $\phi_l$, the relevant quantity is, following the conventions of Ref.~\cite{Roth:1996pd},  
\beqar \label{eq:C0}
C_0(p_k,p_l,M,M_k,M_l)\propto \frac{1}{r_{kl}}\left(\log^2\frac{| r_{kl} |}{M^2}-2 i \pi \Theta(r_{kl})\log\frac{| r_{kl} |}{M^2} \right)\, ,
\eeqar
where $\Theta$ is the Heaviside step function.
It is then clear that rather than starting from $\LrM$ as in eq.~\eqref{eq:angsplit} the correct quantity to be taken into account is
\beqar \label{eq:angsplitnew}
\LrM-2 i \pi \Theta(r_{kl})\lrMwithabs\, .
\eeqar
The difference is an imaginary component that involves a term proportional to $\ls$. For $2\TO2$ processes, this is completely irrelevant and therefore all the results presented for specific processes in Ref.~\cite{Denner:2000jv} are not affected by this additional term. Indeed, since $2\TO2$
tree-level amplitudes are always real (as a consequence of the optical theorem), the imaginary part of the one-loop (or Sudakov-approximated) amplitude drops out when the real 
part of the loop-tree interference is considered. However, this is no longer the case starting from $2\TO3$ processes, and indeed we do find that this imaginary part is not irrelevant. 
We therefore repeat the procedure of eq.~\eqref{eq:angsplit} in order to identify how the impact of the term $2 i \pi \Theta(r_{kl})$ translates into the {\denpoz} algorithm. Moreover, we keep track of the terms that would be otherwise discarded assuming condition \eqref{eq:rijnice}.

Starting from \eqref{eq:angsplitnew} we  obtain 
\beqar 
&&\LrM-2 i \pi \Theta(r_{kl})\lrMwithabs= \nonumber \\
&&=\LsM+2\lsM\left(\lrs- i  \pi \Theta(r_{kl})\right)+ \Lrs-2 i  \pi \Theta(r_{kl}) \lrsalpha  =\nonumber \\
&&=\underbrace{\Ls+2\ls \log\frac{\MW^2}{M^2}}_\text{LSC} +\underbrace{2\ls \left(\lrs- i  \pi \Theta(r_{kl})\right)}_\text{SSC}+ \label{eq:angsplit_new_sm} \\
&&\qquad\underbrace{2\lWM\lrs +\Lrs-2 i  \pi \Theta(r_{kl}) \lrsalpha}_{\SS^{s\TO r_{kl}}} +\cdots \nonumber 
\eeqar
where we have dropped in the splitting of the logarithms only terms involving neither $s$ nor $r_{kl}$.\footnote{These terms are $L(\MW^2,M^2)$ and $-i\pi \Theta(r_{kl}) l(\MW^2,M^2)$, which are indeed neglected unless the vector boson is the photon and $M^2\rightarrow Q^2$. In that case these contributions are retained. The former, together with the term $2\ls \log\frac{\MW^2}{M^2}$ from the LSC, is entering the definition of $\Lemknew$ in eq.~\eqref{eq:Lemknew}. The latter, again only for the photons, enters directly eq.~\eqref{eq:subdl1} together with the term $2\lWM\lrs$ from the $\SS^{s\TO r_{kl}}$. } 
In the third line of eq.~\eqref{eq:angsplit_new_sm} there are terms that are relevant for the formal expansion in LA, {\it i.e.}, the correct expression to be used instead of \eqref{eq:angsplit}. The first two terms in the sum give the $\SC$  logarithms, while the third one contributes to the $\SS$ ones. On the contrary in the fourth line  there are further terms that become relevant when $s\gg r_{kl}\gg M$, {\it i.e.}, departing from condition \eqref{eq:rijnice}. Formally, they do not enter the LA so they cannot be identified neither as LSC nor as SSC. On the other hand,  since they depend on $r_{kl}$, we will take into account their contribution in the expression of the $\SS$ logarithms (Sec.~\ref{sec:SSC}). For this reason we have denoted them in  eq.~\eqref{eq:angsplit_new_sm} as $\SS^{s\TO r_{kl}}$. 

As we will discuss in more detail in Sec.~\ref{sec:rijvss}, even taking into account the $\SS^{s\TO r_{kl}}$ contribution, the full control  of logarithms  involving the ratios of  $|r_{kl}|$ invariants and $s$ cannot be achieved via the {\denpoz} algorithm. We will discuss the case of a specific process for which this limitation is manifest. On the other hand, several numerical results in Sec.~\ref{sec:rijvss} and Sec.~\ref{sec:Res_MC}  clearly show how the inclusion of the $\SS^{s\TO r_{kl}}$ terms substantially improves the approximation of such class of logarithms and in turn of EW virtual one-loop corrections at high energy.

\subsection{LSC: Leading soft-collinear contributions}

The $\SC$ logarithms
can be rearranged as a single sum over the external legs,
\beq \label{SCsum}
\de^{\SC} \M^{i_1 \ldots i_n} =\sum_{k=1}^n \delta^\SC_{i'_ki_k}(k)
\M_0^{i_1 \ldots i'_k\ldots i_n}\,,
\end{equation} 
where $\de^\SC_{i'_ki_k}(k)$ reads
\beq \label{eq:deSC} 
\de^\SC_{i'_ki_k}(k)=- \frac{1}{2}\left[ C^{\ew}_{i'_ki_k}(k)\Ls -2(I^Z(k))_{i'_ki_k}^2 \log{\frac{\MZ^2}{\MW^2}}\, \ls+\de_{i'_ki_k} Q_k^2\Lemknew \right].
\eeq
In this case, besides the term $\Lemknew$, the expression is the same as Ref.~\cite{Denner:2000jv}. The expressions for the electroweak Casimir operator $C^{\ew}_{i'_ki_k}(k)$, the squared $Z$-boson coupling $(I^Z(k))_{i'_ki_k}^2$, and the charge $Q_k^2$ for a generic particle $k$ and a specific polarisation can be found in Ref.~\cite{Denner:2000jv}. It is important to note that the first two quantities have indexes and can be non-diagonal. We will return to this point discussing the implementation in {\mglong}.
Using DR the electromagnetic DL reads
\beq
\Lemknew\equiv 2\ls\log{\left(\frac{\MW^2}{Q^2} \right)}+\LWlanew-\Lklanew\, ,
\label{eq:Lemknew}
\eeq
with
\beq
\Lklanew\equiv\begin{cases*}
      0 & if $m_k^2=0$ \, ,\\
      L(m_k^2, Q^2)        & otherwise \, .
    \end{cases*}
\eeq

\subsection{SSC: Subleading soft-collinear contributions}
\label{sec:SSC}
Unlike the $\SC$ terms, the $\SS$ ones remain a sum over pairs of external legs of the form  

\beq \label{SScorr}
\de^\SS \M^{i_1 \ldots i_n} =\sum_{k=1}^n
\sum_{l<k}\sum_{\GB_a=A,Z,W^\pm}\delta^{\GB_a,\SS}_{i'_ki_k i'_li_l}(k,l)
\M_0^{i_1\ldots i'_k\ldots i'_l\ldots i_n}\,.
\eeq
This part is the one with the largest differences w.r.t.~Ref.~\cite{Denner:2000jv}.
The exchange of soft neutral gauge
bosons contributes with
\beqar \label{eq:subdl1} 
\de^{A,\SS}_{i'_ki_k i'_li_l}(k,l)&=&
\left[2 \left(\ls+\lWlanew \right)\left(\lrs - i  \pi \Theta(r_{kl}) \right) +\Delta^{s\TO r_{kl}}(r_{kl},\MW^2) \right]I_{i'_ki_k}^A(k)I_{i'_li_l}^A(l),\nl
\delta^{Z,\SS}_{i'_ki_k i'_li_l}(k,l)&=&
\left[2\ls \left(\lrs - i  \pi \Theta(r_{kl}) \right) +\Delta^{s\TO r_{kl}}(r_{kl},\MZ^2) \right] I_{i'_ki_k}^Z(k)I_{i'_li_l}^Z(l),
\eeqar
and charged gauge bosons yields
\beq \label{subdl2} 
\delta^{W^\pm,\SS}_{i'_ki_k i'_li_l} (k,l)=\left[2\ls \left(\lrs - i  \pi \Theta(r_{kl}) \right) +\Delta^{s\TO r_{kl}}(r_{kl},\MW^2)\right] I_{i'_ki_k}^\pm(k)I_{i'_li_l}^{\mp}(l),
\eeq
The quantity $\Delta^{s\TO r_{kl}}(r_{kl},M^2)$ is set equal to zero when the condition $\eqref{eq:rijnice}$ is assumed and the LA is applied in a strict sense, as done in Ref.~\cite{Denner:2000jv}.
Taking instead into account the fact that $s\gg r_{kl} \gg M^2$, this quantity reads 
\beq
\Delta^{s\TO r_{kl}}(r_{kl},M^2)\equiv \Lrs+ 2\lWM\lrs -2 i  \pi \Theta(r_{kl}) \lrsalpha\,, \label{eq:deltasr}
\eeq
and precisely corresponds to the $\SS^{s\TO r_{kl}}$ logarithms of eq.~\eqref{eq:angsplit_new_sm}.

The quantities $I^A$, $I^Z$ and $I^\pm$ are the couplings with respectively the photon, the $Z$ boson and the $W^\pm$ boson, where we have omitted the indices $i'_ji_j$. While $I^A$ is always diagonal in these indices,  $I^Z$ can be non-diagonal and  $I^\pm(k)$ is always off diagonal. The impact of the new imaginary terms proportional to $i  \pi \Theta(r_{kl})$ on results obtained with the {\denpoz} algorithm  is directly connected to the aforementioned off-diagonal structures. Indeed the virtual contribution to NLO EW corrections involves terms of the form $ 2\Re(\M_0\delta\M^*)$, where 
 \beq
 2\Re\left(\M^{i_1 \ldots i_n}_0(\delta\M^{i_1 \ldots i_n})^*\right)\supset 2\Re\left(\M^{i_1 \ldots i_n}_0 \left(\delta^{\GB_a,\SS}_{i'_ki_k i'_li_l}(k,l)
\M_0^{i_1\ldots i'_k\ldots i'_l\ldots i_n}\right)^*\right)\,. \label{eq:imsource}
 \eeq 
 If the $I^{V_a}$ entering eq.~\eqref{eq:imsource} via $\delta^{\GB_a,\SS}$ is diagonal  or both $\M^{i_1 \ldots i_n}_0 $ and $\M_0^{i_1\ldots i'_k\ldots i'_l\ldots i_n}$ are real, like in $2\TO 2$ processes, the 
contributions of imaginary terms proportional to $i  \pi \Theta(r_{kl})$ vanish, otherwise they formally contribute. It is also interesting to note that with DR and  massless photons, setting $Q^2=s$ the entire  $\de^{A,\SS}$ contribution vanishes if we also set $\Delta^{s\TO r_{kl}}(r_{kl},\MW^2)=0$. This can be seen from the definition of $\de^{A,\SS}$ in eq.~\eqref{eq:subdl1}. This argument will also be recalled in Sec.~\ref{sec:QCD}, where the QCD contribution to NLO EW corrections to squared matrix-element is discussed.

\subsection{C: Collinear and soft single logarithms}
\label{sec:soft-or-coll}
In this section we provide the results obtained in Ref.~\cite{Denner:2000jv}, adapting them for the case with massless light-fermions and photons.
The formula for the collinear and soft single logarithms
can be written as a sum over the external particles and polarisations,
\beq\label{subllogfact}
\de^{\cc} \M^{i_1 \ldots i_n} =\sum_{k=1}^n \delta^\cc_{i'_ki_k}(k)
\M_0^{i_1 \ldots i'_k \ldots i_n}\, ,
\eeq
with $ \delta^\cc_{i'_ki_k}(k)$ that depends on the external particle and polarisation
$\varphi_{i_k}$.
 We provide the results in the following. The expressions for all the new terms introduced in the formulas  can be found in Ref.~\cite{Denner:2000jv}.

\subsubsection*{Chiral fermions}
Considering fermions $f^\kappa_\si$ with chirality $\kappa=\rR,\rL$ and isospin indices $\si=\pm$, the result is
\beq \label{eq:deccfer}
\de^{\cc}_{f_\si f_{\si'}}(f^\kappa)=\de_{\si\si'}\left\{\left[\frac{3}{2} \cew_{f^\kappa} -\frac{1}{8\sw^2}\left((1+\delta_{\kappa \rR})\frac{m_{f_\si}^2}{\MW^2}+\delta_{\kappa \rL}\frac{m_{f_{-\si}}^2}{\MW^2}\right)\right]\ls+Q_{f_\si}^2\lemfsi\right\},
\eeq
where the purely electromagnetic logarithms read
\beq \label{eq:lemfnew}
\lemf\equiv\frac{1}{2}\lWfnew +\lWlanew \, ,
\eeq
with
\beq
\lWfnew\equiv\begin{cases*}
      l(\MW^2, Q^2) & if $m_f^2=0$ \, , \\
      l(\MW^2, m_f^2)        & otherwise\, .
    \end{cases*}
\eeq

\subsubsection*{Transverse charged gauge bosons W}
The
result is
\beq \label{deccWT}
\delta^\cc_{W^\si W^{\si'}}(\GB_{\rT})=\de_{\si\si'}\left[\frac{1}{2}\bew_{W}\ls +Q_\PW^2\lemW\right] \, ,
\eeq
where  $\bew_{W}$ is a coefficient of the $\beta$-function.
 
\subsection*{Transverse neutral gauge bosons A,Z}
\newcommand{\antikro}{E}
\newcommand{\kroAA}{\de_{\NB A}\de_{\NB'A}}
\newcommand{\kroZZ}{\de_{\NB Z}\de_{\NB'Z}}

The results for symmetric and antisymmetric parts are expressed in
terms of the coefficients $\bew_{\NB\NB'}$ of the $\beta$-function.  
The result is
\beq \label{deccVVT} 
\de^\cc_{\NB'\NB}(\GB_{\rT})= \frac{1}{2}\left[\antikro_{\NB'\NB}\bew_{AZ} +\bew_{\NB'\NB} \right]\ls +\frac{1}{2}\kroAA \de Z^\elm_{AA}. 
\eeq
where the off-diagonal $\beta$-function $\bew_{\NB'\NB}$ coefficient is entering the expression. Since $E_{AZ}=-E_{ZA}=1$  the off-diagonal components read
\beq
\de^\cc_{AZ}(\GB_{\rT})=\bew_{AZ}\ls,\qquad
\de^\cc_{ZA}(\GB_{\rT})=0.
\eeq
The quantity $Z^\elm_{AA}$ in DR reads
\beq
\de Z^\elm_{AA}=-\frac{4}{3}\sum_{f,i,\si \neq t} \NCf Q_{f_\si}^2 \lWlanew \,.
\eeq

\subsection*{Longitudinally polarised gauge bosons}\label{loggaugebos}

 By means of amplitudes involving Goldstone bosons, the complete collinear corrections  \refeq{subllogfact} for longitudinal gauge bosons is
\beqar
\de^{\cc} \M^{\ldots \PW^\pm_\rL \ldots} &=&\de^{\cc}_{\phi^\pm\phi^\pm}(\Phi)\, \M_0^{\ldots \phi^\pm \ldots}=\de^{\cc}_{\phi^\pm\phi^\pm}(\Phi)\, \M_0^{\ldots \PW^\pm_\rL \ldots},\nl
\de^{\cc} \M^{\ldots \PZ_\rL \ldots} &=&\ri\de^{\cc}_{\chi\chi}(\Phi)\,  \M_0^{\ldots \chi \ldots}=\de^{\cc}_{\chi\chi}(\Phi)\, \M_0^{\ldots \PZ_\rL \ldots}\, ,
\eeqar
with
\beqar \label{longeq:coll} 
\de^\cc_{\phi^\pm\phi^\pm}(\Phi)&=&  \left[2\cew_\Phi-\frac{\NCt}{4\sw^2}\frac{\Mt^2}{\MW^2}\right]\ls   +Q_\PW^2\lemW ,\nl
\de^\cc_{\chi\chi}(\Phi)&=& \left[2\cew_\Phi-\frac{\NCt}{4\sw^2}\frac{\Mt^2}{\MW^2}\right]\ls\, .
\eeqar 

\subsubsection*{Higgs boson}
The complete correction is
\beq
\de^{\cc}_{HH}(\Phi)= \left[2\cew_\Phi-\frac{\NCt}{4\sw^2}\frac{\Mt^2}{\MW^2}\right]\ls\, .
\eeq

\subsection{PR: Logarithms connected to the parameter renormalisation}
\label{sec:parren}
\newcommand{\eff}{\mathrm{eff}}
\newcommand{\gt}{g_{\Pt}}
\newcommand{\gH}{\la}
\newcommand{\rt}{h_{\Pt}}
\newcommand{\rH}{h_{\PH}}

The last ingredient is the logarithms related to the UV renormalisation. In Ref.~\cite{Denner:2000jv} they have been identified via the formula
\beq
\de^\pre \M = \left( \frac{\de\M_0}{\de e}\de e 
+ \frac{\de\M_0}{\de\cw}\de\cw 
+ \frac{\de\M_0}{\de\rt}\de\rt 
+ \frac{\de\M_0}{\de\rH}\de\rH^{\eff} 
\, \right)\Big |_{\mu^2=s}
\, , \label{eq:PRorig}
\eeq
where the quantities
\beq
\rt =\frac{\Mt}{\MW}\,, \qquad  \rH =\frac{\MH^2}{\MW^2}\,, \label{eq:unusedrs}
\eeq
are related to the top-quark Yukawa coupling
and to the scalar self coupling, respectively. All the $\delta$'s are the logarithmic part of the renormalisation counter-terms of the corresponding dimensionless quantities. In the $\delta$'s, regardless of the value of $Q$ chosen for the regularising the IR divergences in the other contributions (LSC, SSC, C),  the  UV regularisation-scale  $\mu$ must be set as $\mu^2=s$.  Indeed, although renormalised amplitudes in an on-shell scheme do not depend on the value of the unphysical UV-regularisation scale $\mu$, the {\denpoz} algorithm has been derived assuming $\mu^2=s$. Therefore, in order to preserve the cancellation of the $\mu$ dependence related to the UV poles, in the logarithmic part of the UV counter-terms it is necessary that $\mu^2=s$.

Here, we rearrange the formula in eq.~\eqref{eq:PRorig} for practical purposes related to the implementation in {\mglong}, discussed in Sec.~\ref{sec:implementation}, but the results are fully equivalent with those of  Ref.~\cite{Denner:2000jv}. In practice we rearrange it into 
\beq
\de^\pre \M = \left(\frac{\de\M_0}{\de \alpha}\de \alpha 
+ \frac{\de\M_0}{\de \MW^2}\de \MW^2 
+ \frac{\de\M_0}{\de \MZ^2}\de \MZ^2 
+ \frac{\de\M_0}{\de\Mt}\de\Mt 
+ \frac{\de\M_0}{\de\ntad}\de\ntad 
\, \right)\Big |_{\mu^2=s}
. \label{eq:PRnew}
\eeq
 It is worth to recall that  the renormalisation of 
masses in  propagators or in couplings with mass dimension is not relevant, because those contribute only to  mass-suppressed amplitudes.
 The parameter $\ntad$ is a technical parameter that has the only purpose of keeping track of the appearances  of the tadpole counter-term. In practice what we do is to modify Feynman rules for three-scalar and four-scalar vertices by rescaling their value by the parameter $\ntad$, which is then set equal to one in the numerical evaluation.

We use the following formulas:
 
\beqar \label{massCT}
\frac{\de\MW^2}{\MW^2} &=&-\left[\bew_{W}-4\cew_\Phi\right]\lu -\frac{\NCt}{2\sw^2}\frac{\Mt^2}{\MW^2}\lu, \nl
\frac{\de\MZ^2}{\MZ^2} &=&-\left[\bew_{ZZ} -4\cew_\Phi\right]\lu -\frac{\NCt}{2\sw^2}\frac{\Mt^2}{\MW^2}\lu,
\eeqar
and
\beqar \label{chargerenorm}
\de \alpha= \frac{2 \de Z_e}{4 \pi}&=&\frac{1 }{4 \pi}\left(-\bew_{AA}\lu+2 \de Z^{\elm}_e\right),
\eeqar
where the purely electromagnetic part reads
\beq
\de Z^{\elm}_e\equiv\begin{cases*}
      -\frac{1}{2}\de Z^\elm_{AA}=\frac{2}{3}\sum_{f,i,\si \neq t} \NCf Q_{f_\si}^2 \lWlanew  & in the $\alpha(0)$ scheme\, , \\
      ~~0       & in the $G_\mu$ or $\alpha(\MZ)$ scheme.
    \end{cases*}
\eeq
In this work, all the results are presented by adopting the $G_\mu$ scheme, where in the place of $\alpha$ the input parameter is $G_\mu$, which is related to $\alpha$ via the tree-level relation $G_\mu=\pi \alpha/(\sqrt{2}\MZ^2 \cw^2 \sw^2)$. This translates into the substitution
\beq
\frac{\de\M_0}{\de \alpha}\de \alpha \TO \frac{\de\M_0}{\de G_\mu}\de G_\mu~~{\rm with}~~\de G_\mu=\frac{\de G_\mu}{\de \alpha}\de \alpha + \frac{\de G_\mu}{\de \MZ^2}\de \MZ^2+ \frac{\de G_\mu}{\de \MW^2}\de \MW^2\, ,
\eeq
in eq.~\eqref{eq:PRnew}.

The remaining terms are 
\newcommand{\Qt}{Q_{\Pt}}
\beq
\frac{\de \Mt}{\Mt} =
\biggl[\frac{1}{4\sw^2}+\frac{1}{8\sw^2\cw^2}+\frac{3}{2\cw^2}\Qt -\frac{3}{\cw^2}\Qt^2
+\frac{3}{8\sw^2}\frac{\Mt^2}{\MW^2}\biggr]\lu\,, 
\eeq
where on-shell renormalisation for the mass is assumed, and finally
\beqar
\de\ntad &=& \frac{e}{2\sw}\frac{\de t}{\MW\MH^2}\, ,
\eeqar
with the contribution from the tadpole renormalisation reading
 \beqar
\de t &=& -T =
\frac{1}{e\sw\MW}\biggl[-\frac{3}{2}\MW^2\left(\frac{\MZ^2}{\cw^2}+2\MW^2\right)
\nl &&{}
-\frac{\MH^2}{4}(2\MW^2+\MZ^2+3\MH^2)+2\NCt\Mt^4\biggr]\lu\, .
\eeqar

\section{Sudakov logarithms and NLO EW corrections}

\label{sec:NLOew}

In the previous section we have revisited the {\denpoz} algorithm, which allows the calculation of electroweak DL and SL in LA for virtual scattering {\it amplitudes}. On the other hand, for collider results and in general for the calculation of physical observables, the relevant quantities are amplitudes that are either squared or interfered among them. In particular in this work our final goal is  the NLO EW corrections to LO cross sections.

 For any differential or inclusive cross section $\Sigma$, adopting the notation already used in Refs.~\cite{Frixione:2014qaa, Frixione:2015zaa, Pagani:2016caq, Frederix:2016ost, Czakon:2017wor, Frederix:2017wme, Frederix:2018nkq, Broggio:2019ewu, Frederix:2019ubd,Pagani:2020rsg, Pagani:2020mov, Pagani:2021iwa}, the different contributions from the expansion in powers of $\as$ and $\alpha$  can be denoted as:
\begin{align}
\Sigma^{}_{\LO}(\as,\alpha) &= \Sigma^{}_{\LO_1} + \cdots  + \Sigma^{}_{\LO_k}\, , \label{eq:blobs_LO_general} \\
 \Sigma^{}_{\NLO}(\as,\alpha) &=  \Sigma^{}_{\NLO_1} + \cdots  + \Sigma^{}_{\NLO_{k+1}}\, , \label{eq:blobs_NLO_general} 
\end{align}
 with  $k$ being  process dependent and  $k\ge1$.

 Each $\Sigma^{}_{\LO_i}$ denotes a specific $\as^n \alpha^m$ perturbative order that can be present at LO, {\it i.e.}, arising from tree-level diagrams only. On the contrary, each $\Sigma^{}_{\NLO_i}$ denotes a specific NLO perturbative order to which the interferences between different classes of tree-level and one-loop diagrams can contribute.
 For a given process, the values of $n$ and $m$ vary for each $\Sigma^{}_{\LO_i}$, but the sum $n+m$ is constant. Moreover, if $\Sigma^{}_{\LO_i}\propto \as^n \alpha^m$ then $\Sigma^{}_{\LO_{i+1}}\propto \as^{n-1} \alpha^{m+1}$, $\Sigma^{}_{\NLO_{i}}\propto \as^{n+1} \alpha^{m}$ and $\Sigma^{}_{\NLO_{i+1}}\propto \as^{n} \alpha^{m+1}$. 
 
 It is easy to understand that if the perturbative order of each $\Sigma^{}_{\NLO_{i}}$ is  denoted as $\ord(\Sigma^{}_{\NLO_{i}})$ then 
 \beq
 \ord(\Sigma^{}_{\NLO_{i}})=\ord(\Sigma^{}_{\LO_{i}})\times \as = \ord(\Sigma^{}_{\LO_{i-1}})\times \alpha\,. \label{eq:orders}
 \eeq
 
 Equation \eqref{eq:orders} implies something that is very well known and, {\it e.g.}, has been discussed in detail in Ref.~\cite{Frixione:2014qaa}.
 If $\Sigma^{}_{\NLO_{i}}$ involves EW corrections ($i>1$) and it is not the term with the possibly highest $\alpha$ power at NLO ($i<k+1$), then both QCD and EW loops on top of tree-level amplitudes can enter into the game. Even worse, this separation into ``QCD loops'' and  ``EW loops'' is artificial and especially cannot be rigorously defined. Since one of the main features of our  implementation of the {\denpoz} algorithm is the possibility of comparing the DL and SL terms in LA against the exact result for NLO EW corrections, the contribution of such ``QCD loops'' cannot be ignored.
 
In LA, the contribution from one-loop corrections to the quantity $\Sigma^{}_{\NLO_{i}}$, denoted as $\Sigma^{\rm virt}_{\NLO_{i}}$ can be written in the form
\beq
(\Sigma^{\rm virt}_{\NLO_{i}})\Big|_{\rm LA}=\Sigma^{}_{\LO_{i-1}} \deltaEW + \Sigma^{}_{\LO_{i}} \deltaQCD \, . \label{eq:QCDEWcomb}
\eeq
 
In the following section we provide the necessary ingredients for taking into account single and double-logarithmic enhanced contribution of QCD origin in the computation of  $\Sigma^{}_{\NLO_2}$, what is typically dubbed in the literature as ``NLO EW corrections''. The case of the Complete-NLO, {\it i.e.}~the complete set of $\Sigma^{}_{\NLO_i}$ contributions, is left for future work. In practice, what is discussed in this work is sufficient for both the case of $\Sigma^{}_{\NLO_2}$ and $\Sigma^{}_{\NLO_{k+1}}$, since the latter never receives contributions from ``QCD loops'', as can be seen from eq.~\eqref{eq:orders}.

The quantity $\deltaEW$ is what is calculated via the {\denpoz} algorithm revisited in Sec.~\ref{sec:revisitation} and summarised in eqs.~\eqref{eq:M0}--\eqref{eq:deltatodeltas}. For the case of $\Sigma^{}_{\NLO_2}$, or equivalently the case $\Sigma^{}_{\NLO_{k+1}}$, if $\M_0$ is the amplitude that once squared leads to $\Sigma^{}_{\LO_{1}}$, or equivalently $\Sigma^{}_{\LO_{k}}$, then
\beq
\deltaEW\equiv \frac{2\Re(\M_0\delta\M^*)}{|\M_0|^2}\,. \label{eq:deltaEW}
\eeq
As we said, we leave the case of the Complete-NLO for future work. In that case  also eq.~\eqref{eq:deltaEW} would receive modifications since a generic $\Sigma^{}_{\LO_{i-1}}$ term with $1<i<k$  can itself arise from the interference of amplitudes factorising different powers of $\as$ and $\alpha$.

\subsection{Contributions from QCD loops}
\label{sec:QCD}
If $\Sigma_{\LO_{i}}\propto \as^{n}  \alpha^{m}$, since $\Sigma_{\LO_{i+1}}\propto \as^{n-1}  \alpha^{m+1}$ then $\Sigma_{\LO_{i+1}}$ originates from either a squared amplitude $|\Mew_0 |^2$ with
\beq
\Mew_0 \propto \as^{(n-1)/2}  \alpha^{(m+1)/2}\, ,
\eeq
 or an interference $\Mew_{0,1}\Mew_{0,2}^*$ of two amplitudes $\Mew_{0,1}$ and $\Mew_{0,2}$  with
 \beqar
 \Mew_{0,1}&\propto& \as^{(n-1+j)/2}  \alpha^{(m+1-j)/2}\,,\\
 \Mew_{0,2} &\propto& \as^{(n-1-j)/2}  \alpha^{(m+1+j)/2}\,.
 \eeqar
 with $j$ being an integer in the range $0<j\le \min(n-1,m+1)$.
 Similarly to  eqs.~\eqref{eq:M0}--\eqref{eq:deltatodeltas}, where starting from the amplitude $\M_{0}$ the logarithmic-enhanced EW corrections are denoted as $\delta \M$, we can denote the logarithmic-enhanced QCD corrections to $\Mew_0$ as $\delta \Mew$.  This implies that
 \beq
 \Sigma^{}_{\LO_{i+1}}\propto |\Mew_0|^2 ~~ {\rm or} ~~ \Sigma^{}_{\LO_{i+1}}\propto 2\Re({\Mew_{0,1} \Mew_{0,2}^*}) \label{eq:mew0}\,,
 \eeq
and respectively
\beq
\Sigma^{}_{\LO_{i+1}} \deltaQCD \propto 2\Re(\Mew_0\delta\Mew^*)  ~~ {\rm or} ~~ \Sigma^{}_{\LO_{i+1}} \deltaQCD \propto 2\Re(\Mew_{0,1}\delta\Mew_2^*+\Mew_{0,2}\delta\Mew_1^*) \, .\label{eq:dmew}
 \eeq

In principle, following the same steps of Sec.~\ref{sec:revisitation}, one could derive a general algorithm for obtaining $\delta \Mew$ starting form a generic $\Mew_0$. Indeed, besides the case of non-abelian gluon vertices, the DL and SL logarithms can be identified by looking at the purely QED part of the expressions of Sec.~\ref{sec:revisitation}. In practice, this would lead to non-trivial terms involving colour-linked amplitudes, which are not {\it per se} problematic, but still avoidable via two simple assumptions on the value of $Q^2$ and  $\Delta^{s\TO r_{kl}}(r_{kl},M^2)$.

As already mentioned at the end of  Sec.~\ref{sec:SSC}, if one sets $Q^2=s$ and $\Delta^{s\TO r_{kl}}(r_{kl},M^2)=0$,  as in the formal derivation of Ref.~\cite{Denner:2000jv}, then the SSC contribution from purely QED origin vanishes. Similarly, simplifications in the rest of the expressions of  Sec.~\ref{sec:revisitation} happen. Setting $Q^2=s$   and  $\Delta^{s\TO r_{kl}}(r_{kl},M^2)=0$, these simplifications are present also for the case of QCD. Especially, the SSC contribution vanishes also in the case of QCD corrections. As we will see later in Sec.~\ref{sec:weak}, these two assumptions are innocuous for what concerns $\deltaQCD$ in LA when physical observables are considered.  

With these two assumptions,  if $\Mew_0 \propto \as^{n_{\as}} \alpha^{n_\alpha}$ we can write
\begin{equation}
\delta\Mew\equiv \Mew_0  \left[ \Big(n_t \,\Ltop + n_{\as} \,\Las{\mu_R^2}-n_{g}\, \Las{s}  \Big)+  \frac{\de \Mew_0}{\de \Mt} \dmtQCD  \right]\;,
\label{eq:dQCD}
\end{equation}
where $n_t$ and $n_g$ are the number of top quarks 
and gluons in the external legs, respectively.
The quantities  $\Ltop$, $\Las{\mu^2} $ and $\dmtQCD$ are defined as
\begin{eqnarray}
\Ltop& \equiv &\frac{C_F}{2} \frac{\as}{4 \pi}  \left( \log^2 \frac{s}{\Mt^2} +   \log \frac{s}{\Mt^2}  \right)\, , \label{eq:Lt} \\
 \Las{\mu^2} &\equiv& \frac{1}{3}~ \frac{\as}{4 \pi} \log \frac{\mu^2}{\Mt^2}\, , \\
 \dmtQCD&\equiv& -3 C_F \frac{\as}{4 \pi}    \log \frac{s}{\Mt^2}\, , \label{eq:dmtQCD}
\end{eqnarray}
with $C_F=4/3$, and  they have a very different origin, as explained in the following.

The terms proportional to $\Ltop$ can be obtained by performing the substitution
\beq
Q^2_t \frac{\alpha}{4\pi}\TO C_F\frac{\as}{4\pi} \label{eq:fromatoas} \, ,
\eeq 
in the purely electromagnetic component of the LSC and C contributions for top quarks (eqs.~\eqref{eq:deSC} and \eqref{eq:deccfer}). The reason why the top quark is special is that we are understanding the use of the five-flavour scheme. If other fermions $f$ are treated as massive ($m_f\ne 0$), the corresponding logarithms with $t\TO f$ should be also taken into account. This is true also for the remaining contributions discussed in this section. Conversely, for all the other massless quarks, if one sets $Q^2=s$ and $\Delta^{s\TO r_{kl}}(r_{kl},M^2)=0$, not only the SSC but also the LSC and C contributions to $\deltaQCD$ vanish.  

The term proportional to $\Las{s}$ can be derived from the diagonal C contribution for the photon by applying the substitution \eqref{eq:fromatoas}. These logarithms are the virtual counterpart of the quasi-collinear logarithms emerging from the $g\TO t \bar t$ splittings. The term proportional to $\Las{\mu_R^2}$ has instead a different origin; it is connected to the $\MSbar$ renormalisation of $\as$.
While the renormalisation of the EW sector can be performed without introducing a renormalisation-scale dependence, this is unavoidable in QCD. 
With five active flavours, the logarithmic-enhanced part of the $\as$ counter-term reads
\begin{align}
\label{dzgmsbar}
\frac{\delta \as}{\as} = 
  \frac{\as}{4 \pi}\Big[\beta_{0}\log\frac{\mu_R^2}{Q^2}
+\frac{2}{3}\log\frac{\mu_R^{2}}{m_{t}^{2}}\Big]\, ,
\end{align}
where  $\mu_R$ is the renormalisation scale and
the quantity $\beta_{0}=11-\frac{2}{3}n_f$ is the leading term of the QCD $\beta$ function in the SM ($n_f=6$). We are assuming $Q^2=s$ and it is  reasonable to assume also $\mu_R^2 \sim s$, which let us to ignore the term proportional to $\log\frac{\mu_R^2}{Q^2}$ in eq.~\eqref{eq:dQCD}. 

While the contribution from $\as$--renormalisation to the LA can be expressed via algebraic formulas, this is not possible in the case of $\Mt$--renormalisation (or equivalently for any other quark that would be considered as massive), where the derivative  $\de \Mew_0 /\de \Mt$ is entering the expression.\footnote{By further expanding eqs.~\eqref{eq:blobs_LO_general} and \eqref{eq:blobs_NLO_general} in powers of $h_t$ (see eq.~\eqref{eq:unusedrs}) this would be possible, but it is an unnecessary complication of notation in this context and especially it is not a feature that is easily automatable. It is interesting to note that also the contribution from $\as$--renormalisation could be expressed via the derivative $\frac{\de \Mew_0}{\de \as}$ instead of simply $n_{\as}$; this is precisely what is done in eq.~\eqref{eq:PRnew} for $\alpha$.  } The formula in eq.~\eqref{eq:dmtQCD} has been obtained in the on-shell scheme, consistently with the EW case.

Via eqs.~\eqref{eq:dQCD} and \eqref{eq:dmew} we can finally write a compact formula for $\deltaQCD$ entering eq.~\eqref{eq:QCDEWcomb}. If $\Sigma_{\LO_{i}}\propto \as^{n}  \alpha^{m}$ and therefore $\Sigma^{}_{\LO_{i+1}}\propto \as^{n-1} \alpha^{m+1}$, then
\begin{equation}
\deltaQCD\equiv  2 \left[ n_t \,\Ltop + \left(n-1\right)\Las{\mu_R^2}-n_{g}\, \Las{s}  \right]+ \frac{1}{\de \Sigma^{}_{\LO_{i}}} \frac{\de \Sigma^{}_{\LO_{i}}}{\de \Mt} \dmtQCD \, .
\label{eq:dQCDfinal}
\end{equation}

\section{Sudakov logarithms and physical cross sections} 
\label{sec:SudMC}

What has been discussed up to this point concerns the LA of  one-loop ``EW corrections''  (Sec.~\ref{sec:revisitation}) and one-loop ``QCD corrections'' (Sec.~\ref{sec:QCD}) to amplitudes and their combination for the LA of the virtual contribution to the perturbative orders $\Sigma_{\NLO_{i}}$ with $i=2$ or $i=k+1$ (eq.~\eqref{eq:QCDEWcomb}), in the perturbative expansion of the cross section $\Sigma$.

Both cases, amplitudes or virtual contributions, are unphysical and cannot be directly used for theoretical predictions of physical quantities. Since electromagnetic contributions are included (one-loop QED corrections), the {\denpoz} algorithm leads to the LA of a quantity that is IR divergent and must be combined at least with the LA of the IR-divergent real-emission contributions. Alternatively, the {\denpoz} algorithm can be slightly modified by excluding the QED contribution, which is the only one leading to unphysical quantities. While virtual QED SL and DL involve the unphysical quantity $Q$, the remaining contributions of purely weak origin involve the physical masses. Clearly, also these logarithms can be partially canceled by their real-emission counter part, the heavy-boson-radiation (HBR) of an extra $W$, $H$ or $Z$ boson, but these cancellations strongly depend on the specific set-up and the degree of inclusiveness of the observable considered, see {\it e.g.} Ref.~\cite{Bell:2010gi}. In other words, while photon and gluon emissions and real radiation of light quarks are unavoidable contributions for obtaining IR-finite predictions, the HBR is not necessary for the sake of IR finiteness. The contribution of HBR may be also relevant for the LA of the entire $\Sigma_{\NLO_{i}}$ prediction, but this critically depends on the process and the set-up considered.

In the literature, {\it e.g.}~in the recent work of Ref.~\cite{Bothmann:2020sxm}  or in Ref.~\cite{Chiesa:2013yma}, this problem has been circumvented by dropping the contributions tagged as ``em'' in the {\denpoz} algorithm, namely those involving the ratios $\MW^2/\lambda^2$ or $\MW^2/m_f^2$, in the original formulation, or equivalently the ratio $\MW^2/Q^2$ in this work. We believe this approach is artificial and based on a wrong interpretation of the role of $\MW$ in the {\denpoz} algorithm. While the DL and SL induced by $W$ and $Z$ boson loops (the $\LrM$ and $\lrMwithabs$  terms with $M=\MW,\MZ$) are physical, in the case of QED $\MW$ is only a technical separator used in order to split the logarithms according to the logic discussed in Sec.~\ref{sec:logsplit}. Bypassing the problem of IR finiteness by simply removing the logarithms involving $\MW$ and the IR scale is therefore  an approach mostly driven by simplicity.

We propose a more rigorous approach for avoiding IR sensitivity in the implementation of the {\denpoz} algorithm for physical cross sections. We will denote this approach as  $\rm SDK_{\rm weak}$, where SDK is an abbreviation for Sudakov. The $\rm SDK_{\rm weak}$ approach is based on the idea of selecting only the DL and SL of purely weak origin, excluding the contributions of QED corrections. Actually, for a large class of processes this approach leads to predictions that are much closer to the exact NLO EW corrections than in the case of approaches based on the removal of all ``em'' terms, what we will denote from now on also as $\rm SDK_{0}$ approach. Indeed, in sufficiently inclusive observables, most of the logarithms of QED origin cancel against their real-emission counterparts.

 The {\denpoz} algorithm has been formulated in Ref.~\cite{Denner:2000jv} for one-loop amplitudes and generalised in Sec.~\ref{sec:NLOew} for their interference with tree-level amplitudes. From now on, we will also denote the latter case as simply the $\rm SDK$ approach. One should keep in mind that, at variance with the $\rm SDK_{\rm 0}$ and $\rm SDK_{\rm weak}$ approaches, the $\rm SDK$ one  leads to IR-divergent quantities, which approximate correctly the virtual contribution to NLO EW cross sections, but that cannot be used for physical observables. We describe in the following how expressions of Secs.~\ref{sec:revisitation} and \ref{sec:NLOew} should be modified in order to adapt the {\denpoz} algorithm, which has been formulated so far for the $\rm SDK$ approach, to the $\rm SDK_{\rm weak}$ approach. 
\subsection{$\rm SDK_{\rm weak}$: Purely weak LA for cross sections} 

\label{sec:weak}

In general, when $W$ bosons are not involved in a process, virtual EW corrections can be divided into a QED and a purely weak component in a gauge-invariant way. QED corrections consist of  all loops involving QED interactions between fermions and photons, excluding the vacuum-polarisation diagrams.\footnote{  The relevant renormalisation conditions for fermion masses and wave-functions lead to counter-terms that are derived only taking account the same class of loops.} The purely weak part consists of all the rest of contributions, including also the vacuum-polarisation diagrams and the  renormalisation of the photon wave-function and of the fine-structure constant  $\alpha$. 

In order to isolate purely weak effect in DL and SL, we exclude all contributions induced by fermions and photons interactions in all formulas of Sec.~\ref{sec:revisitation}, with the exception of those from parameter renormalisation (PR). Moreover, we exclude also DL and SL related to photons interacting with $W$ bosons as external legs. While the classification of the $W$-$\gamma$ interaction as either a purely weak or QED effect is ambiguous, the identification of the terms in the expressions of Sec.~\ref{sec:revisitation} that originate  from such interaction is unambiguous. 

The purely weak version of the {\denpoz} algorithm, $\rm SDK_{\rm weak}$, can be obtained following these steps:
\begin{enumerate}

\item Calculate the $\de^\pre$ in eq.~\eqref{eq:deltatodeltas} as in the standard SDK approach. \label{step1}

\item For each external particle $\varphi_{i_k}$ in \eqref{eq:process}, set 
\begin{equation}
Q_k=I^A(k)=0\,. \label{eq:qto0}
\end{equation}
This step alone has the effect of eliminating all the terms tagged as ``em'', with the exception of $ \de Z^\elm_{AA}$. It also eliminates all the SSC terms and C terms  that lead to SL originating from photons, with the exception of those related to transverse $W$ bosons.  \label{step2}

\item  For each external particle $\varphi_{i_k}$ in \eqref{eq:process}, perform the replacement 
\begin{equation}
C^{\ew}_{i'_ki_k}(k)\longrightarrow C^{\ew}_{i'_ki_k}(k) -Q^2_k \, ,
\end{equation}
with the value of $Q^2_k$ before enforcing eq.~\eqref{eq:qto0}.
 This, in combination with eq.~\eqref{eq:qto0},  has the effect of eliminating the DL due to photons. \label{step3}

\item Perform the replacement
\begin{equation}
 \bew_{W} \longrightarrow \bew_{W}-11/3\,. 
 \end{equation}
This  has the effect of eliminating for the transverse $W$ bosons the C terms  that lead to SL originating from photons. \label{step4}

\item Set 
\begin{equation}
\de Z^\elm_{AA}=0\, ,
\end{equation}
 and perform the replacement 
 \begin{equation}
\bew_{AA}\longrightarrow \bew_{AA}+\frac{4}{3}\sum_{f,i,\si \neq t} \NCf Q_{f_\si}^2=\bew_{AA}+80/9\,.
\end{equation}
This  has the effect of eliminating, for  the photons, the C terms  that lead to SL originating from light fermions. \label{step5}

\item Calculate the remaining terms in eq.~\eqref{eq:deltatodeltas} with the new redefinitions of steps \ref{step2}--\ref{step5}. \label{step6}

\end{enumerate}

We want to stress that, thank to the step \ref{step1}, the redefinitions of steps \ref{step2}--\ref{step5} do {\it not} apply to all the PR contributions discussed in Sec.~\ref{sec:parren}; for them any QED-like contribution is retained. We remind the reader that also in this context we assume the use of either the $\alpha(\MZ)$ or $G_\mu$-scheme, which both have an IR structure that is  $\MSbar$-like, namely, IR poles are not present in the $\alpha$ counter-term, $\delta\alpha$.\footnote{The algorithm therefore has to be slightly modified for the case of isolated photons in the final state (see also the discussion in Ref.~\cite{Pagani:2021iwa}); we leave this to future work.} This difference of treatment for the PR terms, besides the definition of purely weak and QED introduced before, can also be understood in a different way. Logarithms from PR are related to UV renormalisation and do not involve IR sensitivity, as can be seen from all the equations in Sec.~\ref{sec:parren}, which do not depend on the infrared regulator $Q$. Therefore, in order to achieve  physical predictions and eliminate the $Q$ dependence, there is no need to exclude their components that are related to QED interactions. This is in contrast with the LSC, SSC and C contributions, where the QED contributions are  $Q$-dependent (see eq.~\eqref{eq:deSC} and eq.~\eqref{eq:subdl1} for respectively LSC and SSC contributions, and eq.~\eqref{eq:lemfnew}, which defines the term $l^{\rm em}$ entering most of the C contributions) and therefore IR-sensitive. Moreover, while LSC, SSC and C contributions have real-emission counterparts that, together with PDF counter-terms, lead to IR finiteness and the (partial) cancellation of $Q$ dependence, this is not the case for PR contributions.

As already mentioned, in sufficiently inclusive observables, most of the logarithms of QED origin cancel against their real-emission counterparts. Thus, as we will show in Sec.~\ref{sec:Res_MC}, the  purely weak version of the {\denpoz} algorithm, the $\rm SDK_{\rm weak}$ approach, reproduces very well the logarithmic dependence of NLO EW predictions for differential distributions in several cases. For example, if we consider  leptons or massive particles in the final state, it is easy to understand how  the purely weak version of the {\denpoz} can be superior when predictions for physical observables are considered.

If the real emission of photons is considered in the eikonal approximation and integrated up to the energy $E_\gamma=\sqrt{s}/2$, where $E_\gamma$ is the energy of the photon, large part of the QED logarithms cancel. For example, all the QED contributions from LSC and SSC terms are canceled exactly. There are two classes of collinear SL that are left uncanceled: those associated to final-state radiation and those associated to initial-state radiation.\footnote{We analytically verified this statement for the process $e^+e^-\TO W^+ W^-$ combining the results of the {\denpoz} algorithm together with the eikonal approximation of the real emission of photons \cite{Beenakker:1993tt}.} 

The former class of SL can be eliminated by clustering photons with charged particles. In the case of massless charged particles, the clustering is anyway unavoidable for IR safety ({\it e.g.} the case of dressed leptons) and eliminates the $Q$ dependence. In the case of massive charged particles, namely top quarks and $W$ bosons, clustering is also very reasonable since in a realistic experimental set-up the separation of collinear real radiation from very boosted objects is not feasible and, from a theoretical point of view, it leads to larger uncertainties. The clustering eliminates the physical collinear SL of QED origin, which have the form $l(s,M)$, where $M$ is the mass of the massive charged particle.   
The latter class of SL (initial-state) are those related to the PDFs, which therefore in an exact NLO EW calculation are subtracted by the corresponding PDF counter-terms. 

Both for the case of initial- and final-state collinear SL, a special case is given by the photon. Since the top quark and the $W$ boson are massive, the corresponding collinear SL associated to their contribution to  $\de Z^\elm_{AA}$  are not canceled. Indeed, in the case of the initial state, no corresponding PDF counter-terms are present, since massive particles do not enter the PDF evolution.  In the case of the final state, no $\gamma\TO t \bar t $ or $\gamma\TO W^+ W^- $ radiation is generated for a final state giving the same signature.\footnote{We remind the reader that we are assuming in this work that the $\alpha(\MZ)$ or $G_\mu$ scheme is used. The use of $\alpha(0)$ and therefore the case of isolated photons in the final state is not considered in this context.}

In order to be on the same ground, we also remove from eq.~\eqref{eq:dQCD} the term proportional to $\Ltop$. Indeed this is canceled by the clustering of real emissions of gluons and  top quarks into recombined top-quarks. At this point it is also easy to understand why in Sec.~\ref{sec:QCD} we have said that setting $Q^2=s$   and  $\Delta^{s\TO r_{kl}}(r_{kl},M^2)=0$ is innocuous for what concerns $\deltaQCD$ in LA when physical observables are considered. If we lifted these two assumptions, we would obtain many more contributions to eq.~\eqref{eq:dQCD}, but they would be all canceled by the real-radiation counterparts, together with the PDF counter-terms. The presence of the term $n_{g}\, \Las{s}$ can also be seen now as the gluon QCD-counterpart of the photonic uncanceled SL that have been mentioned in the previous paragraph: being massive, top quarks do not enter in the PDF counter-terms and no $g\TO t \bar t $ is generated. We leave  to future work the exploration of these effects from QCD corrections in numerical results for physical cross sections. 

All in all, the modifications to eq.~\eqref{eq:dQCD}  that have to be implemented in the $\rm SDK_{\rm weak}$ approach in order approximate physical cross sections is simply:
\begin{itemize} 
\item Set 
\begin{equation}
\Ltop=0\, ,
\end{equation}
in eq.~\eqref{eq:dQCD}.
\end{itemize}

Finally, we want to stress that the real emission of photons from electrically charged particles and of gluons from coloured particles cannot be neglected for IR-safe observables, but at high energy even the HBR may be considered and taken into account in inclusive calculations. In that case, the $\rm SDK_{weak}$ approach should be further modified. For instance, taking into account the $Z$ radiation, also the contribution of  the $Z$ boson  should be removed from the {\denpoz} algorithm. We leave the exploration of this approach for future work.

\section{Implementation in {\mglong} }
\label{sec:implementation}

The theoretical framework described in the previous sections has been implemented in {\mglong} \cite{Alwall:2014hca}, specifically in the part of the code that is deputed to the calculation of NLO QCD and EW corrections and more in general Complete-NLO predictions \cite{Frederix:2018nkq, Pagani:2021iwa}. This allows a direct comparison of results in LA and at exact fixed-order, both at amplitude level and for physical observables.

We remind the reader that in  {\mglong}  the IR singularities are dealt with via the FKS method~\cite{Frixione:1995ms,
Frixione:1997np},  automated for the first time in \mf~\cite{Frederix:2009yq,
Frederix:2016rdc}. In  {\mglong} one-loop amplitudes can be evaluated via 
different types of integral-reduction techniques  (the  OPP method~\cite{Ossola:2006us} or
 the Laurent-series expansion~\cite{Mastrolia:2012bu})
and  techniques for tensor-integral reduction~\cite{Passarino:1978jh,Davydychev:1991va,Denner:2005nn},
all automated within the module \ml~\cite{Hirschi:2011pa}. Moreover, the codes \ct~\cite{Ossola:2007ax}, \nin~\cite{Peraro:2014cba,
Hirschi:2016mdz} and \collier~\cite{Denner:2016kdg} are employed within \ml, which has been optimised by taking inspiration from {\sc\small OpenLoops} \cite{Cascioli:2011va} for the integrand evaluation.

As already possible in the code, NLO QCD and EW corrections can be invoked via the syntax  {\tt [QCD]} {\tt [QED]}, see Refs.~\cite{Frederix:2018nkq, Pagani:2021iwa}  for more details. However, now the code allows also for the evaluation of virtual one-loop Sudakov logarithms by adding after the command {\tt generate} or {\tt add process} the flag {\tt --ewsudakov}. As we have said, the code works for the moment for $\ord(\alpha)$ corrections to the $\Sigma_{\LO_i}$ contribution with $i=1$ and $i=k$, according to eqs.~\eqref{eq:blobs_LO_general} and \eqref{eq:blobs_NLO_general}.
In order to implement the {\denpoz} algorithm in {\mglong}, three main technical features had to be implemented:
\begin{enumerate}
\item The generation of all the amplitudes that are necessary for the computation of the DL and SL.
\item The evaluation of the amplitudes, especially the interferences of amplitudes involving different external legs.
\item The evaluation of the derivatives of the amplitudes, which enter the formulas concerning the PR terms. 
\end{enumerate}

In the following subsections we address each of the previous points.

\subsection{Generation of the amplitudes}
\label{sec:generation}
 We start discussing the case of a generic partonic process 
 \beq \label{eq:process_impl}
\varphi_{i_1}(p_1)\varphi_{i_2}(p_2) \rightarrow \varphi_{i_3}(p_3) \dots \varphi_{i_{n}}(p_{n})\,,
\eeq
 and at the end we return to the case of proton--proton collisions.
 
 The formulas of Sec.~\ref{sec:revisitation}, which are given for $n\TO 0$ processes, can be easily reframed in terms of more common $2\TO n-2$ amplitudes via crossing symmetry,
 \beqar
 \M_{i_1 \ldots i_n}(p_1,\ldots, p_n)&\equiv& \M(\varphi_{i_1}(p_1)\dots \varphi_{i_n}(p_n)\rightarrow 0)\nonumber\\
 &=&\M(\varphi_{i_1}(p_1)\varphi_{i_2}(p_2) \rightarrow \bar \varphi_{i_3}(-p_3) \dots \bar\varphi_{i_{n}}(-p_{n}))\,.
 \eeqar
As a first step, the algorithm checks if  longitudinally polarised $Z$ or $W$ bosons are present in the external legs. In such a case all the possible amplitudes that can be obtained with one or more substitutions according to eq.~\eqref{eq:borneet} are generated. In other words, starting from $\M_{i_1 \ldots \{n_W W^{\pm}\} \{n_Z Z\}\dots i_n}$, where $\{n_W W^{\pm}\}$ and  $\{n_Z Z\}$ stand for $n_W$ and $n_Z$ appearances of  $W$ and $Z$ bosons respectively, the amplitudes $\M_{i_1 \ldots \{(n_W-1) W^{\pm} \}\phi^{\pm} \{n_Z Z\}\dots i_n}$ and $\M_{i_1 \ldots \{nW^{\pm} \} \{(n_Z-1) Z\}\chi \dots i_n}$ are recursively generated via the substitutions 
\beqar
Z &\longrightarrow& \chi\, ,\label{eq:sub1GET}\\
W^{\pm}  &\longrightarrow& \phi^{\pm} \,,\label{eq:sub2GET}
\eeqar
up to the point that all $W$ and/or $Z$ bosons are transformed into Goldstone bosons. Clearly, any of the previous substitutions can lead to a process for which no tree-level  Feynman diagram can contribute to the amplitude. Such a case is automatically detected by the code and the amplitude is not generated. 
From this point on, while the original amplitude  $\M_{i_1 \ldots \{n_W W^{\pm}\} \{n_Z Z\}\dots i_n}$ is retained and used for the computation of the LO cross section, the complete set of amplitudes  
\beq \label{eq:processsubGET}
\M_{i_1 \ldots \{(n_W-k_W) W^{\pm} \}\{k_W \phi^{\pm}\} \{(n_Z-k_Z) Z\}\{k_Z \chi\}\dots i_n}\,,
\eeq 
with $0\le k_W\le n_W$ and $0\le k_Z\le n_Z$ is used for the following steps in the generation of the amplitudes.

As discussed in Sec.~\ref{sec:revisitation}, the formulas for the different contributions leading to DL and SL involve amplitudes with external particles that are different from the original ones in $\M_0$.
In particular, starting from the process in \eqref{eq:process} it is necessary to generate the amplitudes for all the processes 
\beq \label{eq:processsub}
\varphi_{i_1}(p_1)\dots \varphi_{i'_k}\dots \varphi_{i_n}(p_n)\rightarrow 0\,,
\eeq
 with $1\le k\le n$ that can be obtained applying the substitution $\varphi_{i_k}\TO\varphi_{i'_k}$ of the form:
\beqar
Z &\longleftrightarrow& A\, ,\label{eq:sub1}\\
H  &\longleftrightarrow& \chi\,.\label{eq:sub2}
\eeqar
With the symbol $\longleftrightarrow$ we understand that the substitution works in the two directions. Substitution \eqref{eq:sub1} is necessary for the off-diagonal components of $\cew$ entering the LSC terms and  of $\bew_{\NB'\NB}$ entering the C terms. Substitution \eqref{eq:sub2} is necessary for the off-diagonal components of $(I^Z)^2$ entering the neutral SSC terms. 
Moreover it is necessary to generate also the amplitudes for the processes
\beq \label{eq:processsub2}
\varphi_{i_1}(p_1)\dots \varphi_{i'_k}\dots \varphi_{i'_l}\dots \varphi_{i_n}(p_n)\rightarrow 0\,,
\eeq
that can be obtained either applying  two substitutions $\varphi_{i_k}\TO\varphi_{i'_k}$ and $\varphi_{i_l}\TO\varphi_{i'_l}$ of the form \eqref{eq:sub2}, again for the off-diagonal components of $(I^Z)^2$ in the neutral SSC terms, or two different $\varphi_{i_k}\TO\varphi_{i'_k}$ and $\varphi_{i_l}\TO\varphi_{i'_l}$ substitutions that together do not violate charge conservation,  each one of them of the form:
\beqar
f_\sigma &\longleftrightarrow& f _{-\sigma}\, ,\label{eq:sub1b}\\
H  &\longleftrightarrow& \phi^{\pm}\, ,\label{eq:sub2b}\\
\chi  &\longleftrightarrow& \phi^{\pm}\, ,\label{eq:sub3b}\\
A &\longleftrightarrow& W^{\pm}\, ,\label{eq:sub4b}\\
Z &\longleftrightarrow& W^{\pm}\, .\label{eq:sub5b}
\eeqar

The substitutions \eqref{eq:sub1b}--\eqref{eq:sub5b} originate from the purely non-diagonal structure of $I^{\pm}I^{\mp}$ entering the charged SSC terms. We remind the reader that both the substitutions \eqref{eq:sub1}--\eqref{eq:sub2} for the processes \eqref{eq:processsub} and \eqref{eq:sub1b}--\eqref{eq:sub5b} for the processes \eqref{eq:processsub2} have to be performed starting from each one of the possible processes in \eqref{eq:processsubGET} that can be obtained from \eqref{eq:process} via the substitutions \eqref{eq:sub1GET}--\eqref{eq:sub2GET}.

For hadronic calculations (protons in the initial state, jets, {\it etc.}) different partonic subprocesses can contribute at the Born level. Therefore, the procedure described so far has to be separately repeated for each of them.

\subsection{Evaluation of the amplitudes}

The evaluation of the amplitudes follows the standard procedure of the {\mglong} framework, which relies on the helicity routines supplied
by {\ALOHA}~\cite{deAquino:2011ub}. Here, the additional complication consists in the evaluation of interferences of amplitudes that can have different particles in the respective initial and/or final state, as shown in eq.~\eqref{eq:imsource}. As discussed in the previous section, there can be one or even two different external particles between the two interfering amplitudes. Consequently, without altering the external momenta, external particles cannot be in general on-shell in both amplitudes.
In order to preserve the on-shell conditions of external legs, external momenta have to be modified for one of the two amplitudes that are interfered. We follow this approach, modifying the external momenta of the amplitude with different external states w.r.t.~the Born one.

From a technical point of view, this approach is
very similar to the momentum-reshuffling techniques discussed in Ref.~\cite{Frixione:2019fxg}, in the context of the so-called ``Simplified
Treatment of Resonances'' (STR) that are needed to perform, {\it e.g.},  computations in supersymmetric theories.\footnote{STR techniques encompass the so-called 
    diagram-removal and diagram-subtraction ones, see {\it e.g.} Refs.~\cite{Beenakker:1996ch, Frixione:2008yi, Hollik:2012rc, Demartin:2016axk}.} In both cases, on-shell conditions are enforced by modifying part of the external momenta and the remaining ones (possibly a subset) are reshuffled in order to preserve momentum conservation. On the other hand, in this context, not only this procedure has to be applied at the amplitude level and not at the squared-amplitude level, but it is also intrinsically more articulated.
While in the case of Ref.~\cite{Frixione:2019fxg} only one on-shell condition is enforced and involves the invariant mass of two final-state particles, in the present case one or more on-shell conditions have to be enforced and involve the kinematic mass, $\sqrt{p_0^2-|\vec p|^2}$, of one or  two individual external momenta, from the initial and/or final state. 
 Our implementation is based on the one described in detail in Sec.~5.2 of Ref.~\cite{Frixione:2019fxg}, on which we base the following discussion. We use the same notation for describing the technical details.
 
The case of only one on-shell condition for a particle in the final state can be directly derived from Sec.~5.2 of Ref.~\cite{Frixione:2019fxg}. Using the same notation, it
can be summarised as: given a set of momenta $k_i$, generate a new set $\bar k_i$ where a given (final-state) particle, denoted as $\beta$, changes its
mass from $m$ to $m_\beta=M$. The two particles labeled as $\delta$ and $\gamma$ in Ref.~\cite{Frixione:2019fxg} are irrelevant for our case. Among
the infinite number of solutions, two options, dubbed as A and B in Ref.~\cite{Frixione:2019fxg}, have been considered.
In the former the energy-momentum conservation is imposed by modifying all the other final-state particle momenta, while in the latter,  which is the default option in the code, by changing the momenta of the initial-state particles. The case of two  on-shell conditions for  particles in the final state can be achieved by applying the procedure iteratively.

The case of $\beta$ being a particle in the initial state was not relevant for Ref.~\cite{Frixione:2019fxg} and we will briefly present it here.
In this case, we have chosen to change the momentum of only the other initial-state particle, leaving the final-state ones untouched ($\bar k_i = k_i$ for
$i\ge 3$). This implies that
the centre-of-mass energy $s$ is conserved and therefore the procedure is very simple. 
We start with the original initial-state momenta $k_{i}$, with $i=1,2$,
where the one with $i=\beta$ is going to have a new mass. Since $s$ must be conserved and we want the new momenta $\bar k_{i}$ collinear to the beam pipe, in the partonic centre-of-mass frame one has  to simply derive the new quantity $|\bar k_{\beta,z}|=|\bar k_{3-\beta,z}|$ enforcing momentum conservation and on-shell conditions.

We conclude by commenting on the fact that, when masses are modified for both  an initial-state and a final-state particle, the procedure
can again be performed iteratively. However, when the default option B is used for the final-state case, since it assumes massless initial-state momenta,  the case of a new mass in the final state  should be considered first, and only afterwards one should consider the initial-state one.

Before moving to the next section we want to clarify that all this procedure would be unnecessary if an analytical calculation were performed and all the mass-suppressed term were discarded. This is on the other hand not possible in an automated approach. The procedure outlined here leads to a correct evaluation of all the terms that are not mass suppressed. Indeed, all the modifications of the momenta and subsequent reshuffling operations involve only scales connected to the mass of the SM particles. The differences in the kinematics before and after the procedure outlined in this section are mass suppressed themselves, leading to smaller and smaller effects when the energy inrceases. Ambiguities related to the choice of a specific reshuffling technique and to the order in which the reshuffling is performed are also mass suppressed.

\subsection{Derivative of the amplitudes}
\label{sec:derivative}

As can be seen in eqs.~\eqref{eq:PRnew} and \eqref{eq:dQCD}, in order to compute the logarithmic contributions induced by the parameter renormalisation, the derivatives of the amplitudes w.r.t.~part of the input parameters have to be calculated. Although one may in principle use dedicated Feynman rules, such as those used for generating UV counter-terms in an NLO computation, we have opted to calculate the derivatives via numerical methods. In other words, for each phase-space point, we evaluate the quantity 
\beq
\left.\frac{\delta{\mathcal M}}{\delta x} \right |_{x=\bar x}\equiv \frac{( {\mathcal M} |_{x=\bar x(1+\delta_{\bar x})}-{\mathcal M} |_{x=\bar x(1-\delta_{\bar x})})}{2\delta_{\bar x}}  \, ,
\eeq
where $x$ is any of the variables for which the derivative has to be performed ($\MW, \MZ$, {\it etc.}), $\bar x$ is its numerical value when the amplitudes are evaluated and $\delta_{\bar x}$ is a small value, which has been set to $ \delta_{\bar x}=10^{-5}$ for the results presented in this work. The same procedure is done for $\tilde \M$ in eq.~\eqref{eq:dQCD}. 

We have checked that this procedure has a mild impact on the speed of the code  and the choice $ \delta_{\bar x}=10^{-5}$ is excellent in terms of both stability and precision. The use of the numerical derivatives  allows also to easily adapt the calculation of PR terms for possible BSM scenarios, where additional particles and couplings would be present. Moreover, at variance with what has been done in the recent automation \cite{Bothmann:2020sxm}  in the {\sherpa} framework, the SL from PR terms are calculated exactly at $\ord(\alpha)$ as all the other type of DL and SL logarithms, without including spurious terms from higher orders in the $\alpha$ expansion. This fact is crucial for the systematic comparisons we are going to carry out in Secs.~\ref{sec:Res_Amp} and \ref{sec:Res_MC} between NLO EW exact results and their Sudakov approximations.

\section{Numerical Results: matrix-element level}
\label{sec:Res_Amp}

In this section we present numerical results obtained via the revisitation and implementation of the {\denpoz} algorithm in {\mglong}, which has been described in the previous sections.
We focus here on the Sudakov approximation of one-loop amplitudes and in particular on their interferences with the corresponding tree-level ones;  results for cross sections of  processes that are relevant at colliders are discussed in Sec.~\ref{sec:Res_MC}. We compare exact results for the finite part of the virtual contribution to the NLO EW corrections with their Sudakov approximation, what is denoted as the ``SDK approach'' following the notation introduced in Sec.~\ref{sec:SudMC}. After having specified the input parameters in Sec.~\ref{sec:inputs}, we start in Sec.~\ref{sec:rijvss} by discussing the effect of the ${\SS^{s\TO r_{kl}}}$ terms in eq.~\eqref{eq:angsplit_new_sm}, which are relevant when the condition \eqref{eq:rijnice} is not satisfied. Next, in Sec.~\ref{sec:Im} we show the numerical relevance of the terms proportional to $2 i \pi \Theta(r_{kl})$, discussed in Sec.~\ref{sec:logsplit}. Then, in Sec.~\ref{sec:QCDres} we show the relevance of the QCD corrections to the subleading LO (see Sec.~\ref{sec:QCD}) in order to compare  NLO EW corrections with their Sudakov approximation.

Throughout this section, we will consider only the finite part of the virtual contribution to the NLO EW corrections. It is worth recalling that, in general, the virtual contribution is IR divergent and non-physical by itself. Since we regularise IR divergences in DR, the finite part depends on the IR-regularisation scale $Q$, which we set here always as $Q^2=s$.

\subsection{Input parameters}
\label{sec:inputs}
The results presented in this section are obtained using the following input parameters:
\begin{align}
m_{Z}&=91.188~\textrm{GeV}\, ,&\quad  m_{W}&=80.385~\textrm{GeV}\, ,&\quad  m_H&=125~\textrm{GeV}\, ,& \nonumber \\
m_{\textrm{t}}&=173.3~\textrm{GeV}\, ,&\quad  G_\mu &= 1.16639 \cdot 10^{-5} ~\gev^{-2}\,, &\quad \as(\MZ)&=0.119 \,.    
   \end{align}
All the other SM particles are treated as massless and all the decay widths are set equal to zero. Consistently with the input parameters, EW interactions are renormalised in the $G_\mu$-scheme and masses and wave-functions in the on-shell scheme.  
 QCD interactions are renormalised  in the $\MSbar$-scheme, with  the renormalisation-group running at two loops and the renormalisation scale $\mu_R$ set to $\mu_R^2=s$.

\subsection{Impact of ${\SS^{s\TO r_{kl}}}$ terms}
\label{sec:rijvss}
\begin{figure}[!t]
\begin{center}
\includegraphics[width=0.32\linewidth]{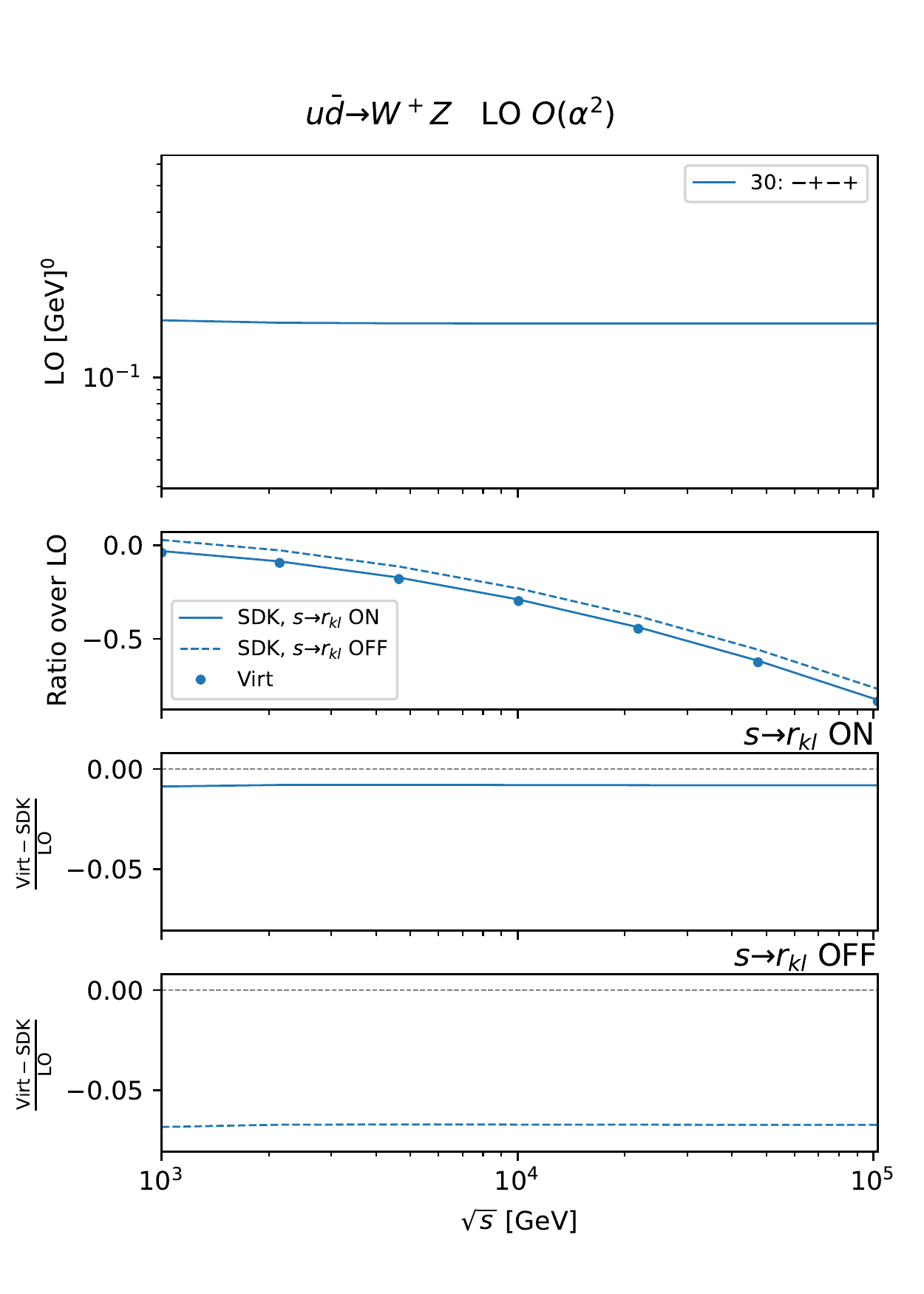}
\includegraphics[width=0.32\linewidth]{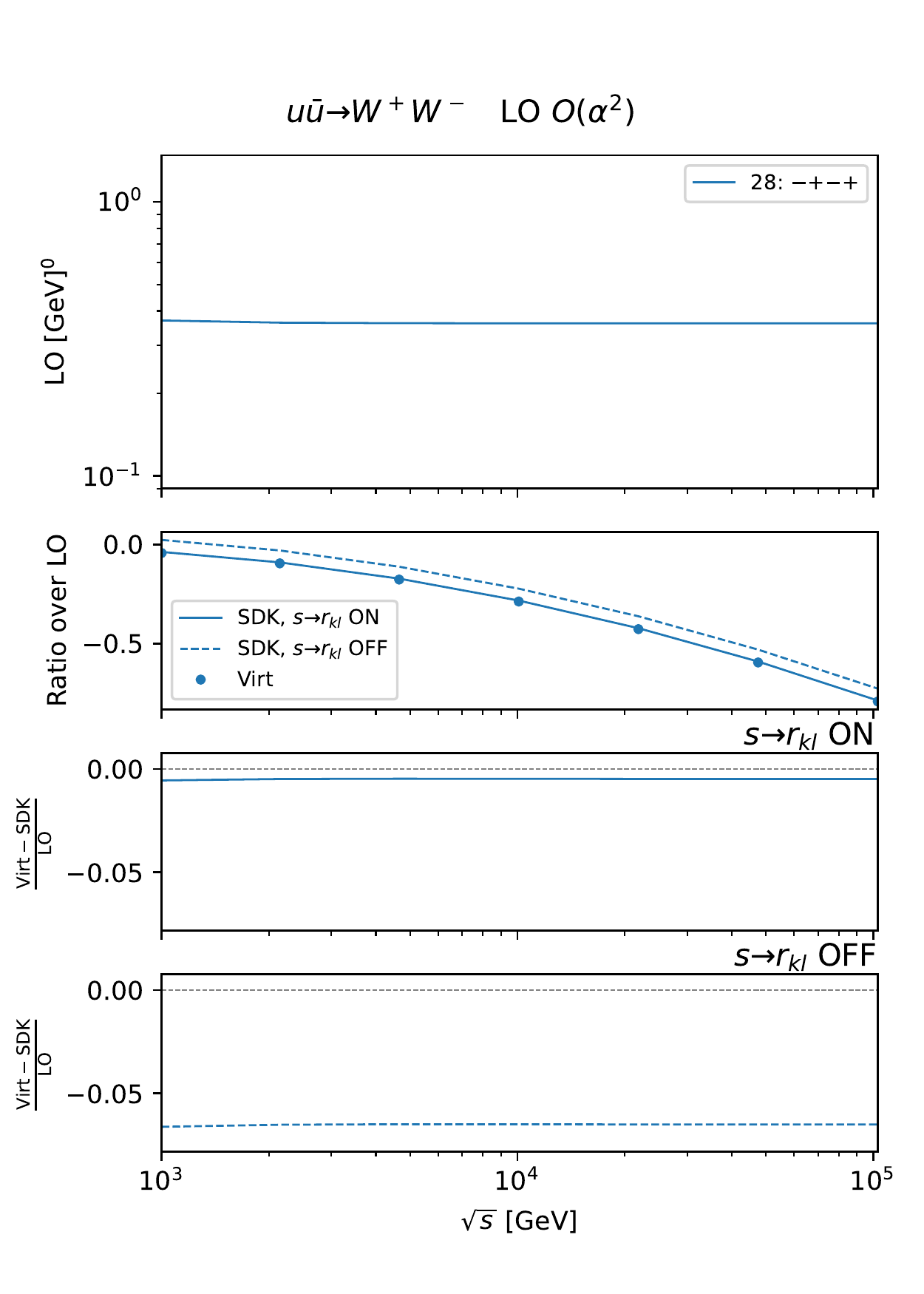}
\includegraphics[width=0.32\linewidth]{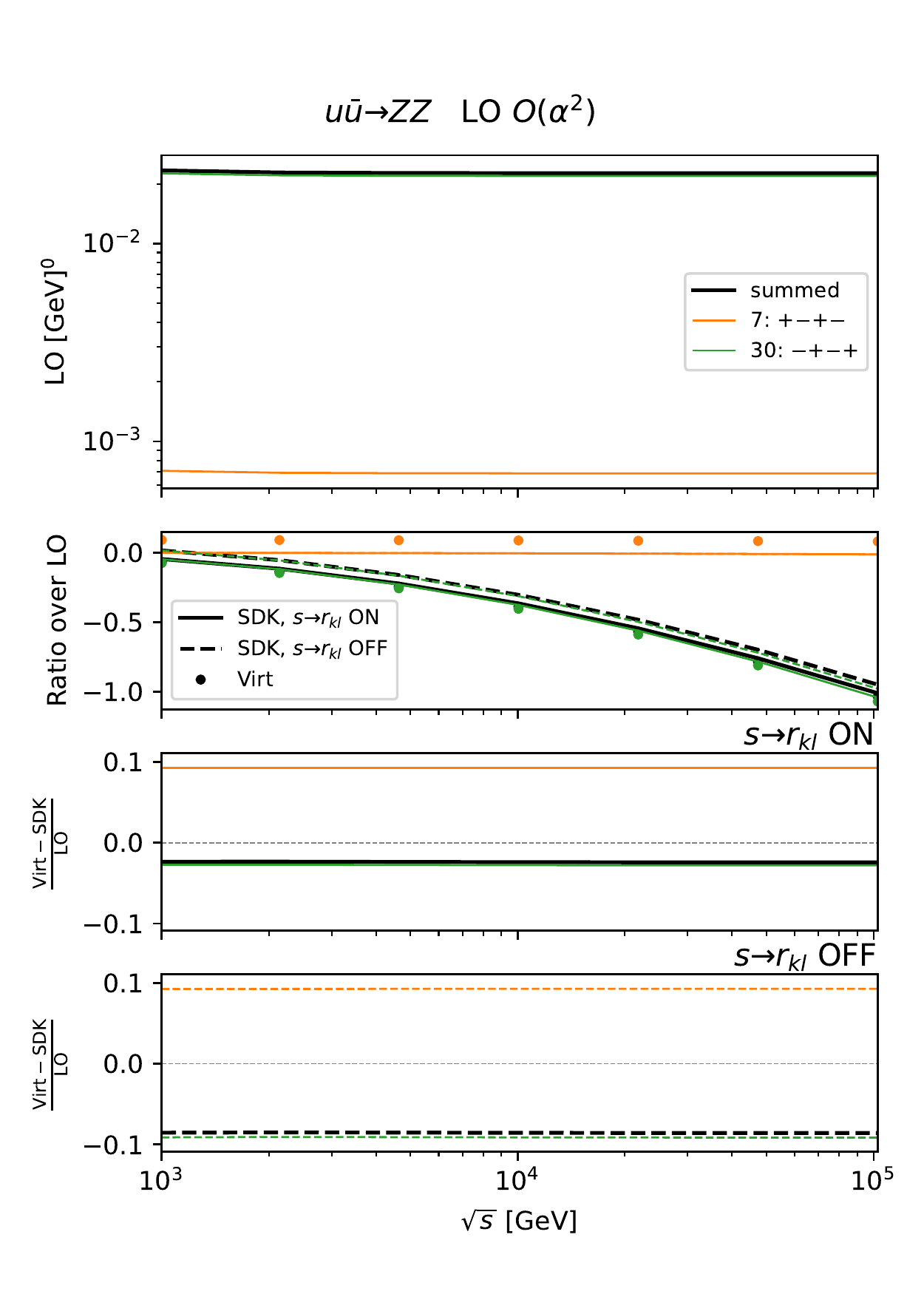}
\includegraphics[width=0.32\linewidth]{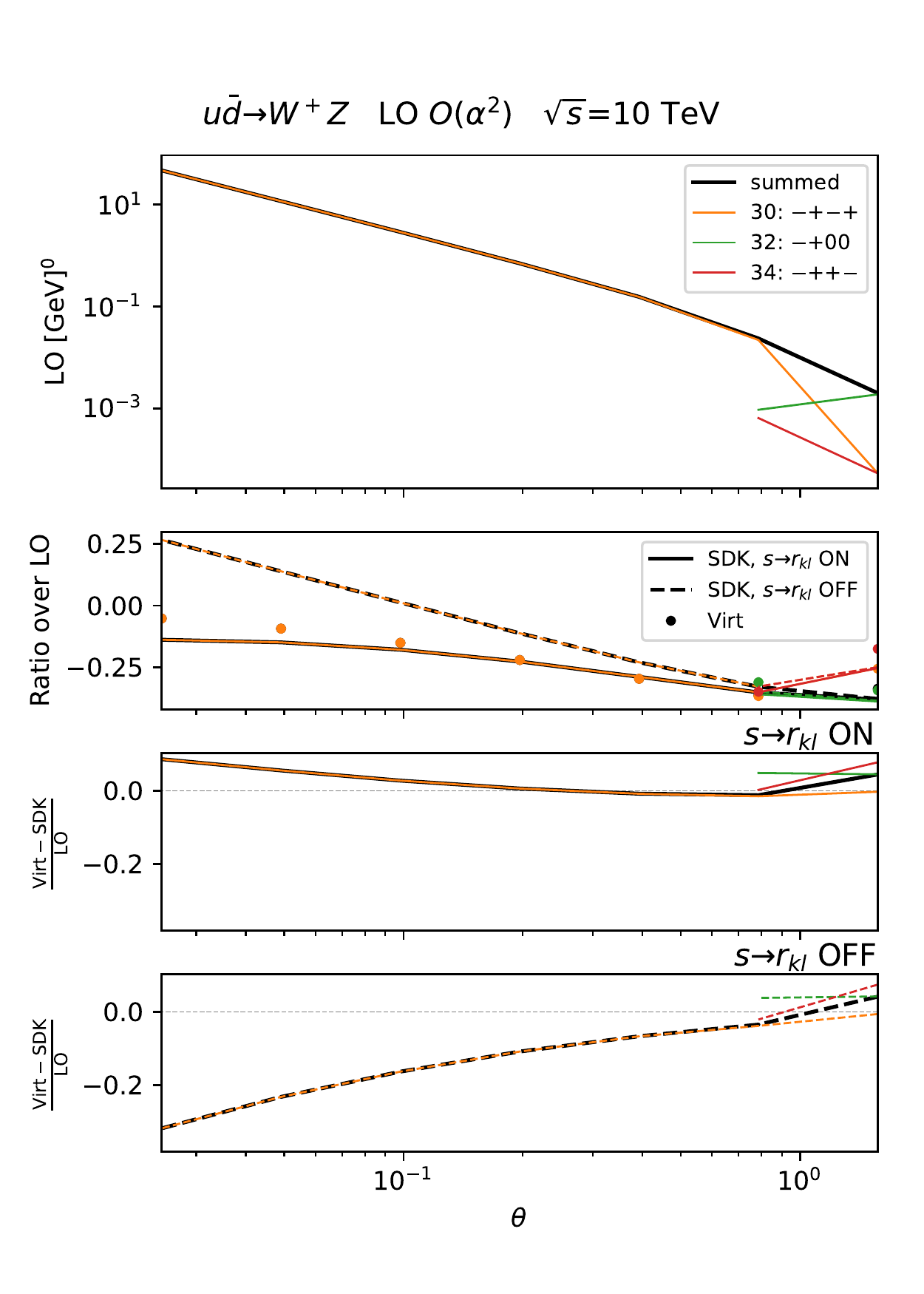}
\includegraphics[width=0.32\linewidth]{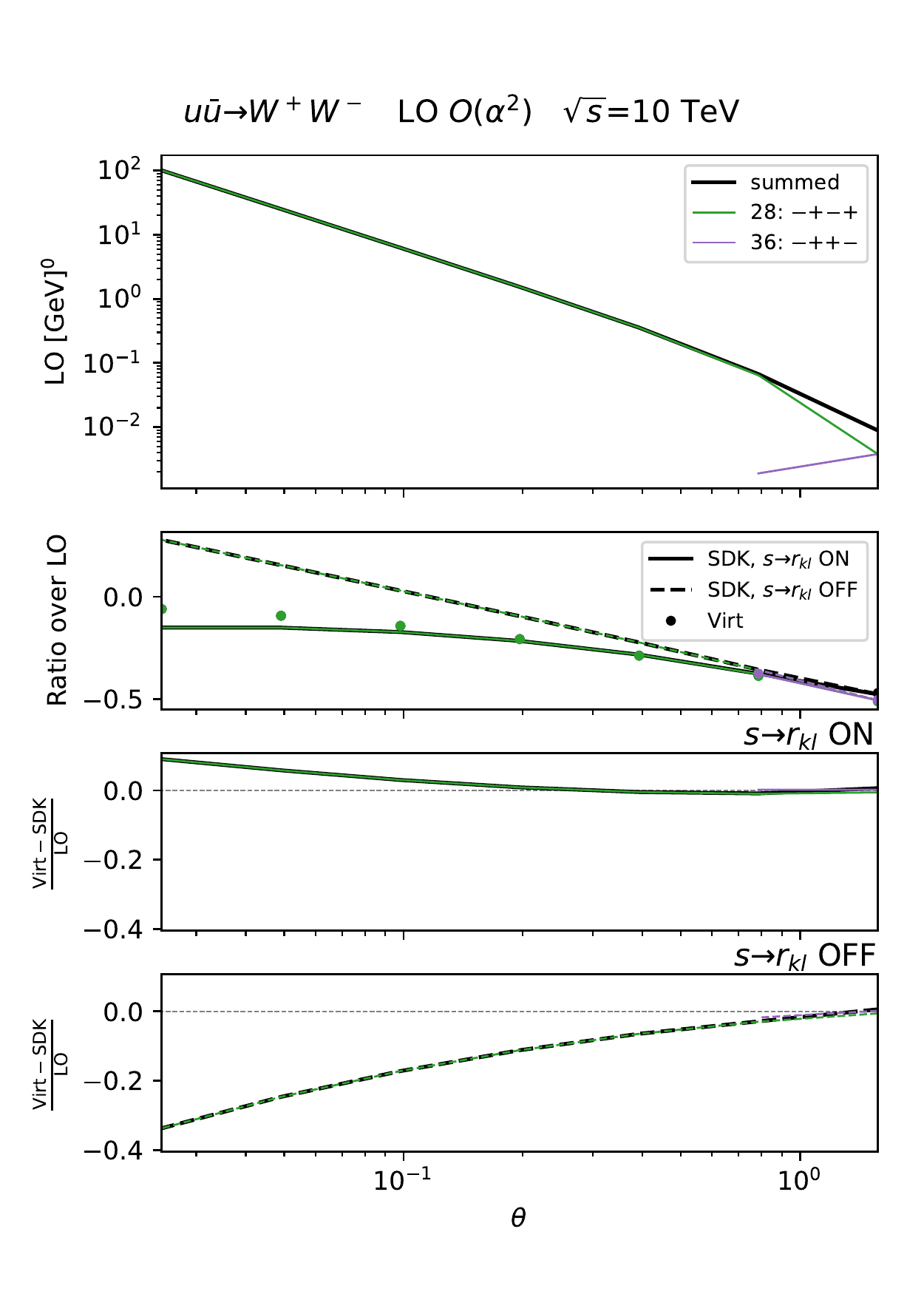}
\includegraphics[width=0.32\linewidth]{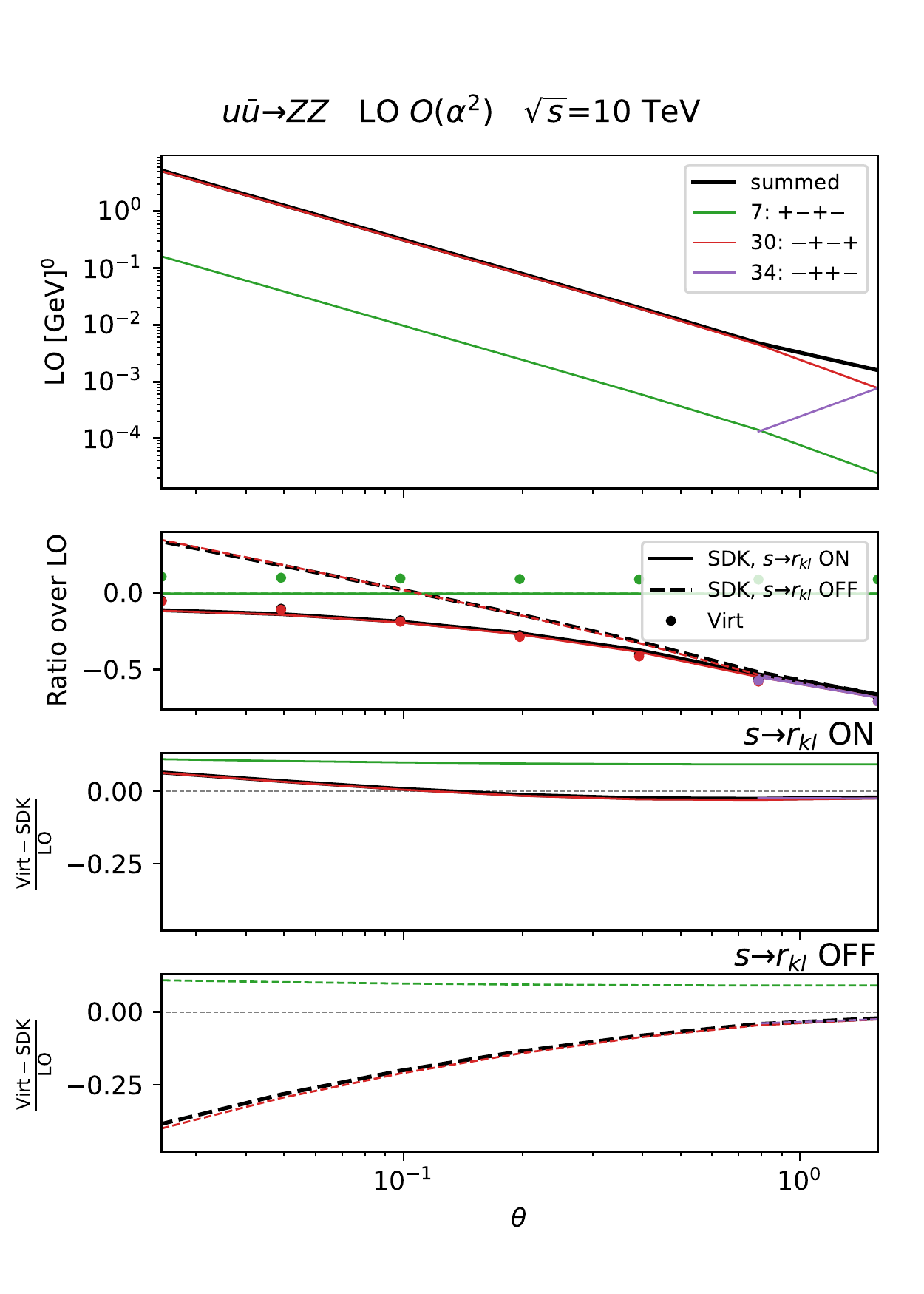}
\end{center}
\caption{Comparison between exact results (dots) for $\ord(\alpha)$ NLO EW virtual corrections and their LA (lines) in the case of squared matrix elements of representative $2\TO2$ processes. Solid lines include the ${\SS^{s\TO r_{kl}}}$ contributions, while dashed lines do not. Upper plots show a scan in energy for a fixed $t/s$ value, while lower plots a scan in the angle $\theta$ between the momenta of the first and third particle. More details are given in the text. \label{fig:rkl_1}}
\end{figure}
\begin{figure}[!t]
\begin{center}
\includegraphics[width=0.32\linewidth]{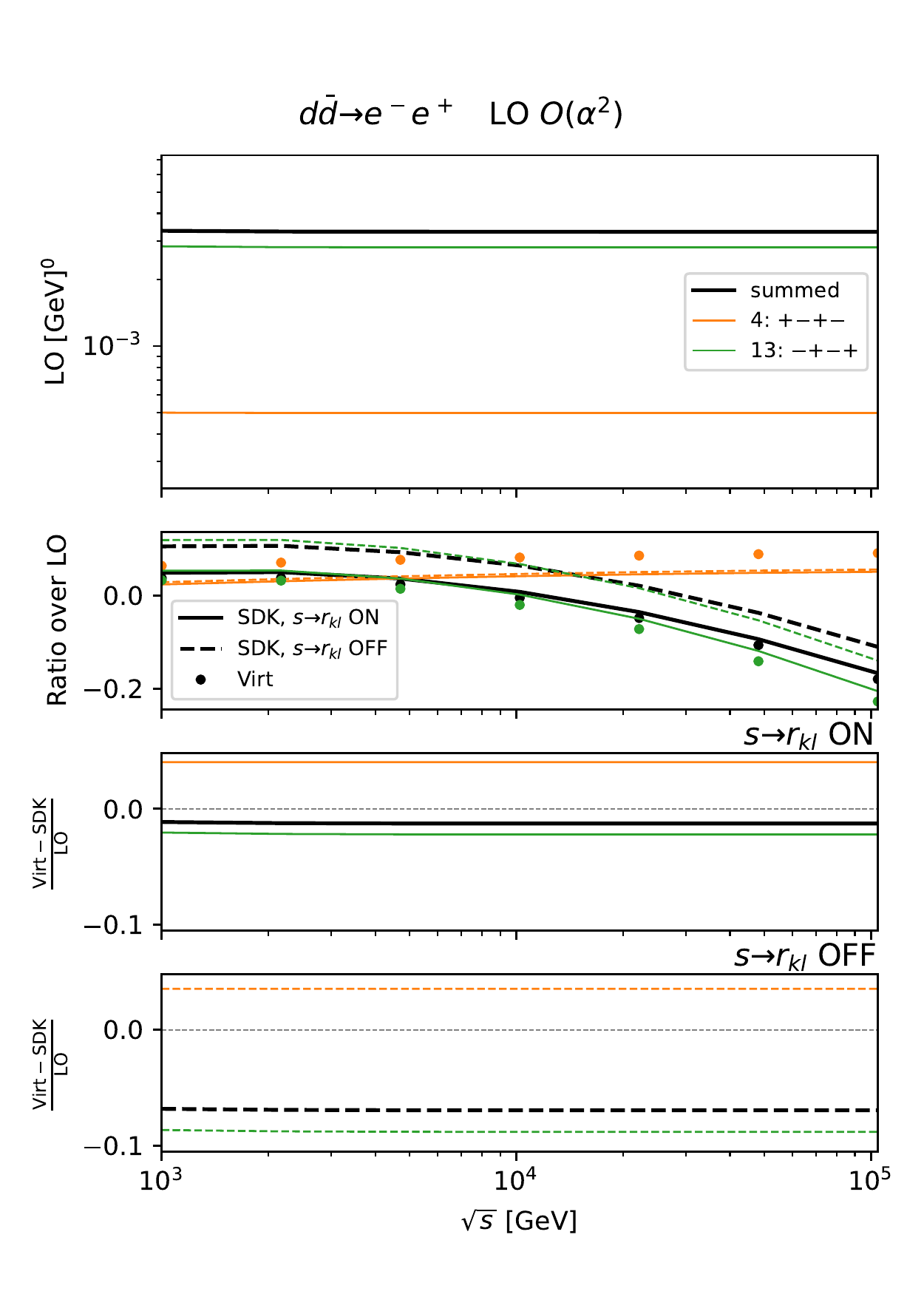}
\includegraphics[width=0.32\linewidth]{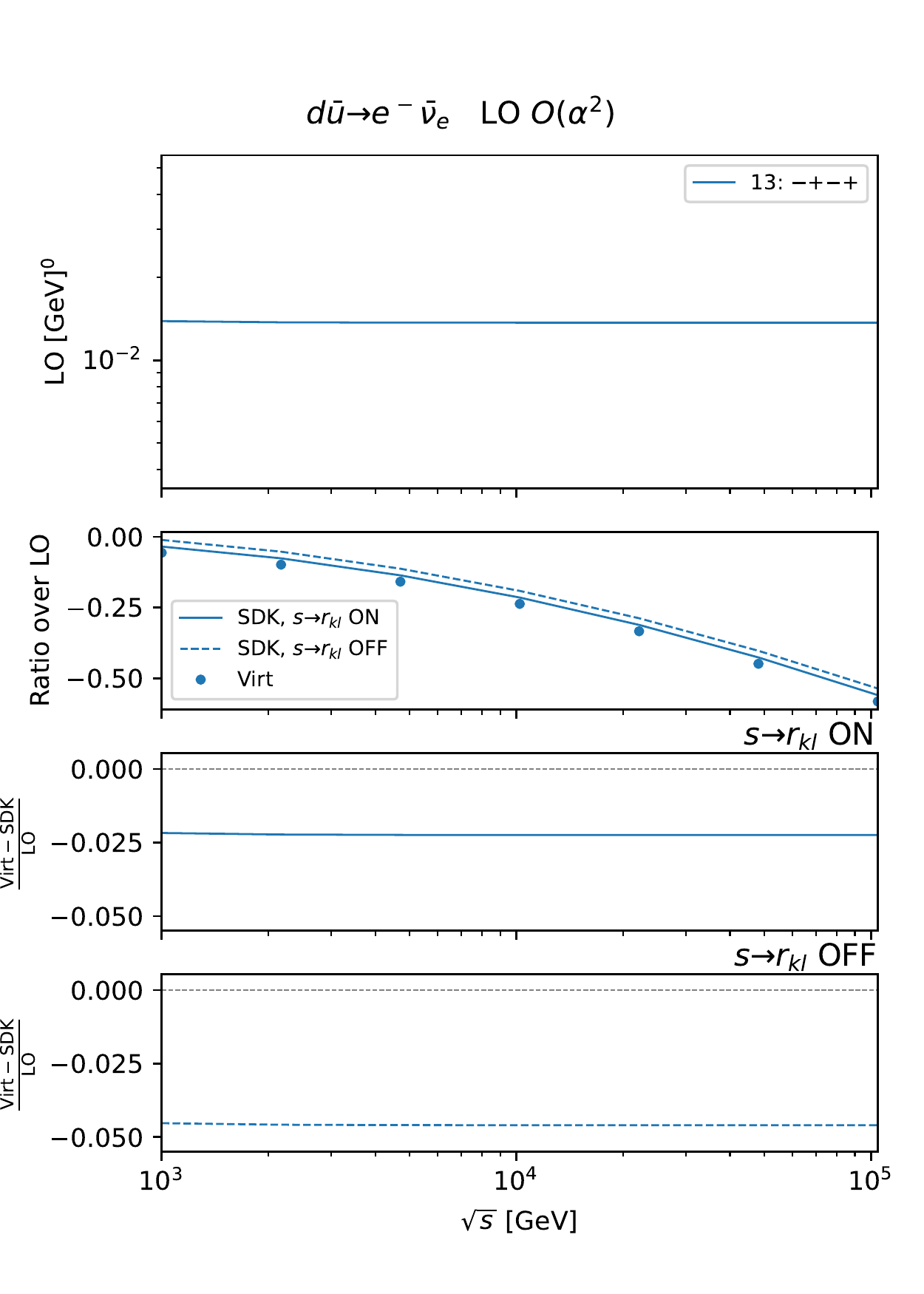}
\includegraphics[width=0.32\linewidth]{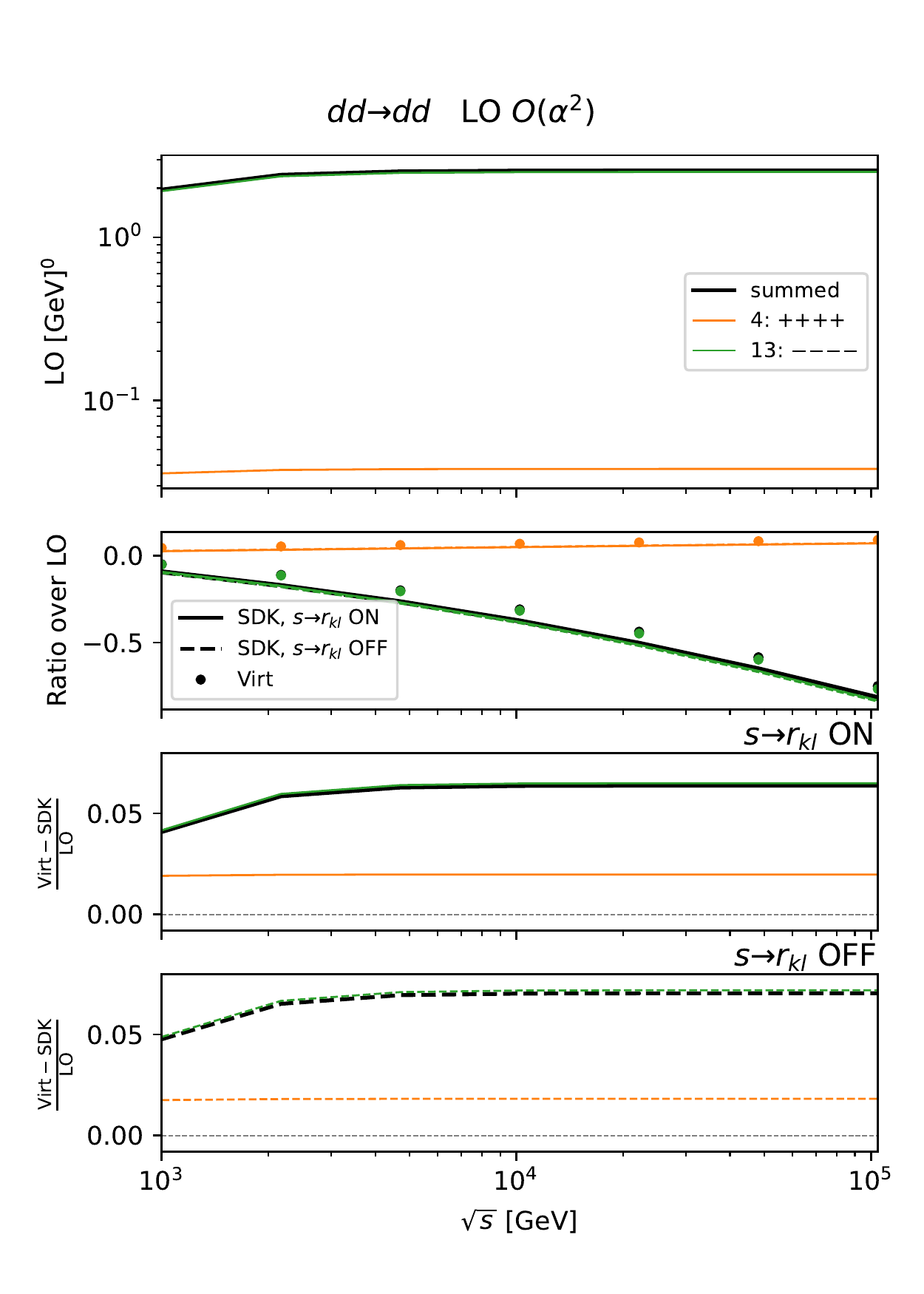}
\includegraphics[width=0.32\linewidth]{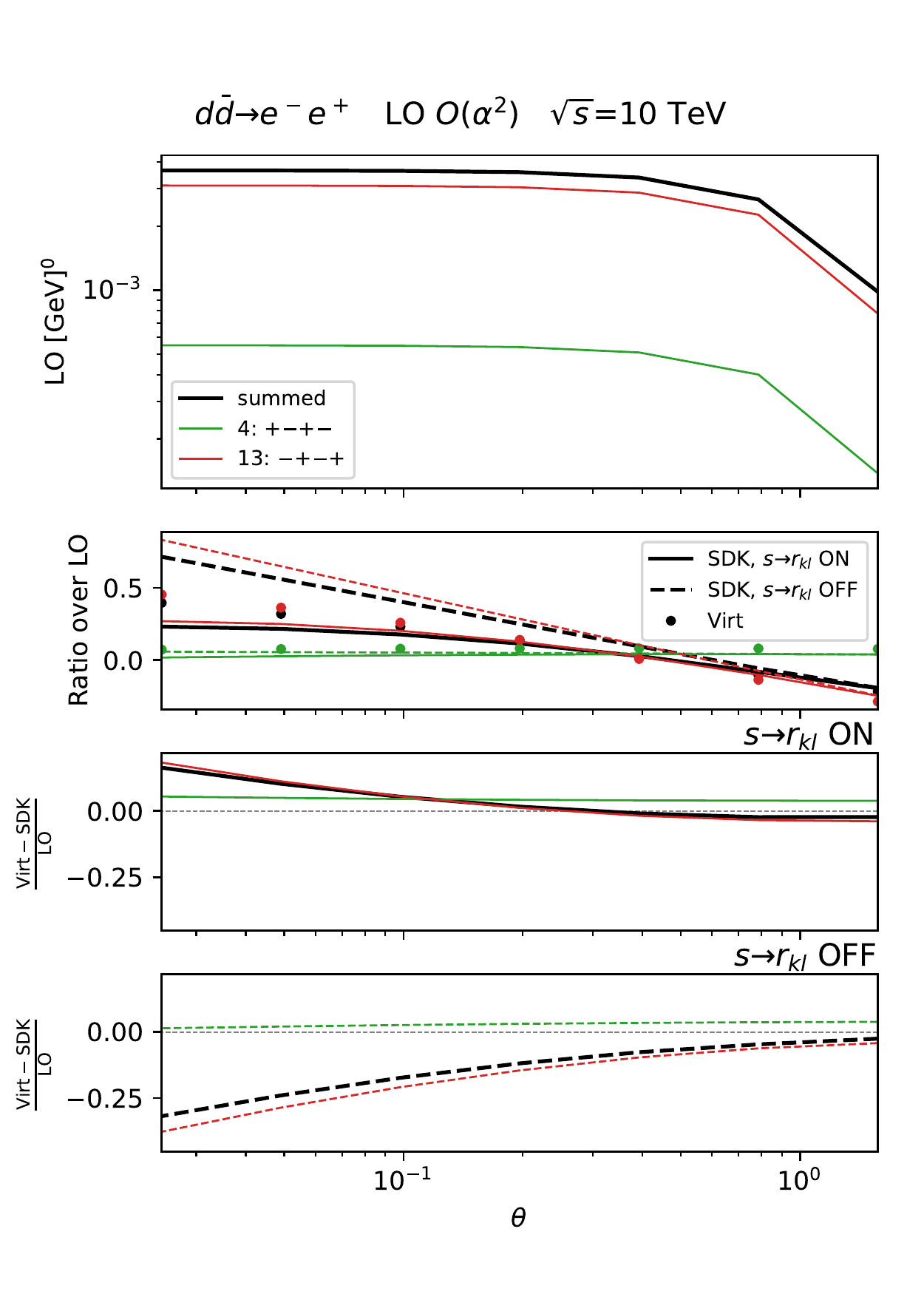}
\includegraphics[width=0.32\linewidth]{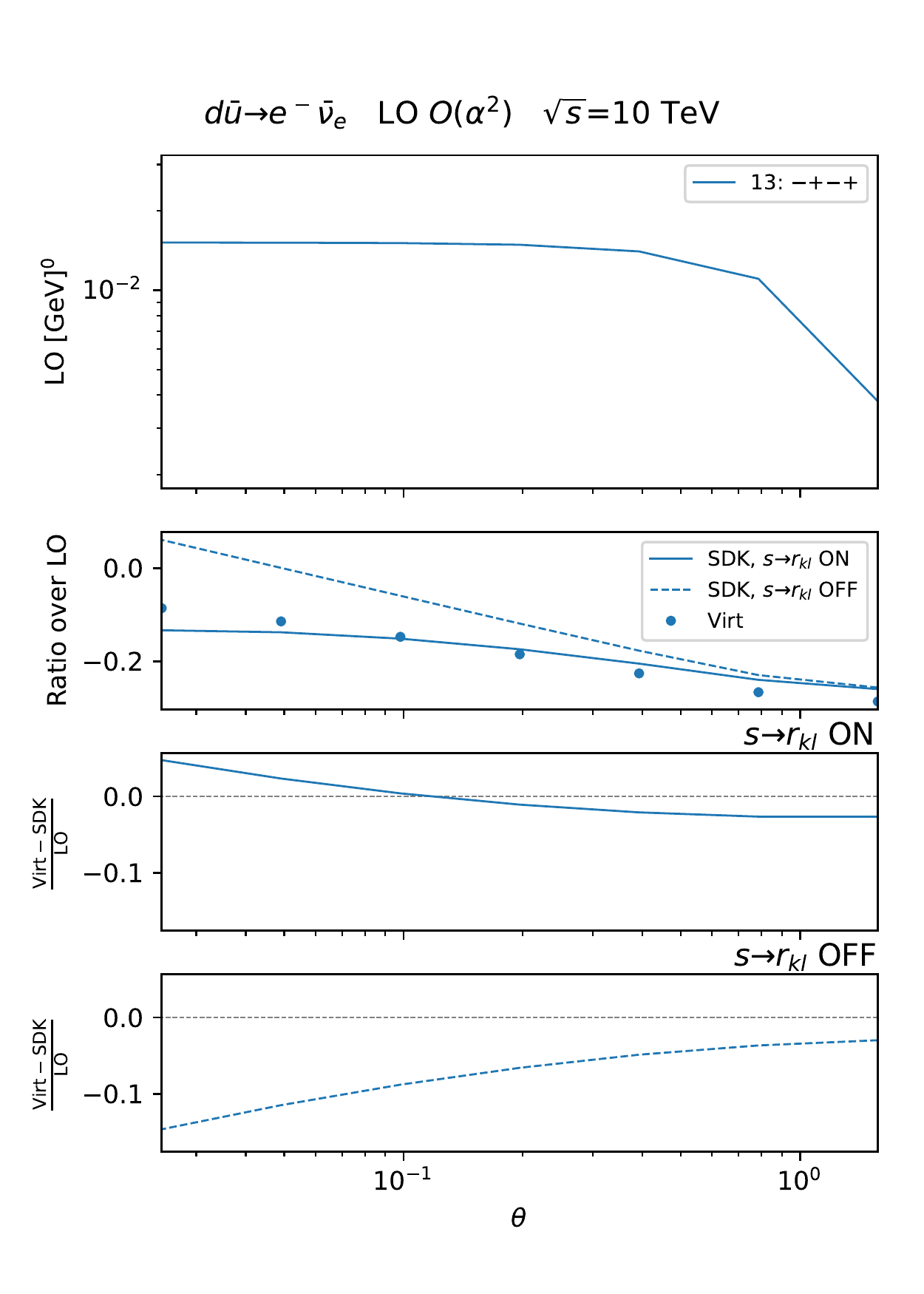}
\includegraphics[width=0.32\linewidth]{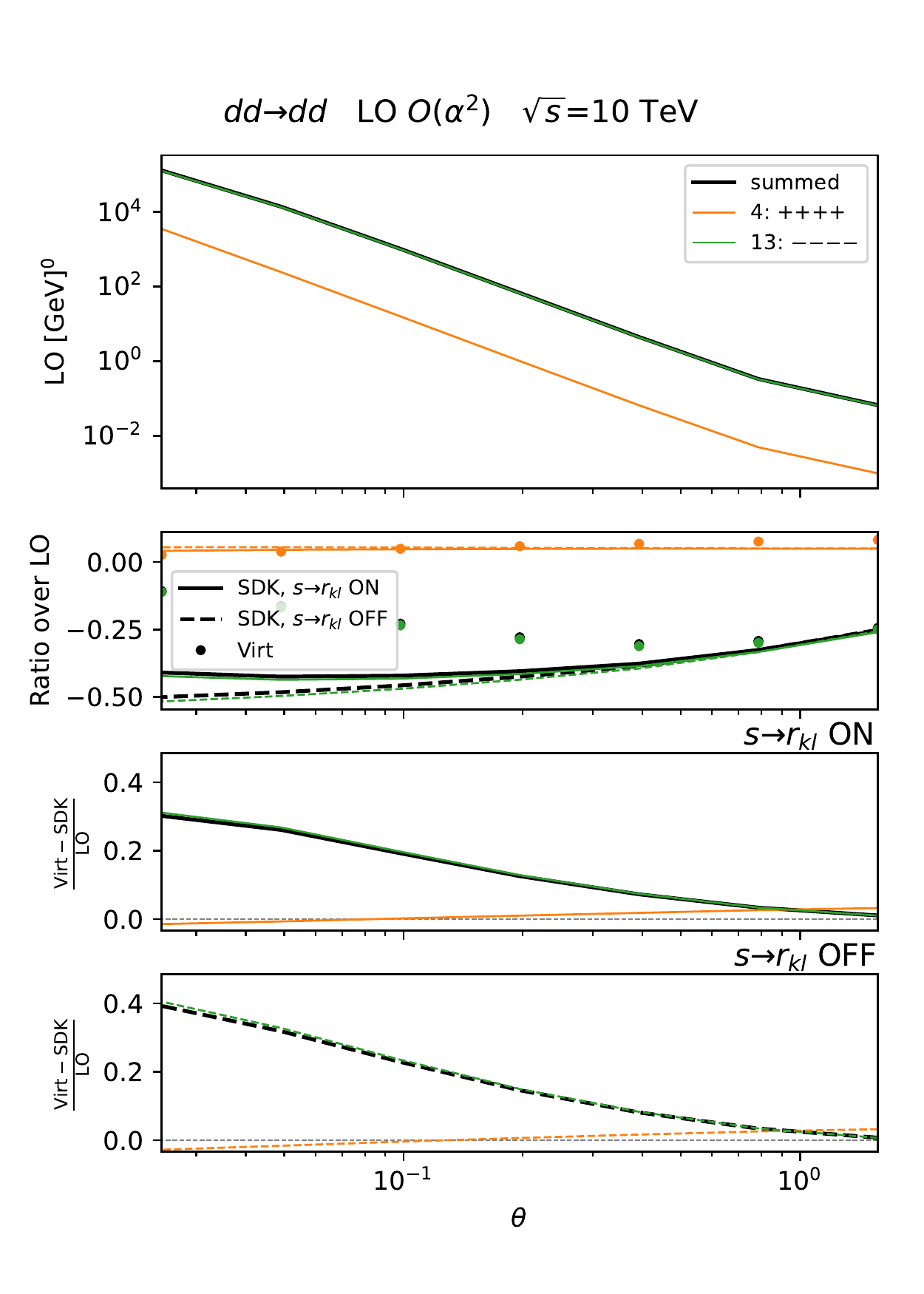}
\end{center}
\caption{Same as Fig.~\ref{fig:rkl_2}, but for a different set of processes. \label{fig:rkl_2}}
\end{figure}
In this section we show numerical results for $2\TO 2$ partonic processes both varying the value of $s$ and the angle $\theta$ between the first and third particle, which in turn parametrises the value of $t$. We select representative processes for which the relevant plots are displayed in Figs.~\ref{fig:rkl_1} and \ref{fig:rkl_2}. In both figures, the plots of each column refer to the same partonic process and the upper plots show the dependence of several quantities on the center-of-mass energy $\sqrt{s}$, while the lower plots show their $\theta$ dependence. In the following, we describe the layout of the plots and how they should be interpreted.

In the first panel we show the value of the LO squared matrix-element, separately for each leading-helicity configuration and possibly their sum if there is  more than one. In order to improve the readability of the legends in the plots, therein we display not only the helicity of any external particle, but also a conventional number associated to the ordering of the helicity configurations within {\mglong}. Conventionally, leading-helicity configurations have been identified as those with a value, for their squared amplitudes, that is at least $10^{-3}$ times the one of the dominant helicity configuration. The main purpose of this conventional choice is to probe and select via a numerical method all the helicity configurations that are not mass suppressed.\footnote{If at least one helicity configuration is not mass suppressed and it is the dominant, the ratio of its squared amplitude and the one of another helicity configuration that is not mass suppressed asymptotically converges to a positive  constant at high energies. Therefore, leading helicities can be present over the entire $s$ range if and only if they are not mass suppressed and this ratio is larger than $10^3$. In other words, if an helicity configuration is mass suppressed, it  is for sure not tagged as leading, while if it is not mass suppressed can be not tagged as leading, but it means its contribution is at less than per-mill level of the dominant-helicity squared amplitude. \label{footnote_a}} In the first inset we show the ratio between the $\ord(\alpha)$ virtual corrections and the LO in different approximations. We display as separate dots the exact results  obtained via {\ml} (Virt)  for selected values of $s$, while as lines\footnote{Lines are obtained via the  interpolation of the results of the LA approximation of Sudakov logarithms obtained  for the same $s$ values for which the exact one-loop results from {\ml}, namely the dots, are calculated.} the LA approximation of Sudakov logarithms that are obtained via the new implementation of the modified {\denpoz} algorithm described in this work. Dashed lines refer to the pure LA  (${\SS^{s\TO r_{kl}}}$ terms not included), denoted in the plots as ``${\rm SDK}, s\TO r_{kl} \rm~OFF$'', while the solid lines to the case in which   ${\SS^{s\TO r_{kl}}}$ terms are taken into account, denoted in the plots as ``${\rm SDK}, s\TO r_{kl} \rm~ON$''. As expected, the values of the ratio over LO for both dots and lines are negative and grow in absolute value for large values of $s$. A correct implementation and evaluation of the LA of Sudakov logarithms implies that the differences between each line and the dots converge to a constant value for $s \TO \infty$. Indeed, since all the mass-suppressed terms of $\ord(\alpha)$ corrections  go to zero for large $s$, the terms that survive  are either logarithmic enhanced, those that have to be exactly captured by the LA (lines), or constant for $t/s$ fixed. We therefore separately display the interpolation of the difference between the dots and the solid line (second inset) and between the dots and the dashed line (third inset). These quantities are denoted as (Virt-SDK)/LO in the plots.
The layout of the lower plots of Figs.~\ref{fig:rkl_1} and \ref{fig:rkl_2} is very similar to the one of the upper plots, however, in this case the $x$-axis refers to the angle $\theta$  between the first and third particle, which in turn parametrises the value of $t$, in the range $10^{-2}\lesssim\theta\lesssim\pi/2$. We have fixed the value of $s$ to $\sqrt{s}=10~\rm{TeV}$ for all lower plots.

 In order to produce the upper plots, the scan in $\sqrt{s}$ with $t/s$ fixed, we have performed the following procedure. We start by generating the  momenta for a phase-space point with $\sqrt{s}=10^3~\gev$ and $t/s=-1/20$ for the specific process considered. Then, we iteratively repeat the following steps for increasing the value of $\sqrt{s}$ by keeping fixed the $t/s$ ratio within an error of the order of permille. First, we rescale the trimomenta of the outgoing particles by a common factor. Second, we impose on-shell conditions for the outgoing particles in order to obtain their energies. Finally, we impose momentum conservation for determining the momenta of the initial state. In this way, we can generate several phase-space points by scanning the $\sqrt{s}$ range and keeping the ratio $t/s$ very stable. Each one of the phase-space points obtained is then used as input for evaluating the exact virtual NLO EW corrections of $\ord(\alpha)$ as well the LA with and without the inclusion of the ${\SS^{s\TO r_{kl}}}$ terms. The  ${\rm SDK}, s\TO r_{kl} \rm~ON$ and the ${\rm SDK}, s\TO r_{kl} \rm~OFF$ lines are the interpolation of these LA results.

As can be seen in both Figs.~\ref{fig:rkl_1} and \ref{fig:rkl_2}, all the second and third insets of upper plots show perfectly horizontal lines for large values of $s$, for each individual helicity configuration. We have shown here only representative processes, but we did not see any exception in all cases that we have checked. This is a clear sign of a correct implementation of the LA of Sudakov logarithms.

 In order to rigorously check the last statement, we have fitted the quantities (Virt-SDK)/LO via a function of the form 
\begin{equation}
A \log_{10}(\sqrt{s}/[1~\gev])+B \label{eq:fit}\,,
\end{equation}
with the method of least squares. While the coefficient $B$ has been found in general of the order of few percents for the plots shown here, the quantity $A$ is in general of the order of $10^{-4}$  and compatible with 0 due to the associated statistical error,\footnote{We remind the reader that statistical errors also include effects induced by the numerical method that is used for performing the derivatives, which is discussed in Sec.~\ref{sec:derivative}, as well as by possible instabilities of the evaluation of exact virtual amplitudes. } therefore supporting our previous statement about the correct implementation of the LA of Sudakov logarithms.

If we consider Fig.~\ref{fig:rkl_1}, comparing the second and third inset of the upper plots we can also appreciate how  ${\SS^{s\TO r_{kl}}}$ terms further reduce the gap between the LA and the exact value of the virtual. Being $|t|/s=1/20$, these terms are constant when varying $s$, but still non-negligible. In other words, their inclusion preserves the LA and further improves the approximation of the exact result at high energy. In the lower plots, the impact of the ${\SS^{s\TO r_{kl}}}$ terms can be better understood. 
By varying $\theta$, the difference between the strict LA and the exact result is not expected to be a constant times the LO, because the condition \eqref{eq:rijnice} is assumed. This can be observed in the second and third insets. However, although in both insets lines are not perfectly horizontal, one can notice how the ${\SS^{s\TO r_{kl}}}$ terms substantially reduce the slope. In other words, as can be also seen in the first inset, they lead to a better approximation of the exact results also at small angles. Clearly, at very small angles power-suppressed terms may not be negligible, since the $\MW^2/|t|$ ratio increases and this in general leads to non-horizontal lines even with  the inclusion of the ${\SS^{s\TO r_{kl}}}$ terms. These effects are enhanced for the left and central plots of  Fig.~\ref{fig:rkl_2} and even more for the right one, the $\LO_3 \propto \alpha^2$ of the $dd\TO dd$ process,  which deserves some more comments. 

By looking at the right-upper plot, it is clear that the LA works well and both the second and third insets show perfectly horizontal lines, however, by looking the lower plots it is manifest that both with or without the inclusion of ${\SS^{s\TO r_{kl}}}$ terms the $\theta$ dependence cannot be correctly captured, although better approximated in the former case. First, it is important to note, as can also be seen in the main panel of the bottom-right plot, that the LO cross section diverges for $\theta\TO 0$, indeed it is proportional to $1/t^2$. Actually, if condition \eqref{eq:Sudaklim} is not assumed, not only terms proportional to $1/t^2$ (the photon $t$-channel propagator) are present, but also terms proportional to $1/(t^2-\MZ^2)$ (the $Z$-boson $t$-channel propagator) appear. Approaching the regime with $|t|=(1-\cos(\theta))s/2\simeq \MZ^2$, which implies $\theta\TO \sim 0.02$ for $\sqrt{s}=10$ TeV,  condition \eqref{eq:Sudaklim} is not valid and therefore even including ${\SS^{s\TO r_{kl}}}$ terms a good approximation is not expected.\footnote{We have noticed an additional interesting behaviour in the $\theta$ dependence, which is not shown here in the plot. When even smaller angles are considered, $|t|\ll \MZ^2$, while the ${\rm SDK}, s\TO r_{kl} \rm~OFF$ predictions depart even more from the exact result, the ${\rm SDK}, s\TO r_{kl} \rm~ON$ approximation considerably improves. This is due to the fact that, for the $\LO_3 \propto \alpha^2$ of the $dd\TO dd$ process,  the QED contribution is dominant in this regime and furthermore it does not involve neither $\MZ$ nor $\MW$. } Second, the ${\SS^{s\TO r_{kl}}}$ correctly takes into account the value of the different $r_{kl}$ invariants entering eq.~\eqref{eq:angsplitnew}, but this equation derives from eq.~\eqref{eq:C0}, which by itself does not depend  on ratios of different invariants. This means, for instance for $2\TO2$ processes, that additional corrections that involve the $(s/t)$ ratio and have a functional form different from the terms included in the {\denpoz} algorithm cannot be recovered, even with the inclusion of ${\SS^{s\TO r_{kl}}}$ terms. In other words, while the correctness of LA is guaranteed and ${\SS^{s\TO r_{kl}}}$ terms further improve the approximation keeping track of the correct dependence on the different $r_{kl}$ invariants in eq.~\eqref{eq:angsplitnew}, this does not mean that the full dependence is retained. In order to do that, the information on the internal structure of the Feynman diagrams would be necessary. 
Indeed, the starting point in the derivation of the  ${\SS^{s\TO r_{kl}}}$ terms is the $C_0$ function in \eqref{eq:C0}, which is associated to simply the masses and the invariant mass of two external particles involved in the process. However, already with $2\TO 2 $ processes, $D_0$ functions can appear in virtual corrections, involving also at high energies more than one invariant and leading to additional terms when the condition \eqref{eq:rijnice} is not satisfied.

\subsection{Impact of the imaginary component}
\label{sec:Im}

\begin{figure}[!t]
\begin{center}
\includegraphics[width=0.32\linewidth]{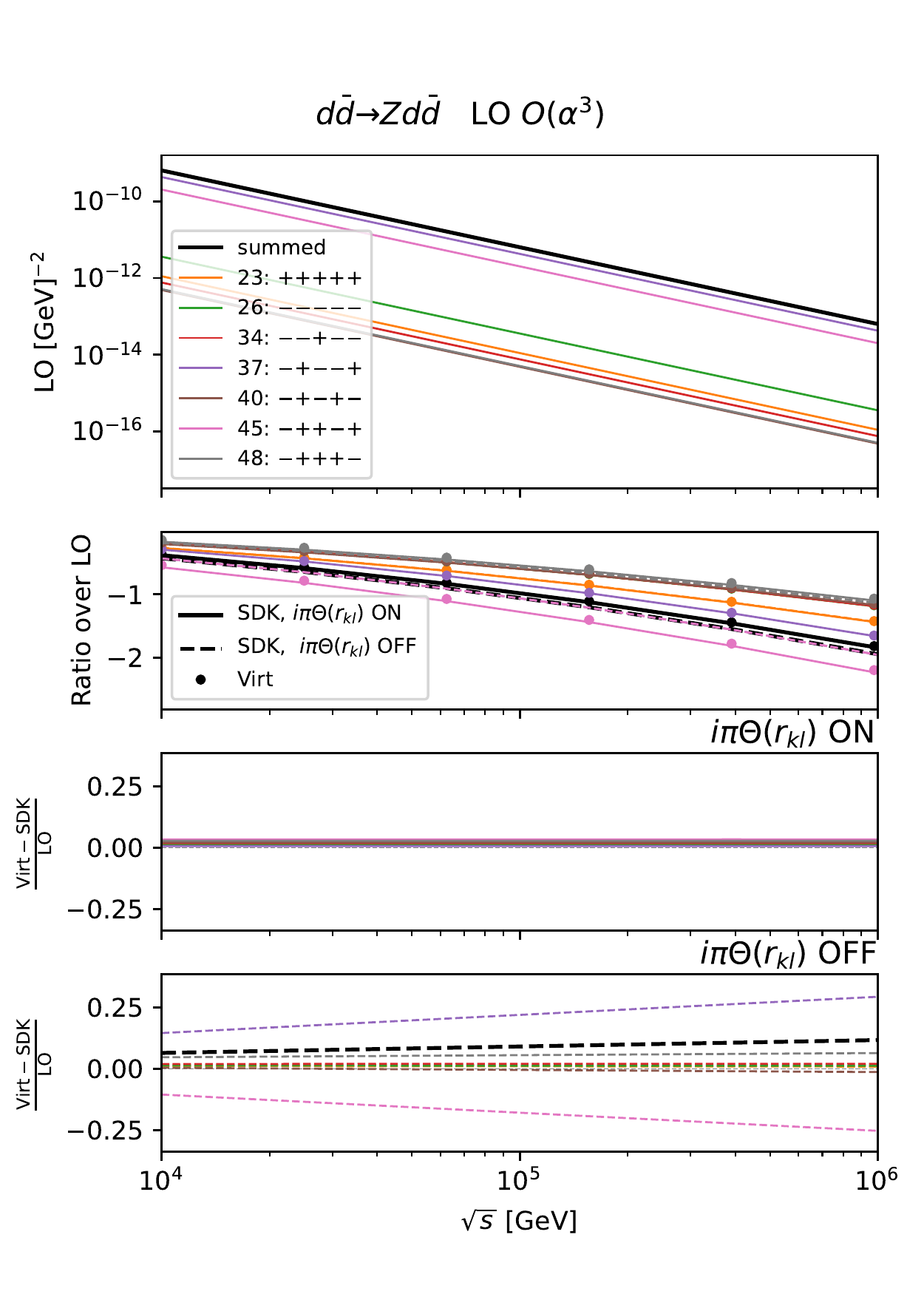}
\includegraphics[width=0.32\linewidth]{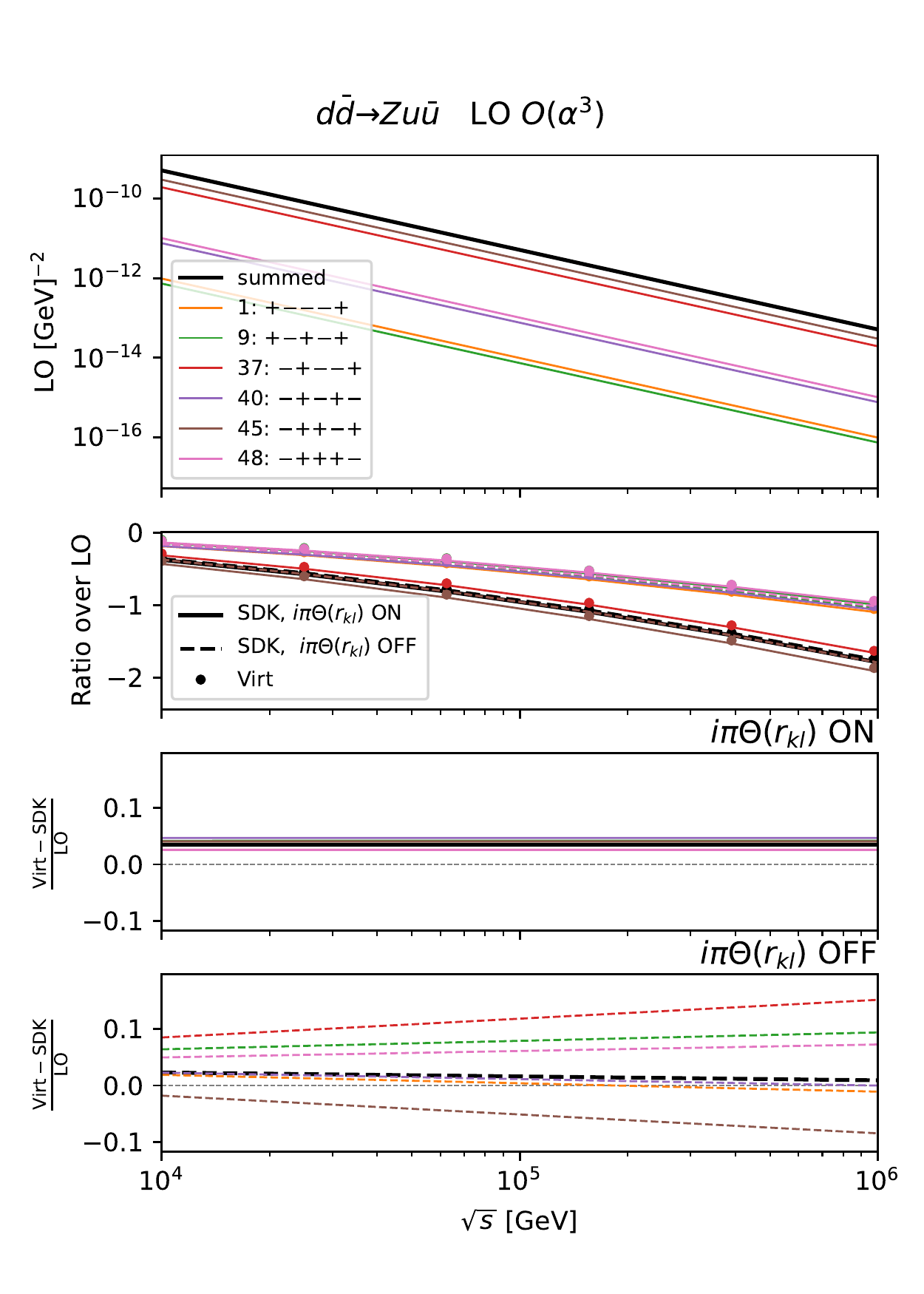}
\includegraphics[width=0.32\linewidth]{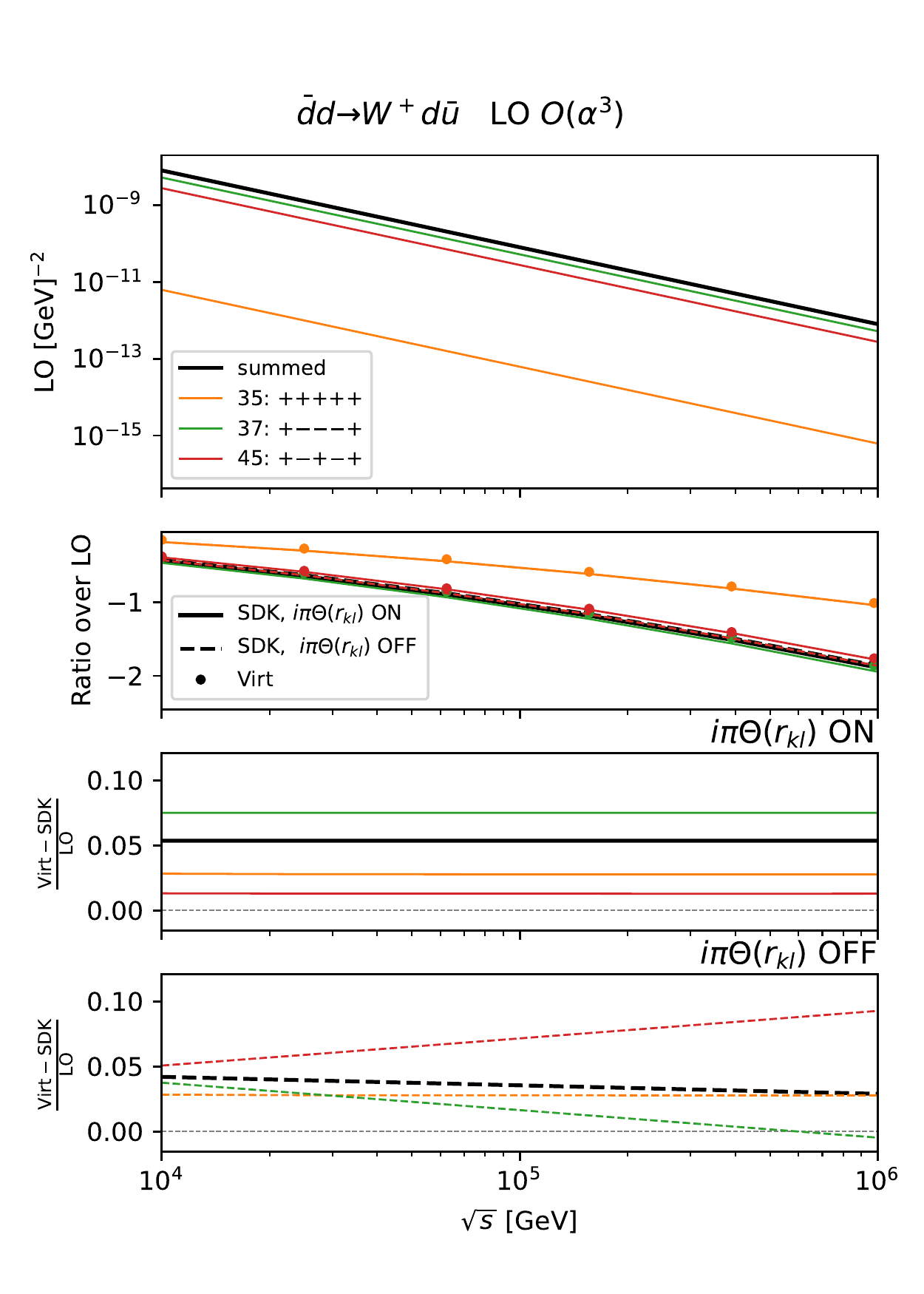}
\includegraphics[width=0.32\linewidth]{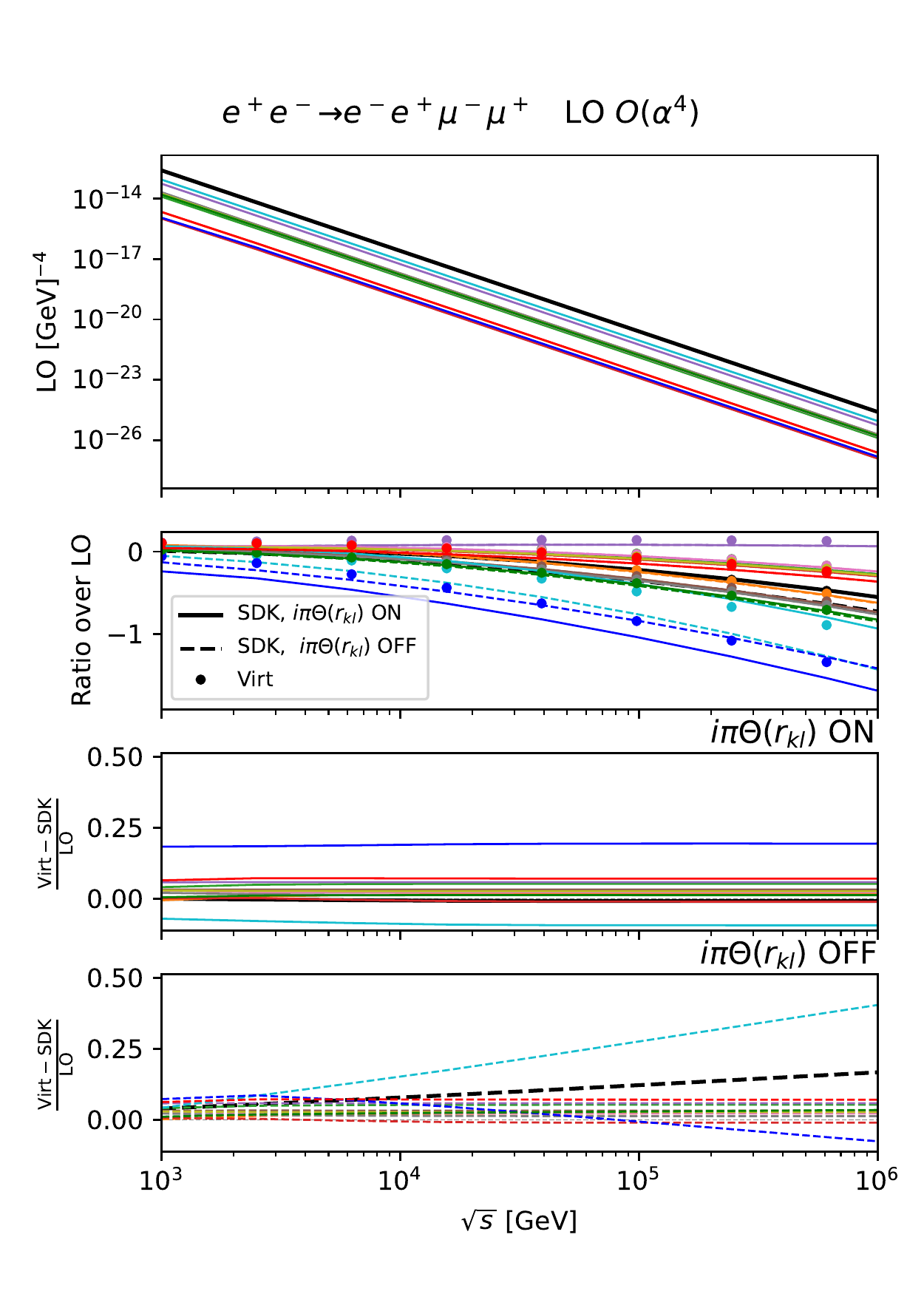}
\includegraphics[width=0.32\linewidth]{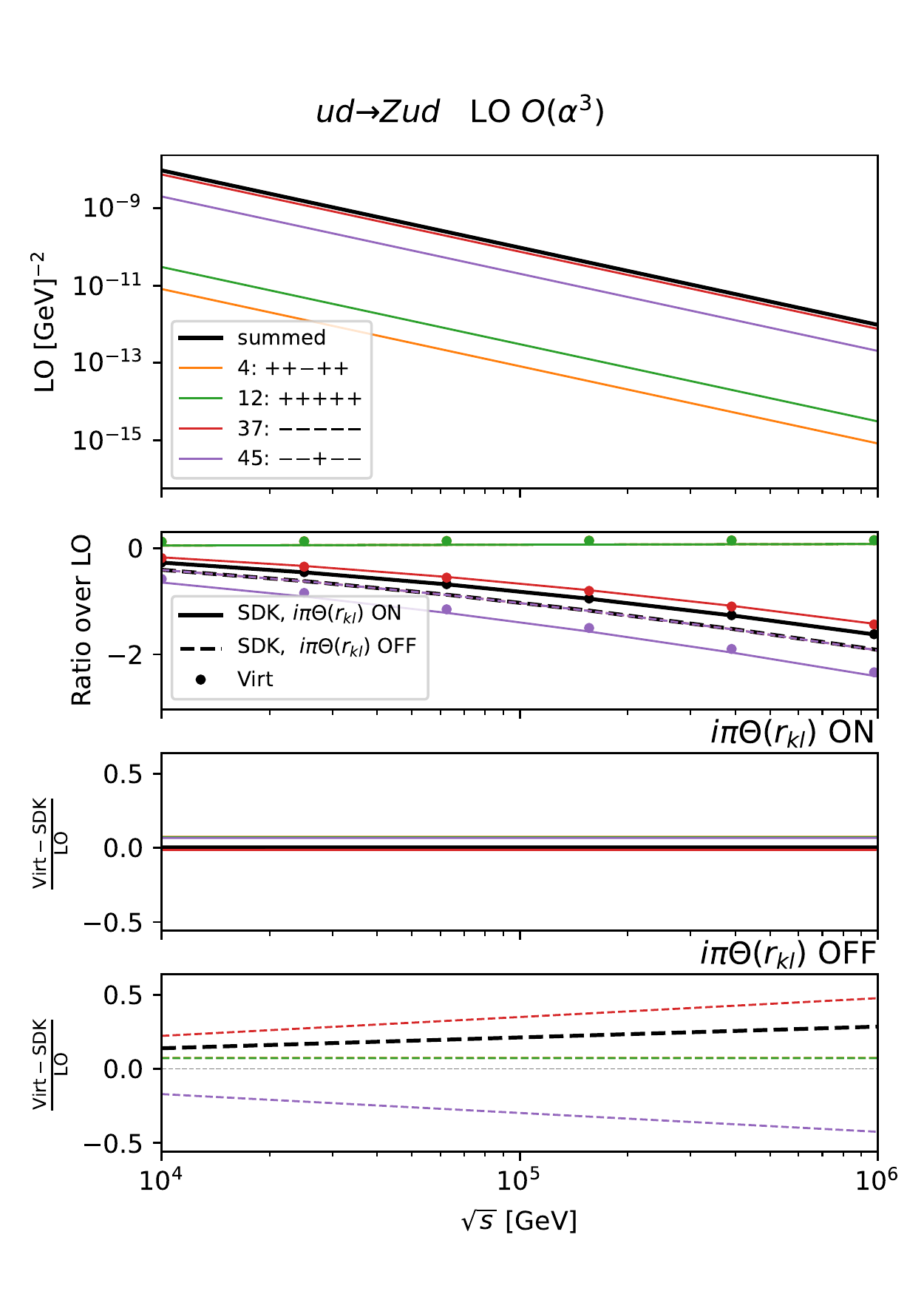}
\includegraphics[width=0.32\linewidth]{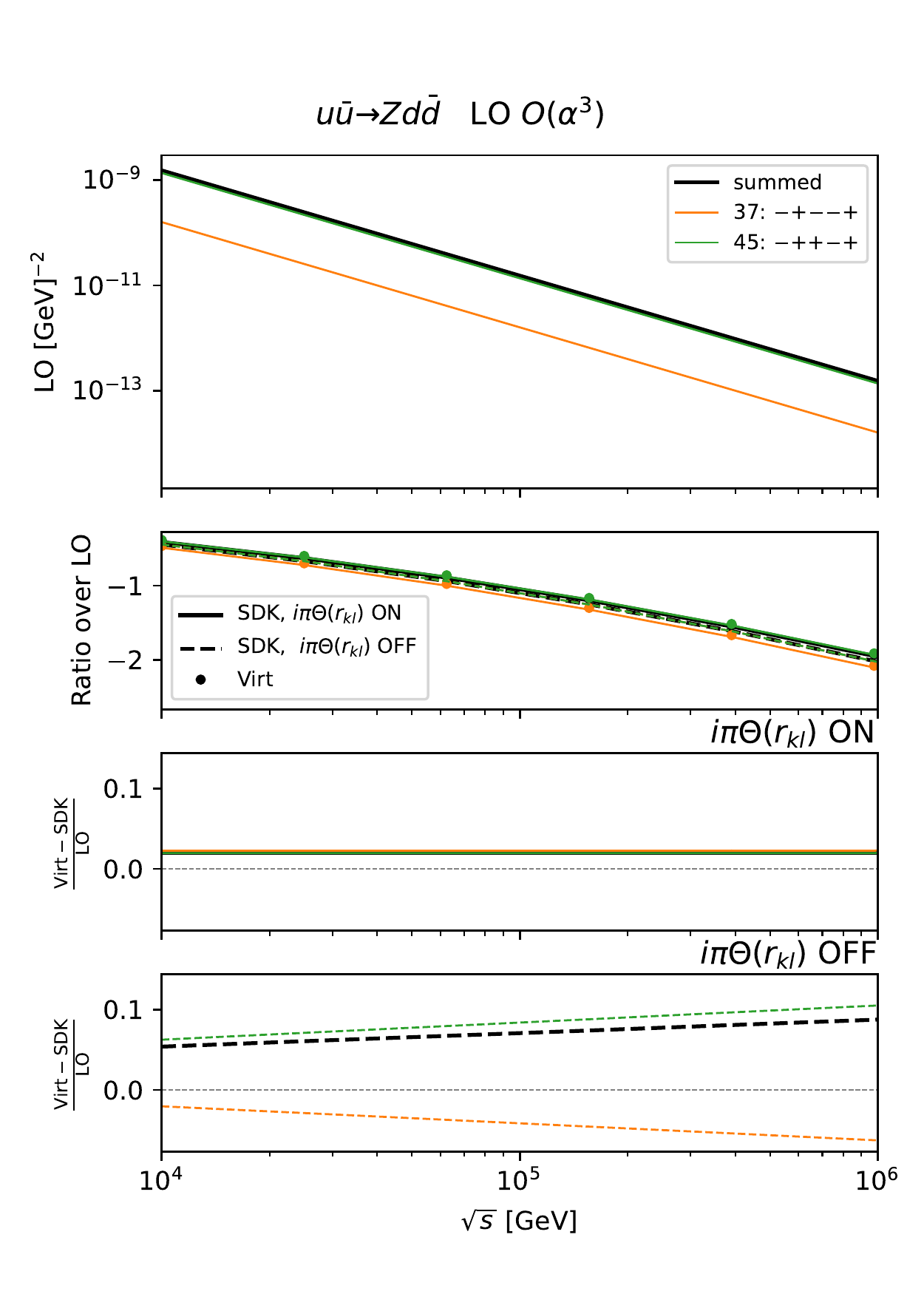}

\end{center}
\caption{Comparison between exact results (dots) for $\ord(\alpha)$ NLO EW virtual corrections and their LA (lines) in the case of squared matrix elements of representative $2\TO n$ processes with $n=3,4$. Solid lines include the contributions proportional to $i\pi \Theta(r_{kl})$, while dashed lines do not. Plots show a scan in energy for fixed $r_{kl}/s$ ratios. More details are given in the text.  \label{fig:Im}}

\end{figure}
As explained in Sec.~\ref{sec:logsplit}, in the original work of Ref.~\cite{Denner:2000jv} an imaginary component has been omitted in the formulas. On the other hand, this component affects results only for $2\TO n$ processes with $n > 2$.
In this section we show numerical results about this aspect, for $2\TO n$ partonic processes with $n=3,4$.\footnote{For all the $2\TO3$ processes we considered as the Born the $\LO_i$ with the highest possible value of $i$, since it  receives a contribution  due to the term $i  \pi \Theta(r_{kl})$ that is larger than in the case of the $\LO_1$.}  Again, we select representative processes for which the relevant plots are displayed in Fig.~\ref{fig:Im}. Each plot shows the dependence on $s$ of several quantities, and the layout is very similar to the one of the upper plots of Figs.~\ref{fig:rkl_1} and \ref{fig:rkl_2}. Here, the  LA always includes the ${\SS^{s\TO r_{kl}}}$ terms,\footnote{For brevity, in this section we will write in the plots only SDK and not ${\rm SDK}, s\TO r_{kl} \rm~ON$ as in Sec.~\ref{sec:rijvss}.} but we distinguish the case in which the terms proportional to $i  \pi \Theta(r_{kl})$ in eqs.~\eqref{eq:subdl1}--\eqref{eq:deltasr} are excluded, as in the original {\denpoz} algorithm in Ref.~\cite{Denner:2000jv}, or retained. The former are displayed as dashed lines ($i  \pi \Theta(r_{kl})$ OFF) and the latter as solid lines ($i  \pi \Theta(r_{kl})$ ON). For each leading helicity configuration, we also show in the second and third inset the difference between the LA and the exact result both normalised to the LO, respectively with and without taking into account the imaginary component.

In order to produce the plots, scanning in $\sqrt{s}$, we have performed a  procedure similar to the one explained in the previous section for the upper plots in Figs.~\ref{fig:rkl_1} and \ref{fig:rkl_2}. The only difference here is the starting point. For $2\TO n$ partonic processes with $n>2$, besides $s$, there is more than only one independent kinematic invariant that can be built with the external momenta. In order to avoid pathological configurations with an $|r_{kl}|\simeq \MW^2$, we randomly generate the first set of external momenta setting $\sqrt{s}=10^4~\gev$ and requiring
\begin{equation}
   \frac{ |r_{kl}|}{s}>\frac{1}{8}\qquad \forall ~r_{kl}\,. \label{eq:minfrac}
\end{equation}
We remind the reader, as already explained in footnote \ref{footnote1}, that eq.~\eqref{eq:minfrac} is a condition that can be satisfied for  $2\TO 3$ or $2\TO 4$ processes, but not in general for $2\TO n$, for which this lower bound has to be  lowered more and more  increasing the value of $n$, further departing from the condition of eq.~\eqref{eq:rijnice}. 

Looking at Fig.~\ref{fig:Im}, it is manifest how the case including terms proportional to $i  \pi \Theta(r_{kl})$ correctly catches the LA, while the other one does not; perfectly horizontal lines are present in the second inset, while in the third inset a dependence on $s$ is clearly visible. For some of the processes considered, such as $d \bar d \TO Z d \bar d$, this dependence seems to cancel out for the sum over the different helicity configurations. In large part this is correct, but a small dependence is still present and it is simply not visible from the plot. We in general see this feature also for individual helicity configurations, namely the $i  \pi \Theta(r_{kl})$ is often formally relevant but sometimes the numerical effect is very small. For other processes, such as $e^+e^-\TO e^+e^-\mu^+\mu^-$ or $u d  \TO Z u d $,  even for the helicity-summed result the lack of the terms proportional to $i  \pi \Theta(r_{kl})$ leads to sizeable numerical effects.

\renewcommand{\arraystretch}{1.3}

\begin{table}
\scriptsize
\begin{tabular}{c|cc|cc}
 & \multicolumn{2}{c}{$i\pi \Theta(r_{kl})$ ON } & \multicolumn{2}{c}{$i\pi \Theta(r_{kl})$ OFF } \\ 
 \hline
 Helicity & $A$ & $B$   & $A$   & $B$  \\ 
 \hline
 summed   &   $ (  5 \pm  28 ) \cdot 10^{-6} $  &  $ ( 1.55 \pm 0.01 ) \cdot 10^{-2} $    &    $ ( 2.622 \pm 0.004 ) \cdot 10^{-2} $  &  $ ( -4.00 \pm 0.02 ) \cdot 10^{-2} $     \\ 
$ 23: +++++ $  &   $ ( 1.3 \pm 4.2 ) \cdot 10^{-6} $  &  $ ( 1.318 \pm 0.002 ) \cdot 10^{-2} $    &    $ ( -1.07 \pm 0.02 ) \cdot 10^{-3} $  &  $ ( 1.56 \pm 0.01 ) \cdot 10^{-2} $     \\ 
$ 26: ----- $  &   $ ( 2.2 \pm 6.7 ) \cdot 10^{-6} $  &  $ ( 1.380 \pm 0.003 ) \cdot 10^{-2} $    &    $ ( -3.1 \pm 0.2 ) \cdot 10^{-4} $  &  $ ( 1.44 \pm 0.01 ) \cdot 10^{-2} $     \\ 
$ 34: --+-- $  &   $ ( 10 \pm  10 ) \cdot 10^{-5} $  &  $ ( 1.86 \pm 0.05 ) \cdot 10^{-2} $    &    $ ( 4.1 \pm 0.8 ) \cdot 10^{-4} $  &  $ ( 1.80 \pm 0.04 ) \cdot 10^{-2} $     \\ 
$ 37: -+--+ $  &   $ ( 2.7 \pm 6.6 ) \cdot 10^{-5} $  &  $ ( 6.4 \pm 0.3 ) \cdot 10^{-3} $    &    $ ( 7.408 \pm 0.008 ) \cdot 10^{-2} $  &  $ ( -1.503 \pm 0.004 ) \cdot 10^{-1} $     \\ 
$ 40: -+-+- $  &   $ ( -8.0 \pm 8.0 ) \cdot 10^{-5} $  &  $ ( 2.24 \pm 0.04 ) \cdot 10^{-2} $    &    $ ( -8.72 \pm 0.09 ) \cdot 10^{-3} $  &  $ ( 3.89 \pm 0.04 ) \cdot 10^{-2} $     \\ 
$ 45: -++-+ $  &   $ ( -5 \pm  50 ) \cdot 10^{-6} $  &  $ ( 3.42 \pm 0.02 ) \cdot 10^{-2} $    &    $ ( -7.405 \pm 0.005 ) \cdot 10^{-2} $  &  $ ( 1.909 \pm 0.003 ) \cdot 10^{-1} $     \\ 
$ 48: -+++- $  &   $ ( -5.2 \pm 3.3 ) \cdot 10^{-6} $  &  $ ( 2.901 \pm 0.001 ) \cdot 10^{-2} $    &    $ ( 8.631 \pm 0.009 ) \cdot 10^{-3} $  &  $ ( 1.251 \pm 0.004 ) \cdot 10^{-2} $     
\end{tabular}
\caption{Result of the fit of the quantity (Virt-SDK)/LO using the method of least squares and the function \eqref{eq:fit} for the representative process $d \bar d \TO Z d \bar d$ at $\ord (\alpha^3)$. The case including(excluding) the contribution of $i\pi \Theta(r_{kl})$ corresponds to the quantities shown in the second(third) inset of the upper-left plot of Fig.~\ref{fig:Im}.  \label{tab:zdd}}
\end{table}
In order to provide a more quantitative statement, we list in Tab.~\ref{tab:zdd} the results of the fit of (Virt-SDK)/LO for each leading-helicity configuration (and their sum) of the process $d \bar d \TO Z d \bar d$. We have used again the method of least squares and the functional form of eq.~\eqref{eq:fit}. As can be seen in the third column of Tab.~\ref{tab:zdd}, all helicities exhibit a non-vanishing slope when  the terms proportional to $i\pi \Theta(r_{kl})$ are turned off. Notably, as anticipated before, this happens also for the sum over the helicities, which for this particular process and kinematic configuration (condition \eqref{eq:minfrac}) leads to a cumulative error of $2.6\%$ in the LA for every factor of 10 in increase of the  energy. The error is process dependent and  can also be larger, as can be seen in the bottom-left plot of Fig.~\ref{fig:Im} for the  $e^+e^-\TO e^+e^-\mu^+\mu^-$ process.

Finally, given Fig.~\ref{fig:Im}, we would like to stress  how well the LA of Sudakov terms can work  in the high-energy regime. All these processes receive corrections of the order of $-200\%$ and the difference between the LA and the exact result is always (well) below the $10\%$.

\subsection{Impact of the corrections of QCD origin}

\label{sec:QCDres}

\begin{figure}[!t]
\begin{center}
\includegraphics[width=0.32\linewidth]{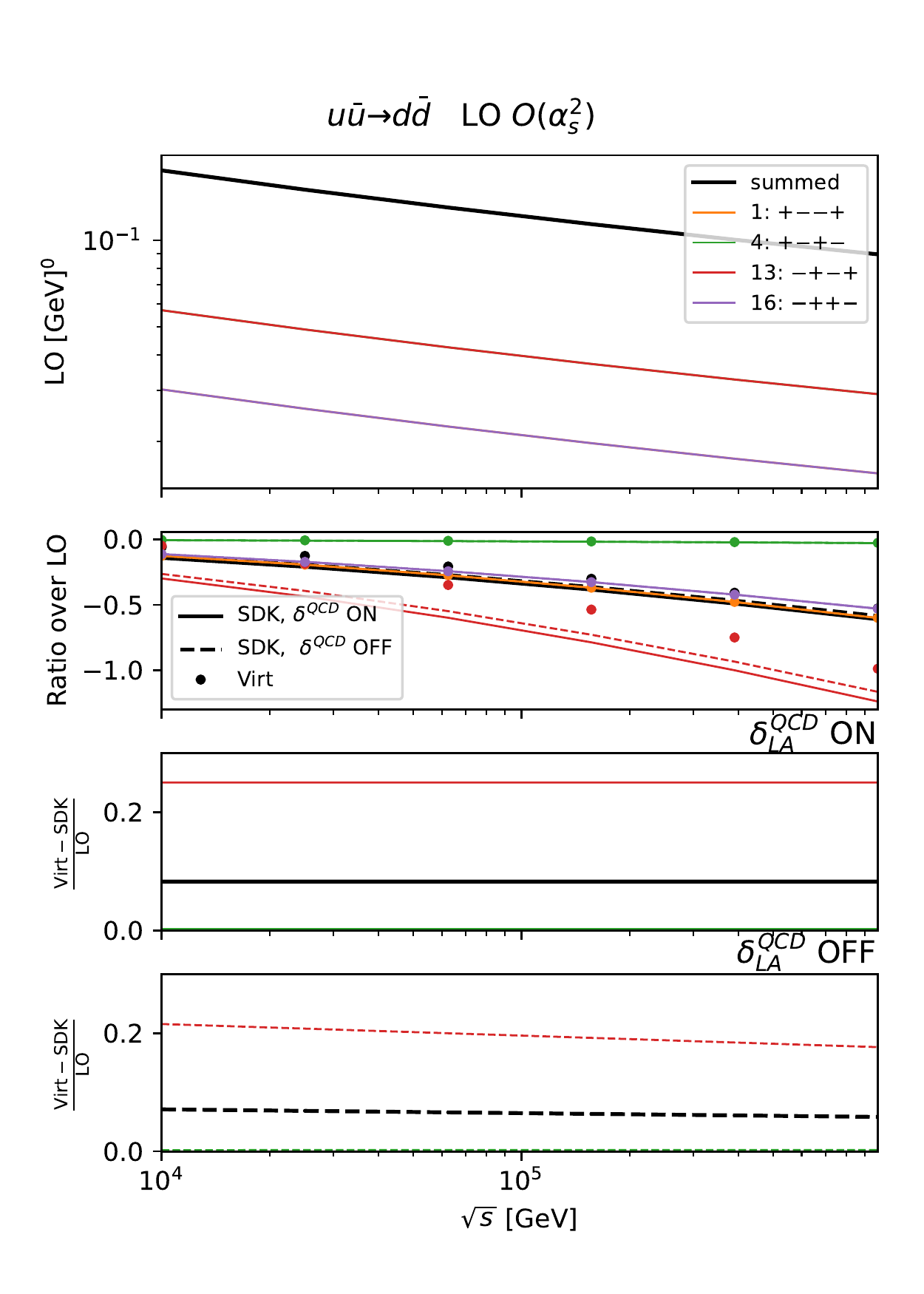}
\includegraphics[width=0.32\linewidth]{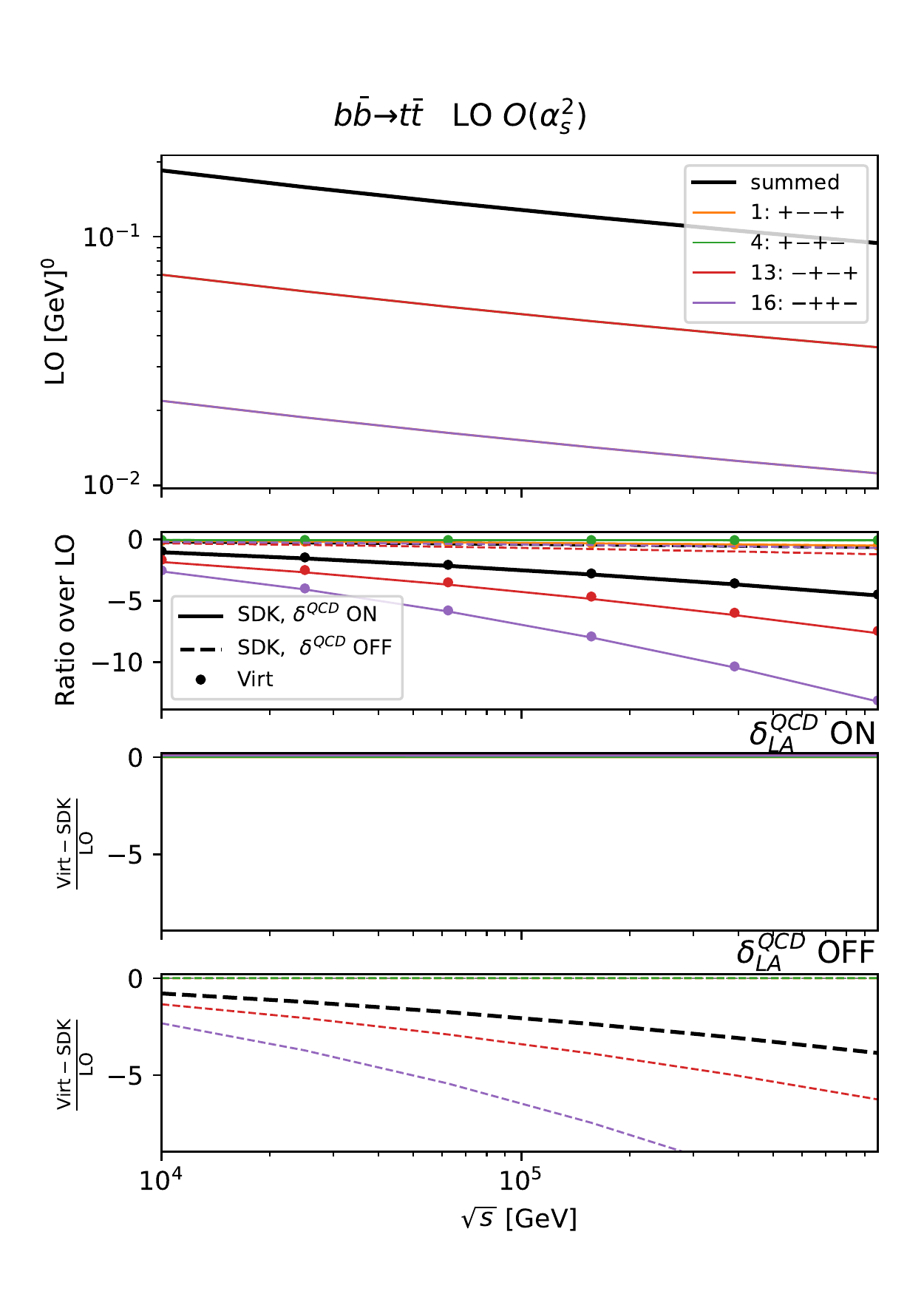}
\includegraphics[width=0.32\linewidth]{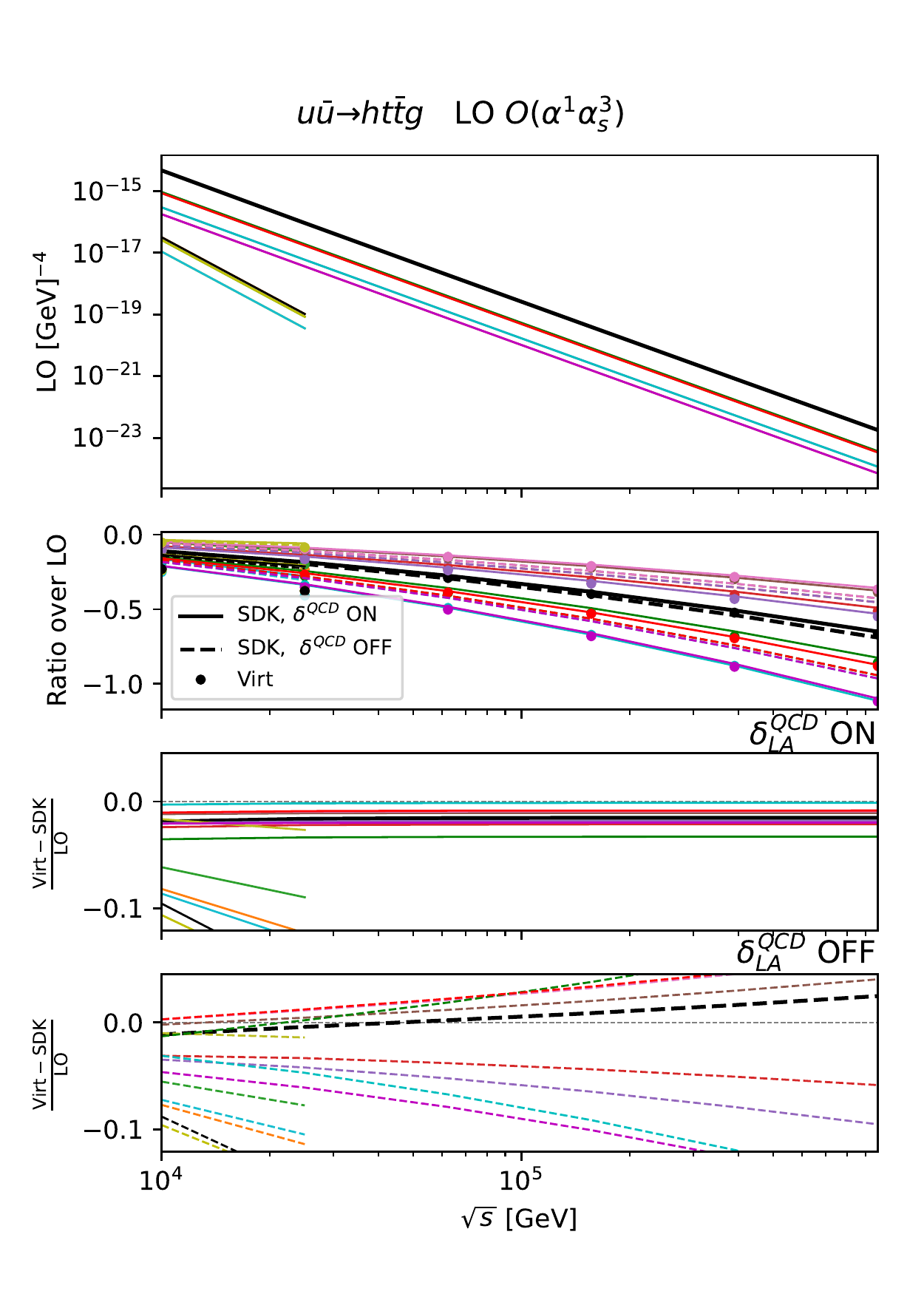}
\includegraphics[width=0.32\linewidth]{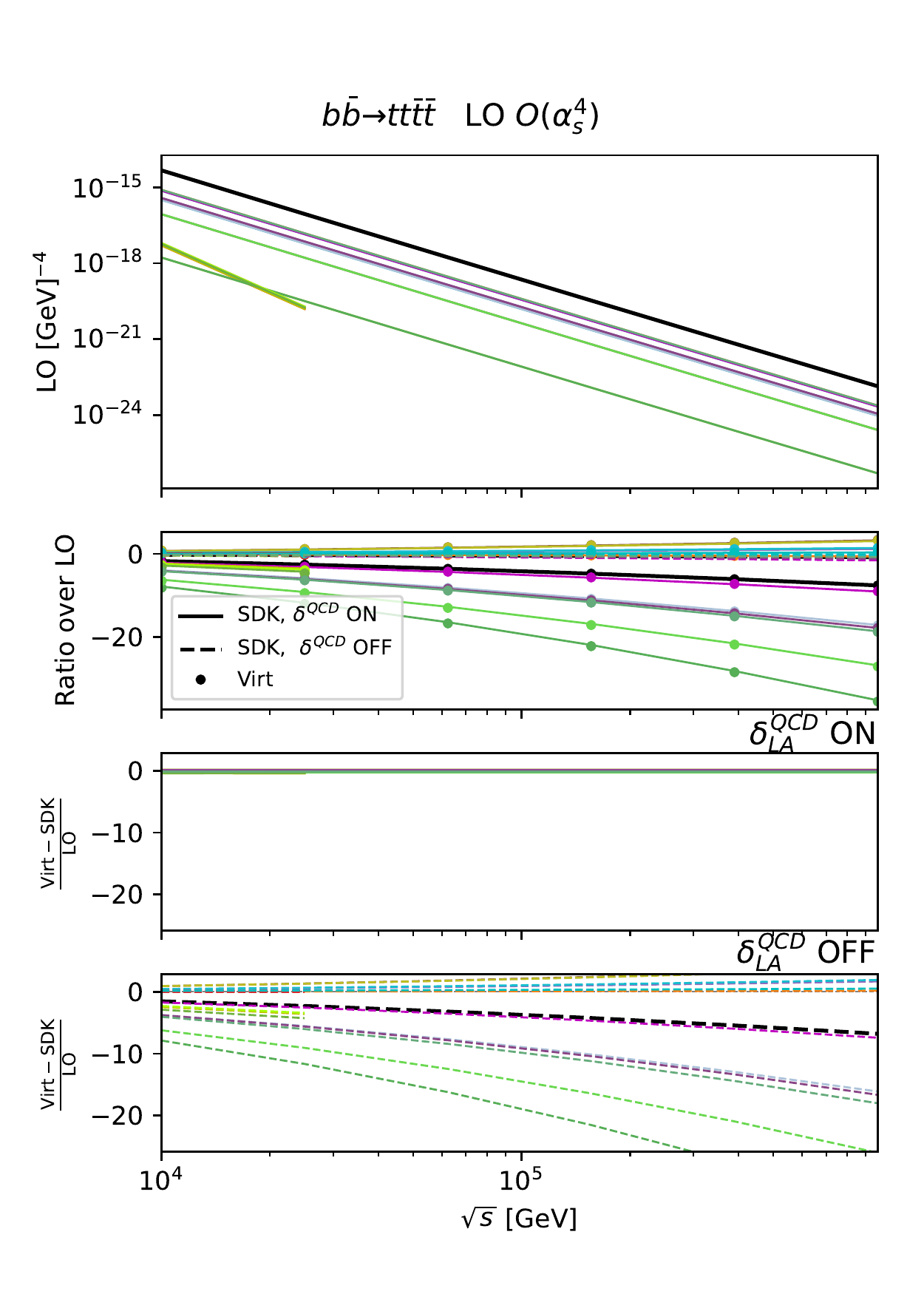}
\includegraphics[width=0.32\linewidth]{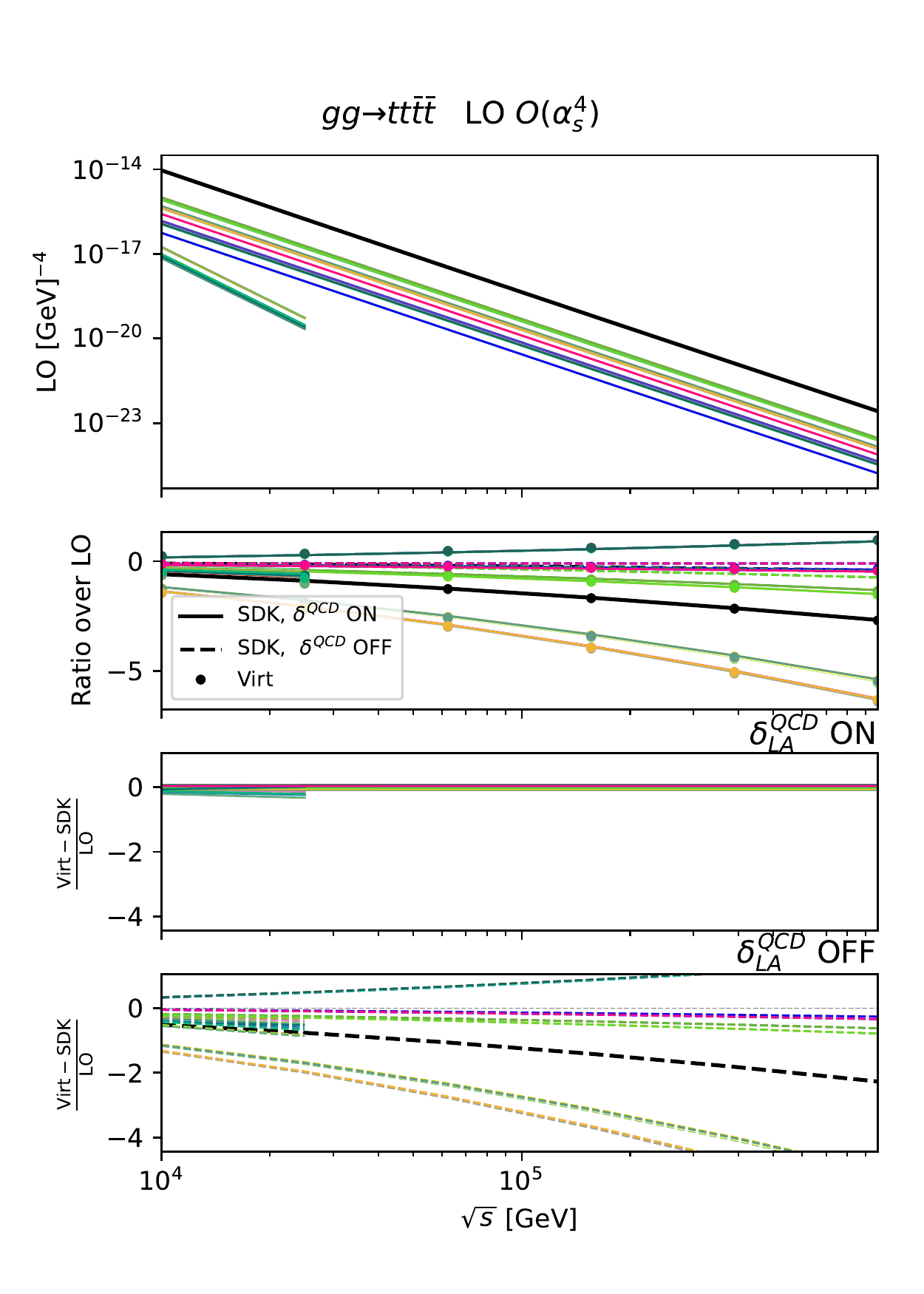}
\includegraphics[width=0.32\linewidth]{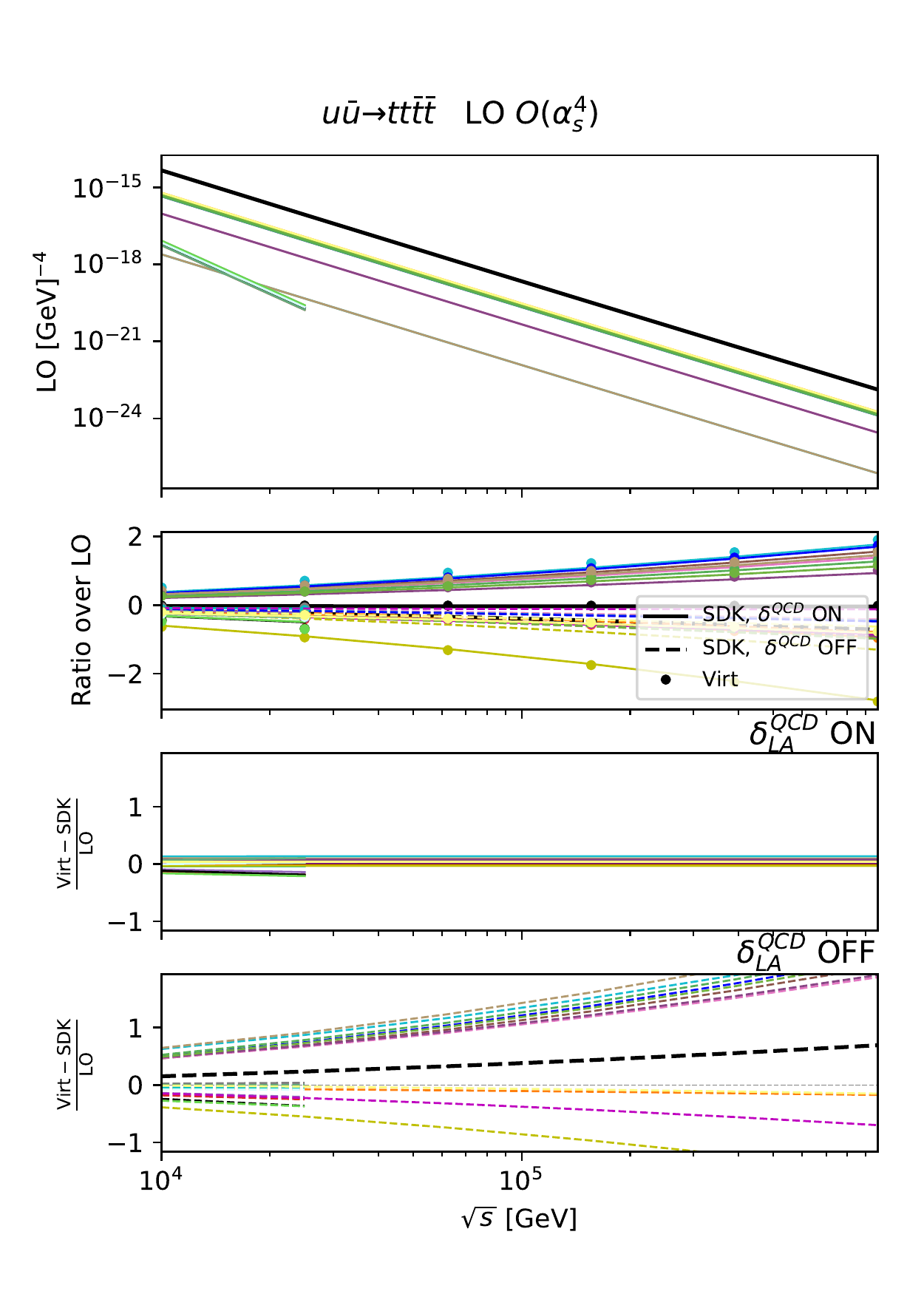}
\end{center}
\caption{Comparison between exact results (dots) for $\ord(\alpha)$ NLO EW virtual corrections and their LA (lines) in the case of squared matrix elements of representative $2\TO n$ processes with $n=2,3,4$. Solid lines include the $\deltaQCD$ contribution, while dashed lines do not. Plots show a scan in energy for fixed $r_{kl}/s$ ratios. More details are given in the text. \label{fig:QCD}}
\end{figure}

As explained in Sec.~\ref{sec:NLOew}, NLO EW corrections can receive contribution of both EW and QCD origin (see discussion therein for details). In this section we show numerical results on this aspect, for $2\TO n$ partonic processes with $n=2,3,4$.  We again select representative processes for which the relevant plots are displayed in Figs.~\ref{fig:QCD}. Each plot shows the dependence on $s$ of several quantities, and the layout is very similar to the one of the plots of Fig.~\ref{fig:Im} in the previous section. Phase-space points have also been generated following the same procedure described in the previous section and according to the condition \eqref{eq:minfrac}. 
As in Sec.~\ref{sec:Im} the  LA always includes the ${\SS^{s\TO r_{kl}}}$ terms in $\deltaEW$, however here we distinguish the cases in which the term $\Sigma^{}_{\LO_{i}} \deltaQCD$ in eq.~\ref{eq:QCDEWcomb} is excluded or retained.
 The former are displayed as dashed lines (SDK, $\deltaQCD$ OFF) and the latter as solid lines (SDK, $\deltaQCD$ ON).  For each leading helicity configuration, we also show in the second and third inset the difference between the LA and the exact result both normalised to the LO, respectively with and without taking into account the $\deltaQCD$  component.

As can be seen in Fig.~\ref{fig:QCD}, perfectly horizontal lines are displayed only in the second inset.\footnote{Some lines do not span the entire $\sqrt{s}$-range. These are related to helicity configurations that are actually mass suppressed but still contribute more than $10^{-3}$ of the dominant helicity configuration for sufficiently ``low'' energies. See also the  definition of leading helicity configurations at the beginning of Sec.~\ref{sec:rijvss} and in footnote \ref{footnote_a}.} For each plot, lines in the third inset have a slope, which largely depends on the process considered. The slope is directly connected to the term $\Sigma^{}_{\LO_{i}} \deltaQCD$, therefore its size depends on both the size of $\deltaQCD$ (see eq.~\eqref{eq:dQCDfinal}) and, since in the plots shown in Figs.~\ref{fig:QCD} we have $i=2$, on the $\LO_{2}/\LO_{1}$ ratio.

Considering simple $2\TO 2 $ processes, one can see how different is the impact of  $\Sigma^{}_{\LO_{2}} \deltaQCD$ in the top-left and top-center plots; in the case of $b \bar b \TO t \bar t$ both $\deltaQCD$ and the $\LO_{2}/\LO_{1}$ ratio are larger. The top-right plot refers to the process $u\bar u \TO t \bar t g h $, the simplest process for which all the terms of eq.~\eqref{eq:dQCDfinal} are non-vanishing. The lower plots refer to different partonic processes entering the process $pp\TO t \bar t t \bar t$. As already discussed in Ref.~\cite{Frederix:2017wme} a large part of the NLO EW corrections are of QCD origin, and this can be observed also in the lower plots of Fig.~\ref{fig:QCD}.

Given the large number of leading helicity configurations for processes with $t \bar t t \bar t $ in the final state, we did not list them in the legend of the corresponding plots of  Fig.~\ref{fig:QCD}. On the other hand, in Tab.~\ref{tab:tttt} we provide, as an example,  the results of the fit of (Virt-SDK)/LO for all leading-helicity configurations (and their sum) for the process $gg \TO t \bar t t \bar t $. We have used again the method of least squares and the function \eqref{eq:fit} as done for Tab.~\ref{tab:zdd}, but in this case we have considered two different scenarios: the inclusion or the exclusion of the $\deltaQCD$ contribution. By comparing the numbers in the first and third column of Tab.~\ref{tab:tttt}, one can further see how the $\deltaQCD$ contribution is essential in flattening the (Virt-SDK)/LO curve.

\begin{table}
\scriptsize
\begin{tabular}{c|cc|cc}
 & \multicolumn{2}{c}{$\deltaQCD$ ON } & \multicolumn{2}{c}{$\deltaQCD$ OFF } \\ 
 \hline
 Helicity & $A$ & $B$   & $A$   & $B$  \\ 
 \hline
 summed   &   $ ( 9.9 \pm 8.4 ) \cdot 10^{-4} $  &  $ ( -2.7 \pm 0.4 ) \cdot 10^{-2} $    &    $ ( -8.9 \pm 1.2 ) \cdot 10^{-1} $  &  $ ( 3.1 \pm 0.6 ) \cdot 10^{0} $     \\ 
$ 1: ----++ $  &   $ ( 2.7 \pm 1.6 ) \cdot 10^{-4} $  &  $ ( -2.56 \pm 0.08 ) \cdot 10^{-2} $    &    $ ( -2.9 \pm 0.4 ) \cdot 10^{-1} $  &  $ ( 9.9 \pm 1.8 ) \cdot 10^{-1} $     \\ 
$ 6: ---++- $  &   $ ( -5.5 \pm 5.4 ) \cdot 10^{-3} $  &  $ ( 6.6 \pm 2.7 ) \cdot 10^{-2} $    &    $ ( 5.3 \pm 0.7 ) \cdot 10^{-1} $  &  $ ( -1.8 \pm 0.4 ) \cdot 10^{0} $     \\ 
$ 7: ---+-+ $  &   $ ( 4.3 \pm 6.7 ) \cdot 10^{-4} $  &  $ ( 5.8 \pm 0.3 ) \cdot 10^{-2} $    &    $ ( 5.3 \pm 0.7 ) \cdot 10^{-1} $  &  $ ( -1.8 \pm 0.3 ) \cdot 10^{0} $     \\ 
$ 10: --+-+- $  &   $ ( 6.5 \pm 6.1 ) \cdot 10^{-4} $  &  $ ( 5.3 \pm 0.3 ) \cdot 10^{-2} $    &    $ ( 5.3 \pm 0.7 ) \cdot 10^{-1} $  &  $ ( -1.8 \pm 0.3 ) \cdot 10^{0} $     \\ 
$ 11: --+--+ $  &   $ ( -5.9 \pm 5.8 ) \cdot 10^{-3} $  &  $ ( 8.2 \pm 2.9 ) \cdot 10^{-2} $    &    $ ( 5.3 \pm 0.7 ) \cdot 10^{-1} $  &  $ ( -1.8 \pm 0.4 ) \cdot 10^{0} $     \\ 
$ 16: --++-- $  &   $ (  5 \pm  14 ) \cdot 10^{-5} $  &  $ ( 3.53 \pm 0.07 ) \cdot 10^{-2} $    &    $ ( -1.4 \pm 0.2 ) \cdot 10^{-1} $  &  $ ( 5.4 \pm 0.9 ) \cdot 10^{-1} $     \\ 
$ 17: -+--++ $  &   $ ( 4.3 \pm 6.3 ) \cdot 10^{-4} $  &  $ ( -4.6 \pm 0.3 ) \cdot 10^{-2} $    &    $ ( -2.2 \pm 0.3 ) \cdot 10^{-1} $  &  $ ( 7.3 \pm 1.4 ) \cdot 10^{-1} $     \\ 
$ 22: -+-++- $  &   $ ( -5.5 \pm 3.5 ) \cdot 10^{-4} $  &  $ ( -6.4 \pm 0.2 ) \cdot 10^{-2} $    &    $ ( -2.3 \pm 0.3 ) \cdot 10^{0} $  &  $ ( 8.1 \pm 1.5 ) \cdot 10^{0} $     \\ 
$ 23: -+-+-+ $  &   $ ( 1.2 \pm 0.6 ) \cdot 10^{-3} $  &  $ ( -6.3 \pm 0.3 ) \cdot 10^{-2} $    &    $ ( -1.9 \pm 0.3 ) \cdot 10^{0} $  &  $ ( 6.9 \pm 1.3 ) \cdot 10^{0} $     \\ 
$ 26: -++-+- $  &   $ ( 1.2 \pm 0.9 ) \cdot 10^{-3} $  &  $ ( -5.9 \pm 0.4 ) \cdot 10^{-2} $    &    $ ( -2.3 \pm 0.3 ) \cdot 10^{0} $  &  $ ( 8.2 \pm 1.5 ) \cdot 10^{0} $     \\ 
$ 27: -++--+ $  &   $ ( -3.7 \pm 5.0 ) \cdot 10^{-4} $  &  $ ( -8.1 \pm 0.3 ) \cdot 10^{-2} $    &    $ ( -2.0 \pm 0.3 ) \cdot 10^{0} $  &  $ ( 7.0 \pm 1.3 ) \cdot 10^{0} $     \\ 
$ 32: -+++-- $  &   $ ( 3.1 \pm 3.1 ) \cdot 10^{-4} $  &  $ ( 2.7 \pm 0.2 ) \cdot 10^{-2} $    &    $ ( -1.1 \pm 0.1 ) \cdot 10^{-1} $  &  $ ( 4.1 \pm 0.6 ) \cdot 10^{-1} $     \\ 
$ 33: +---++ $  &   $ ( 1.5 \pm 1.2 ) \cdot 10^{-3} $  &  $ ( -4.5 \pm 0.6 ) \cdot 10^{-2} $    &    $ ( -2.2 \pm 0.3 ) \cdot 10^{-1} $  &  $ ( 7.3 \pm 1.4 ) \cdot 10^{-1} $     \\ 
$ 38: +--++- $  &   $ ( 1.4 \pm 1.0 ) \cdot 10^{-3} $  &  $ ( -7.1 \pm 0.5 ) \cdot 10^{-2} $    &    $ ( -2.0 \pm 0.3 ) \cdot 10^{0} $  &  $ ( 7.0 \pm 1.3 ) \cdot 10^{0} $     \\ 
$ 39: +--+-+ $  &   $ ( -1.9 \pm 3.7 ) \cdot 10^{-4} $  &  $ ( -7.4 \pm 0.2 ) \cdot 10^{-2} $    &    $ ( -2.3 \pm 0.3 ) \cdot 10^{0} $  &  $ ( 8.2 \pm 1.5 ) \cdot 10^{0} $     \\ 
$ 42: +-+-+- $  &   $ ( -2.7 \pm 2.5 ) \cdot 10^{-4} $  &  $ ( -8.3 \pm 0.1 ) \cdot 10^{-2} $    &    $ ( -2.0 \pm 0.3 ) \cdot 10^{0} $  &  $ ( 6.8 \pm 1.3 ) \cdot 10^{0} $     \\ 
$ 43: +-+--+ $  &   $ ( 1.5 \pm 0.7 ) \cdot 10^{-3} $  &  $ ( -6.6 \pm 0.4 ) \cdot 10^{-2} $    &    $ ( -2.3 \pm 0.3 ) \cdot 10^{0} $  &  $ ( 8.1 \pm 1.5 ) \cdot 10^{0} $     \\ 
$ 48: +-++-- $  &   $ ( 4.0 \pm 2.3 ) \cdot 10^{-4} $  &  $ ( 2.1 \pm 0.1 ) \cdot 10^{-2} $    &    $ ( -1.1 \pm 0.1 ) \cdot 10^{-1} $  &  $ ( 4.1 \pm 0.6 ) \cdot 10^{-1} $     \\ 
$ 49: ++--++ $  &   $ ( 4.0 \pm 2.5 ) \cdot 10^{-4} $  &  $ ( -2.9 \pm 0.1 ) \cdot 10^{-2} $    &    $ ( -2.9 \pm 0.4 ) \cdot 10^{-1} $  &  $ ( 9.9 \pm 1.8 ) \cdot 10^{-1} $     \\ 
$ 54: ++-++- $  &   $ ( 7.4 \pm 7.8 ) \cdot 10^{-4} $  &  $ ( 4.9 \pm 0.4 ) \cdot 10^{-2} $    &    $ ( 5.3 \pm 0.7 ) \cdot 10^{-1} $  &  $ ( -1.8 \pm 0.3 ) \cdot 10^{0} $     \\ 
$ 55: ++-+-+ $  &   $ ( -5.4 \pm 5.2 ) \cdot 10^{-3} $  &  $ ( 7.4 \pm 2.6 ) \cdot 10^{-2} $    &    $ ( 5.3 \pm 0.7 ) \cdot 10^{-1} $  &  $ ( -1.8 \pm 0.4 ) \cdot 10^{0} $     \\ 
$ 58: +++-+- $  &   $ ( -6.0 \pm 6.0 ) \cdot 10^{-3} $  &  $ ( 7.7 \pm 3.0 ) \cdot 10^{-2} $    &    $ ( 5.3 \pm 0.7 ) \cdot 10^{-1} $  &  $ ( -1.8 \pm 0.4 ) \cdot 10^{0} $     \\ 
$ 59: +++--+ $  &   $ ( 3.7 \pm 4.6 ) \cdot 10^{-4} $  &  $ ( 6.1 \pm 0.2 ) \cdot 10^{-2} $    &    $ ( 5.3 \pm 0.7 ) \cdot 10^{-1} $  &  $ ( -1.8 \pm 0.3 ) \cdot 10^{0} $     \\ 
$ 64: ++++-- $  &   $ ( 10 \pm 141 ) \cdot 10^{-6} $  &  $ ( 3.85 \pm 0.07 ) \cdot 10^{-2} $    &    $ ( -1.4 \pm 0.2 ) \cdot 10^{-1} $  &  $ ( 5.4 \pm 0.9 ) \cdot 10^{-1} $ 
\end{tabular}
\caption{Result of the fit of the quantity (Virt-SDK)/LO using the method of least squares and the function \eqref{eq:fit} for the representative process $gg \TO t \bar t t \bar t$. The case including(excluding) the $\deltaQCD$ contribution corresponds to the quantities shown in the second(third) inset of the lower-central plot of Fig.~\ref{fig:QCD}.  \label{tab:tttt}}
\end{table}

\section{Numerical Results: differential cross sections}
\label{sec:Res_MC}

In this section we consider IR-safe collider observables and we compare exact NLO EW  predictions and their LA  (the approximation given by Sudakov logarithms) obtained via the {\denpoz} algorithm as framed in Sec.~\ref{sec:SudMC}, {\it i.e.},  the ${\rm SDK_{weak}}$ approach. We show differential distributions for proton--proton collisions at 100 TeV, which is one of the possible  experimental set-ups that has been considered as an option for a high-energy future colliders  \cite{Arkani-Hamed:2015vfh, Mangano:2016jyj, Contino:2016spe, Dainese:2016gch}. In this regime, LA is expected to be very efficient in capturing NLO EW effects. By explicit examples, we show how both the inclusion of the $\SS^{s\TO r_{kl}}$ terms in eq.~\eqref{eq:angsplit_new_sm}, which are relevant when the condition \eqref{eq:rijnice} is not satisfied, and the usage of the purely weak LA described in Sec.~\ref{sec:SudMC}, the ${\rm SDK_{weak}}$ approach, has to be  in general  preferred for predictions of physical observables. Indeed, these features not only improve the LA, but they are also instrumental in order to capture the correct logarithmic dependence. 

As already said, we consider cross-section predictions  for differential distributions of IR-safe  observables. Thus, bare leptons, which are treated as massless,  have to be recombined with photons into dressed leptons. A dressed lepton is here obtained by recombining a bare lepton $\ell$ with any photon $\gamma$ that  satisfies the condition 
\begin{equation}
\Delta R(\ell, \gamma) < 0.4\,, \label{eq:recQED}
\end{equation}
 where  $\Delta R(\ell, \gamma) \equiv \sqrt{(\Delta \eta(\ell, \gamma))^2+(\Delta \phi(\ell, \gamma))^2} $, and  $\Delta \eta(\ell, \gamma)$ and $\Delta \phi(\ell, \gamma)$ are the differences of the bare-lepton and photon pseudo-rapidities and azimuthal angles, respectively.\footnote{ In case  the recombination condition is satisfied for more than one bare lepton, the photon is clustered with the bare lepton for which $\Delta R(\ell, \gamma)$ is the smallest.} 
 
 In this context, however, we recombine with photons any electrically charged particle, including top quarks and $W$ bosons, which are massive. This choice is not due to IR safety, but rather to the fact that very energetic massive particles are typically identified as tagged jets, namely a jet which contains the considered particle. The recombination of photons and heavy charged particles is inspired by precisely  this procedure, and similarly the condition \eqref{eq:recQED} leading to a large cone. In one case we will also consider the aforementioned tagged jets, where the clustering is performed  not only with  photons but with all the particles and afterwards tagging the considered heavy particle among the jet constituents. In that case, we will use the anti-$k_T$ algorithm \cite{Cacciari:2008gp}  as implemented in {\sc \small FastJet}~\cite{Cacciari:2011ma},   with $R=0.4$.
 We remind the reader that, as already explained in Sec.~\ref{sec:weak},  photon recombination has an effect on the LA: it cancels the virtual QED contributions associated to the collinear configuration in the final state, which is precisely what is taken into account by the ${\rm SDK_{weak}}$ approach.
 
\begin{figure}[!t]
\begin{center}
\includegraphics[width=0.38\linewidth]{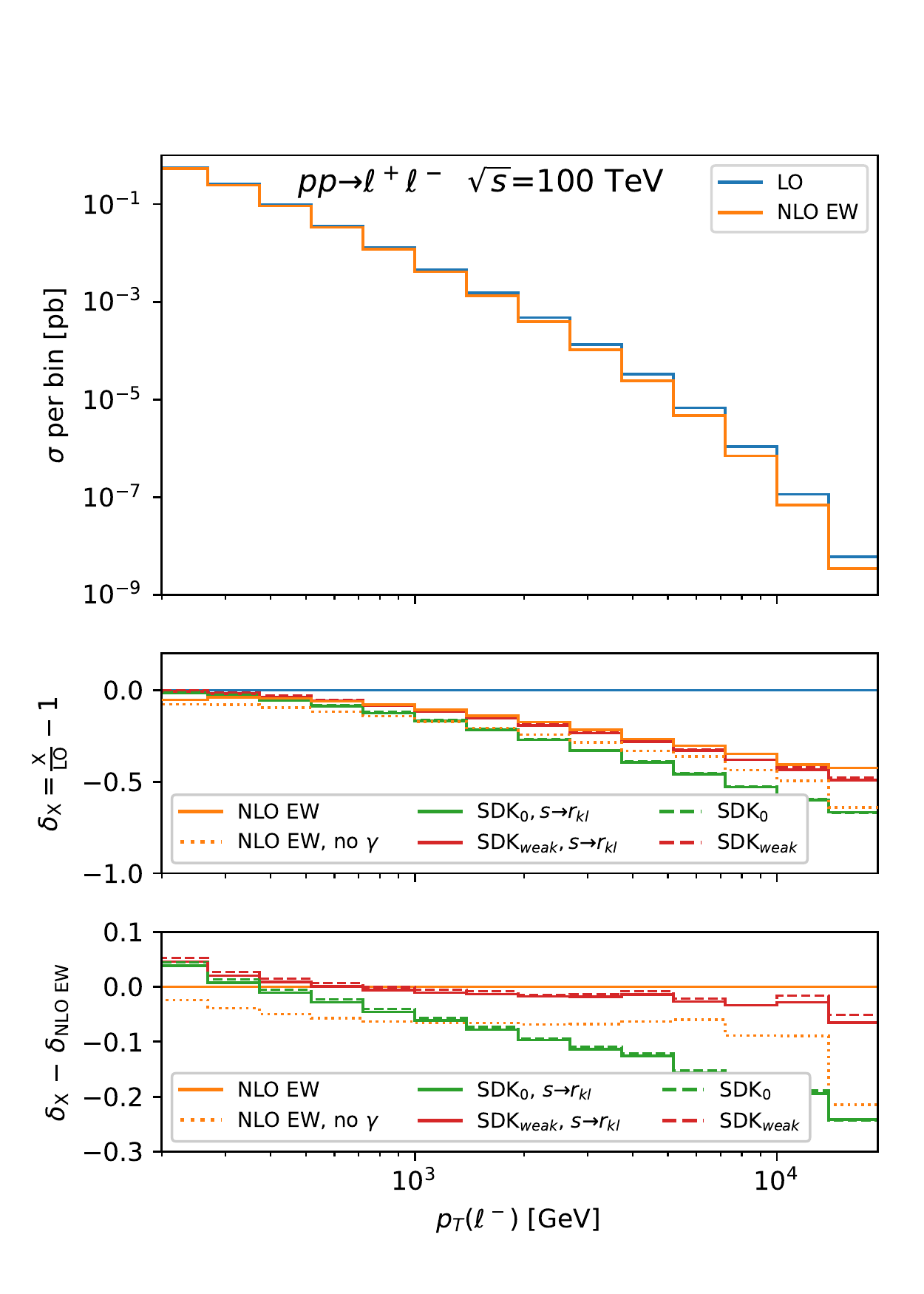}
\includegraphics[width=0.38\linewidth]{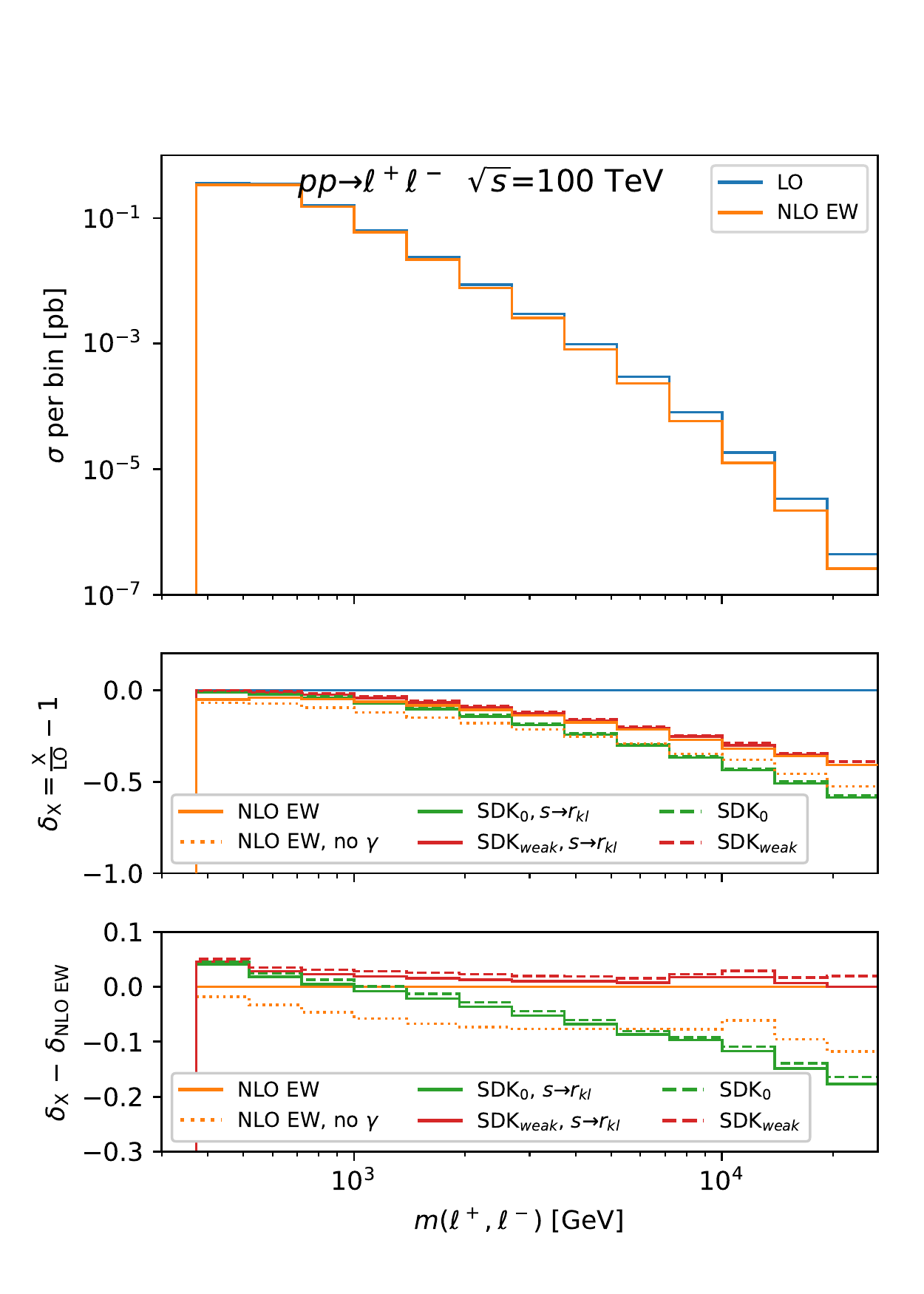}
\includegraphics[width=0.38\linewidth]{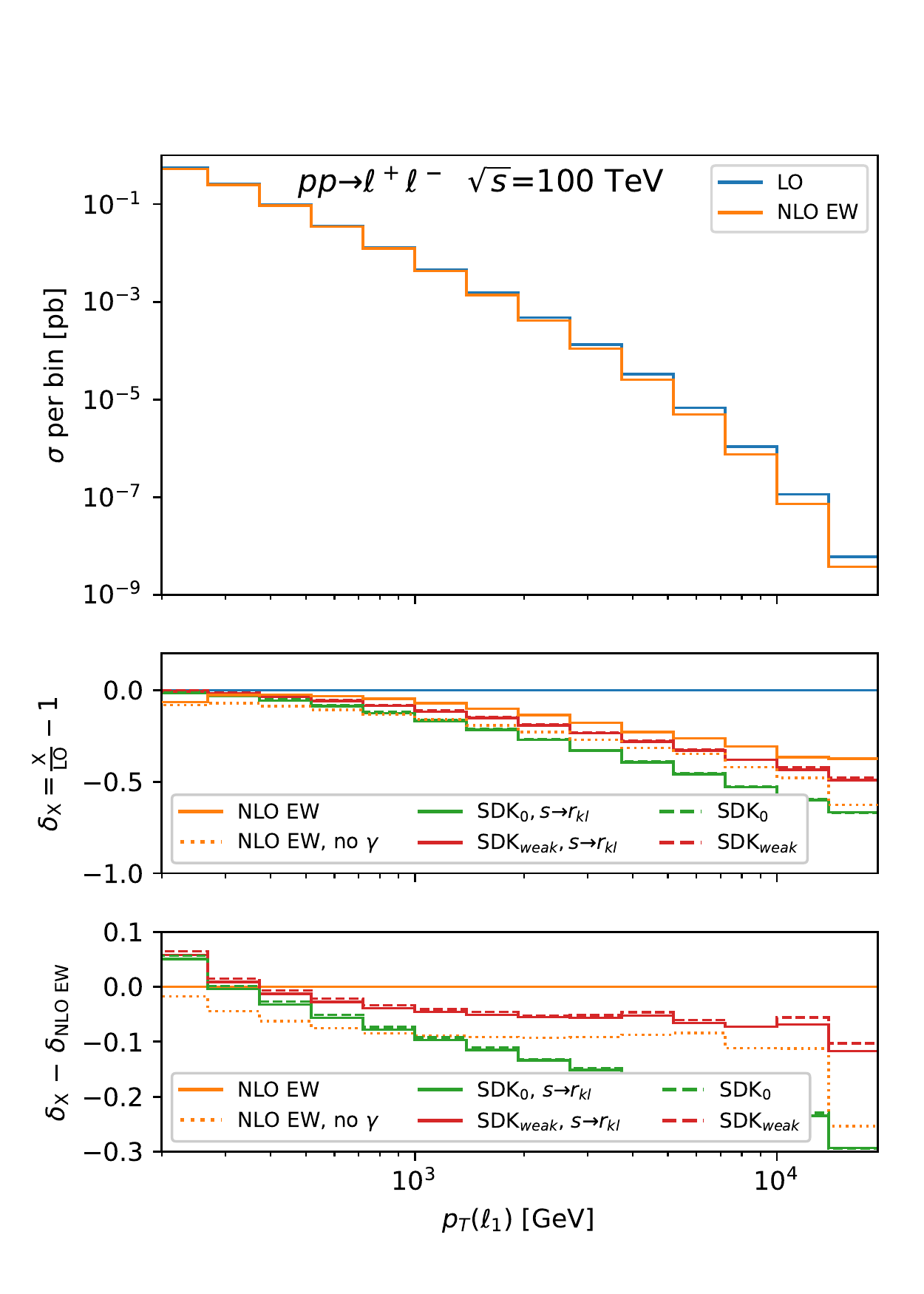}
\includegraphics[width=0.38\linewidth]{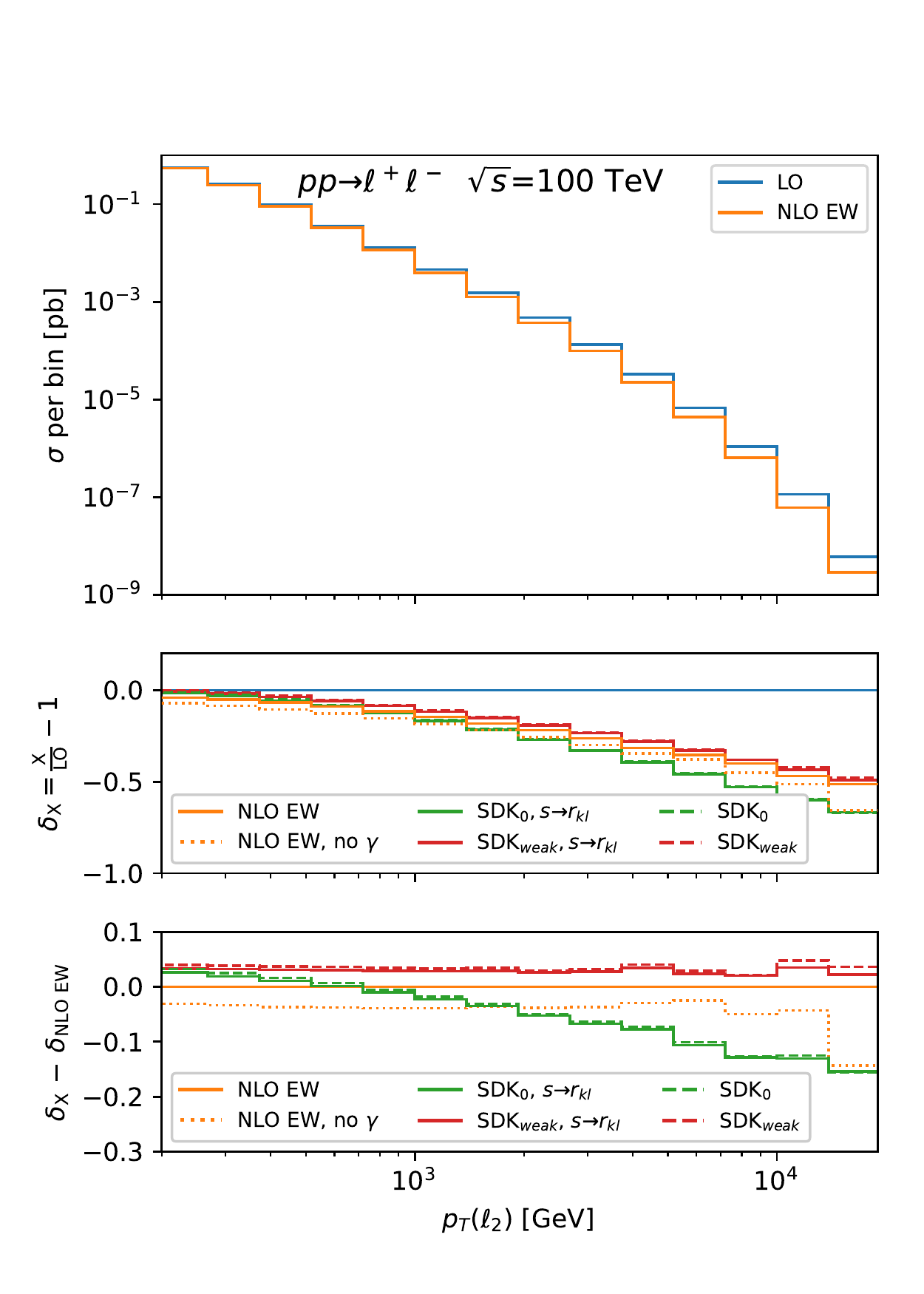}
\end{center}
\caption{Differential distributions for recombined leptons in Drell-Yan production at 100 TeV. Comparison between exact NLO EW predictions  (solid orange) and their LA approximations in the ${\rm SDK_{weak}}$ (red) and ${\rm SDK_{0}}$ (green) approaches, including (solid) or excluding (dashed) the $\SS^{s\TO r_{kl}}$ terms.  More details are given in the text. \label{fig:DY}}
\end{figure}

\subsection{Drell-Yan}

We start discussing the case of Drell-Yan production with leptons in the final state, namely the process $p p \TO \ell^+ \ell^-$.\footnote{ For simplicity, we have computed results for the specific case $\ell=e$.} For all processes that will be considered in this section we have used the same input parameters already listed in Sec.~\ref{sec:inputs}. However, in the case of Drell-Yan, we have employed the complex-mass scheme \cite{Denner:1999gp,Denner:2005fg,Frederix:2018nkq} for the exact NLO EW corrections.\footnote{We used the complex-mass scheme  in order to set the $Z$- and $W$-boson decay widths to a value different from zero, in particular $\Gamma_W=2.49877~\gev$ and $\Gamma_Z=2.092910~\gev$. Given the accuracy of our simulations, the non-zero $Z$-boson decay width is fundamental. As explained later in the text we require an invariant mass larger than $\MZ$ for the dressed lepton pair, but this is not sufficient for using the on-shell scheme. The dressed lepton pair can originate from a configuration where the bare leptons have $m(\ell^+_{\rm bare},\ell^-_{\rm bare}) \simeq \MZ$ and one of them is recombined with a hard photon, leading to $m(\ell^+,\ell^-) \gg \MZ$ and therefore passing the cuts. This configuration is not associated to any enhancement and therefore very rare, but in the on-shell scheme it leads to the evaluation of a resonant $Z$ propagator with zero width and therefore it is inconsistent.  } We use the PDF set {\sc\small NNPDF3.1} \cite{Ball:2017nwa, Bertone:2017bme},  in particular the  \texttt{NNPDF31\_nlo\_as\_0118\_luxqed} distributions, which include NLO QED evolution and especially a photon density  following the   {\sc\small LUXqed} parameterisation \cite{Manohar:2016nzj, Manohar:2017eqh}. The renormalisation ($\mu_R$) and factorisation ($\mu_F$) scales are both set equal to the partonic center-of-mass energy $\sqrt{s}$. This set-up is common with all the other processes discussed in this section.

In the Drell-Yan simulation the following cuts are imposed on the dressed leptons:
\begin{equation}
p_T(\ell^{\pm})>200~\gev\,, \qquad |\eta(\ell^{\pm})|<2.5\,, \qquad m(\ell^+,\ell^-)>400~\gev\,, \qquad \Delta R (\ell^{+},\ell^{-})>0.5\,. 
\label{eq:DYcuts}
\end{equation}
On the one hand, these cuts are imposed in order to resemble realistic experimental cuts for high-energy objects. On the other hand, they avoid additional logarithmic enhancements from collinear splittings appearing in the real radiation processes or even at the Born level. In Fig.~\ref{fig:DY} we show differential distributions for the transverse momentum of the electron, $p_T(\ell^-)$, for the transverse momentum of the leading (trailing) lepton, $p_T(\ell_1)$ ($p_T(\ell_2)$), and for the dilepton invariant mass $m(\ell^+,\ell^-)$.

The layout of each plot in Fig.~\ref{fig:DY}, and in general of each plot in this section\footnote{An important difference is present for Figs.~\ref{fig:WZ} and \ref{fig:WWW} and explained later in the text.}, is the following. In the main panel we show the differential distribution at LO (solid blue line) and NLO EW (solid orange line) accuracy, where the exact $\ord(\alpha)$ corrections are taken into account. If the NLO EW prediction turns negative, meaning that NLO EW corrections are negative and larger than the LO in absolute value, the curve corresponds to its absolute value and is drawn as dashed. In the first inset we show the relative impact of EW corrections,  $\delta_X\equiv{X/\LO-1}$, in different approximations.
The solid orange line corresponds to the one in the main panel with the same style, {\it i.e.} the exact $\ord(\alpha)$ corrections (NLO EW), and the dotted orange line  corresponds to the same case where the photon PDF has been set equal to zero (NLO EW, no $\gamma$). The other curves correspond to results in LA, with different assumptions. First, the solid curves include the $\SS^{s\TO r_{kl}}$ contribution (${\rm SDK}_X, s\TO r_{kl}$), while the dashed ones do not (${\rm SDK}_X$). Second, the green lines are obtained by simply omitting the QED and IR-sensitive terms, which are dubbed as ``em'' in the {\denpoz} algorithm. This is analogous to the approach of {\it e.g.}~Refs.~\cite{Chiesa:2013yma, Bothmann:2020sxm} and dubbed here as $\rm SDK_{0}$. The red lines are instead  obtained by completely removing the QED contribution, namely, following the procedure described in Sec.~\ref{sec:weak}, the $\rm SDK_{weak}$ approach. Both the $\rm SDK_{0}$ and $\rm SDK_{weak}$ predictions, similarly to the NLO EW ones in this section, include also the LO contribution. Needless to say, the closest a line is to the solid orange one, the better is the approximation of the exact NLO EW corrections. Therefore, in order to better judge this characteristic, in the second inset we zoom on the lines by simply plotting for each line in the first inset the difference with the solid orange one. Clearly, the reference prediction in LA is the solid red line, which both includes the $\SS^{s\TO r_{kl}}$ contribution and is obtained via the $\rm SDK_{weak}$ approach. 

We remind the reader that neither the $\rm SDK_{0}$ nor the $\rm SDK_{weak}$ approach are equal to the approach dubbed as SDK in Sec.~\ref{sec:Res_Amp}, which concerns the LA of the interference of Born amplitude and the  IR-divergent virtual amplitude. The SDK prediction cannot be used for IR-safe observables.  Moreover, the  $\rm SDK_{0}$ approach, even when the $\SS^{s\TO r_{kl}}$ contributions are not taken into account, is not exactly equal to the one used so far in the literature, since we do include also in this case the terms proportional to $i\pi \Theta(r_{kl})$. This is particularly relevant for Secs.~\ref{sec:ZZZ} and \ref{sec:WWW}, where $2\TO3$ processes are considered.

Starting from the top-left plot of Fig.~\ref{fig:DY}, the $p_T(\ell^-)$ distribution, we can see how sound is the LA in the high-energy limit. The distribution is spanning two orders of magnitude of $p_T(\ell^-)$, from 200 GeV to 20 TeV, and NLO EW corrections reach $\sim-50\%$ in the tail. As can be seen in the second inset, the solid red line differs from the solid orange line by only a very few percents of the LO. The situation is opposite for the green solid line. In that case the difference with the solid orange line grows logarithmically with   $p_T(\ell^-)$ and reaches $\sim-20\%$ of the LO in the tail. This is a clear example of how the $\rm SDK_{weak}$ approach can be superior to the  $\rm SDK_{0}$ one in the approximating the exact NLO EW corrections for IR-safe collider observables. For this observable the differences between dashed and solid lines of the same colour are negligible. This is not surprising, since large $p_T(\ell^-)$ implies large values of $t$ and therefore $|t|/s\sim\ord(1)$, leading to small contributions from the $\SS^{s\TO r_{kl}}$ terms.
Finally, by looking at the difference between the solid and dotted orange line, we can also appreciate how the photon-initiated processes are relevant for this process and especially unavoidable also in the LA approximation in order to correctly approximate the exact NLO EW effects.

Moving to the top-right plot of Fig.~\ref{fig:DY}, the $m(\ell^+,\ell^-)$ distribution, we see a very similar situation to the  $p_T(\ell^-)$ distribution. In this case there is a visible difference between solid and dashed lines of the same colours, with the solid red line better approximating the exact NLO EW result than the dashed red line. Indeed, large $m(\ell^+,\ell^-)$ values do not imply large values for $|t|$. However, due to the cuts in \eqref{eq:DYcuts} the effect of the $\rm SDK_{weak}$ terms is very mild and also the dashed red line is leading to a very good approximation. 
The lower plots, the $p_T(\ell_1)$ (left) and $p_T(\ell_2)$ (right) distributions, display some differences w.r.t.~the upper ones. For these two observable the LA  is less efficient, indeed, as can be seen in the plots, the difference with the exact results can reach up to 5\% of the LO also for solid red lines. However, this difference is converging to a constant value in the tail, for both distributions. Moreover the difference is of opposite sign for the two complementary observables, $p_T(\ell_1)$  and $p_T(\ell_2)$. This behaviour is due to the indirect cut that is affecting the recoiling  particle when a particular value of $p_T(\ell_1)$ or $p_T(\ell_2)$ is considered, namely the trivial condition   $p_T(\ell_1) >p_T(\ell_2)$. This has nothing to do with EW Sudakov effects, but it is a particular feature of this kind observables: a similar pattern was observed for top-quark pair production in Ref.~\cite{Czakon:2019bcq}. Indeed, we also verified this explicitly for top-quark pair production, which we do not show here since results are very similar to the case of Drell-Yan that we are discussing. The bottom line is that the $\rm SDK_{weak}$ approach including $\SS^{s\TO r_{kl}}$ terms for LA can always have $\ord(\alpha)$ finite discrepancies with the exact result, especially it cannot take into account effects that are not related to a Born-like kinematic. On the other hand, also for these two observables there are not logarithmically enhanced differences with the exact NLO EW prediction, only constant terms are present.  Instead, with the $\rm SDK_{0}$ approach, the logarithmically enhanced differences are clearly visible.

\begin{figure}[!t]
\begin{center}
\includegraphics[width=0.32\linewidth]{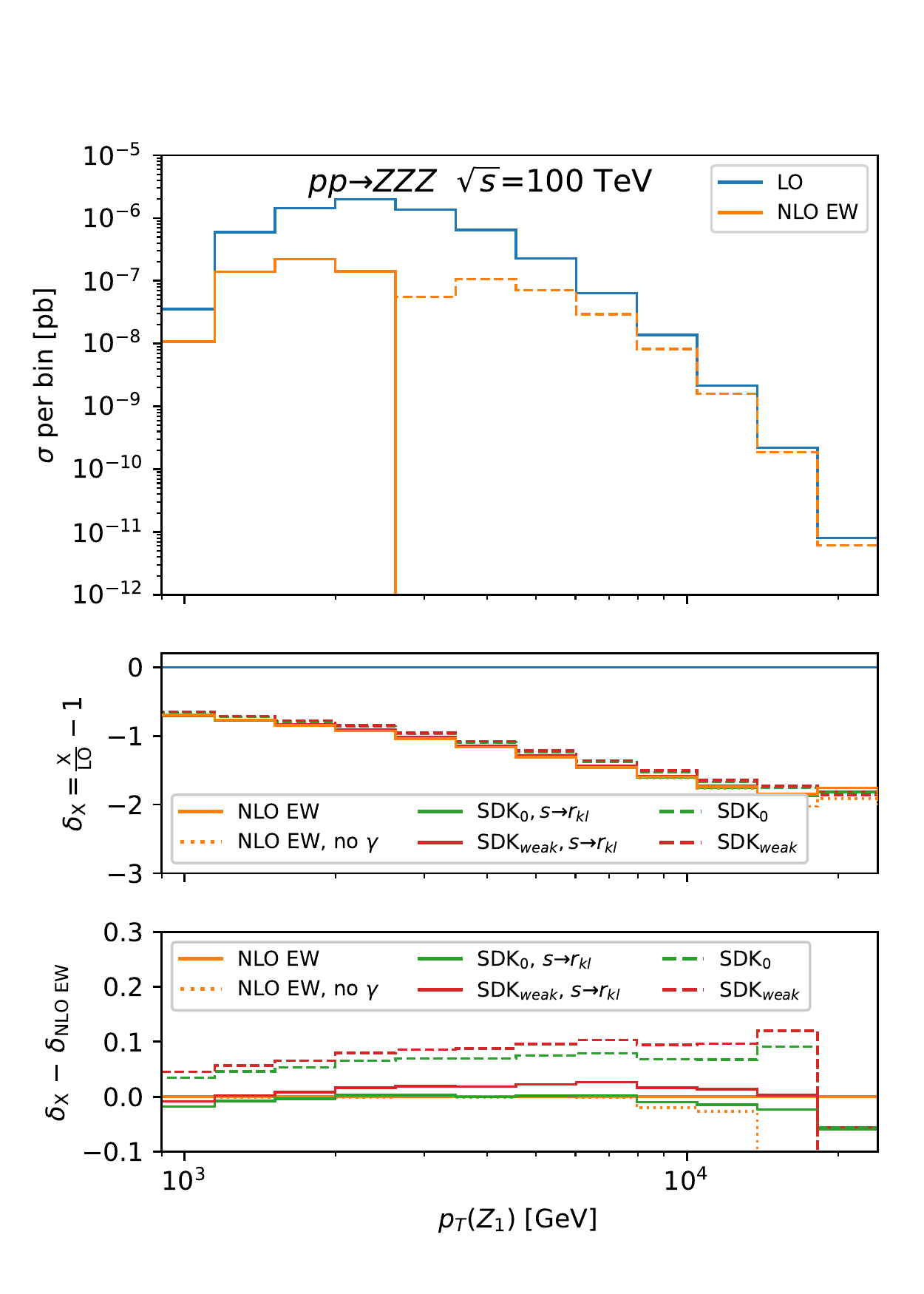}
\includegraphics[width=0.32\linewidth]{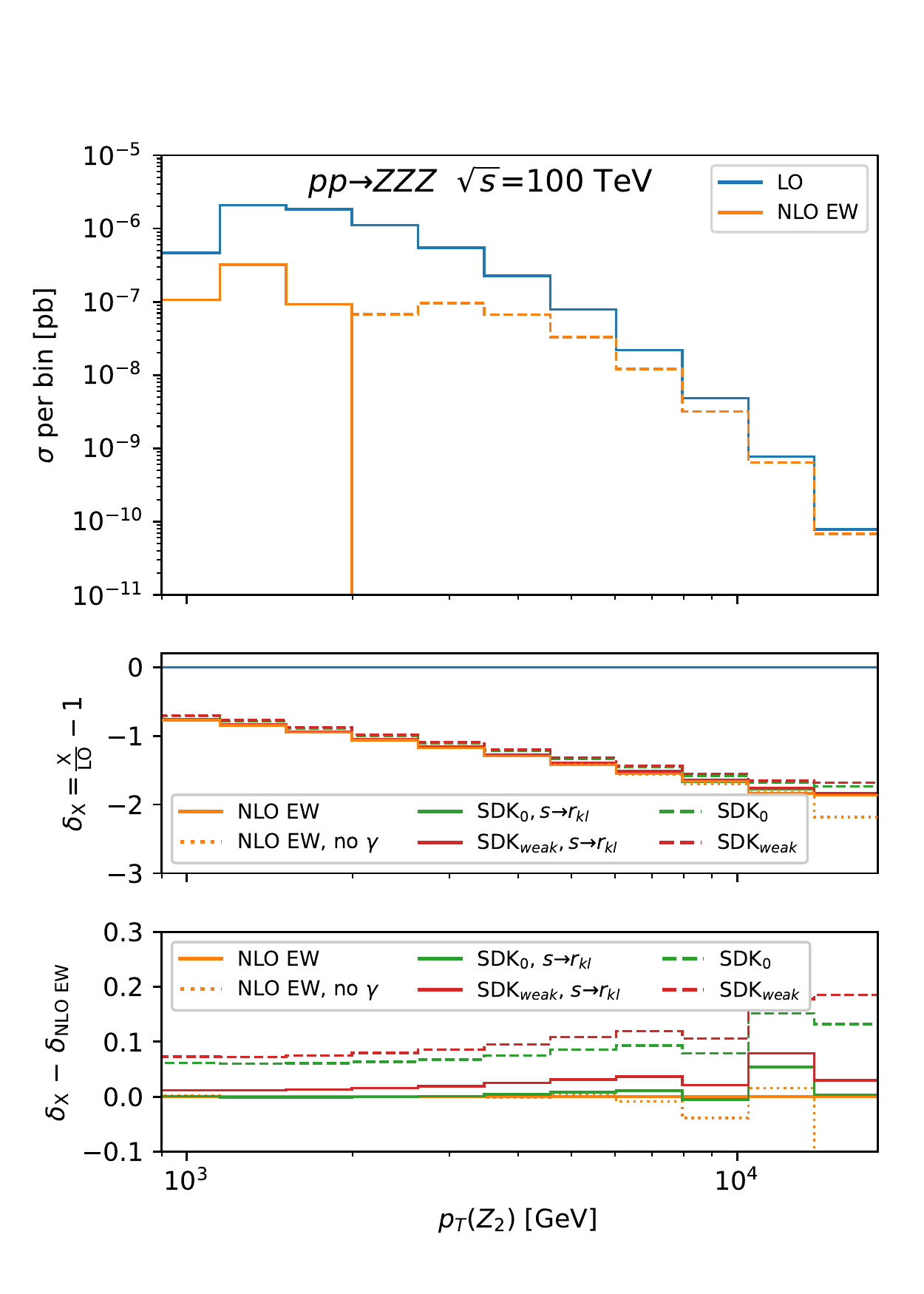}
\includegraphics[width=0.32\linewidth]{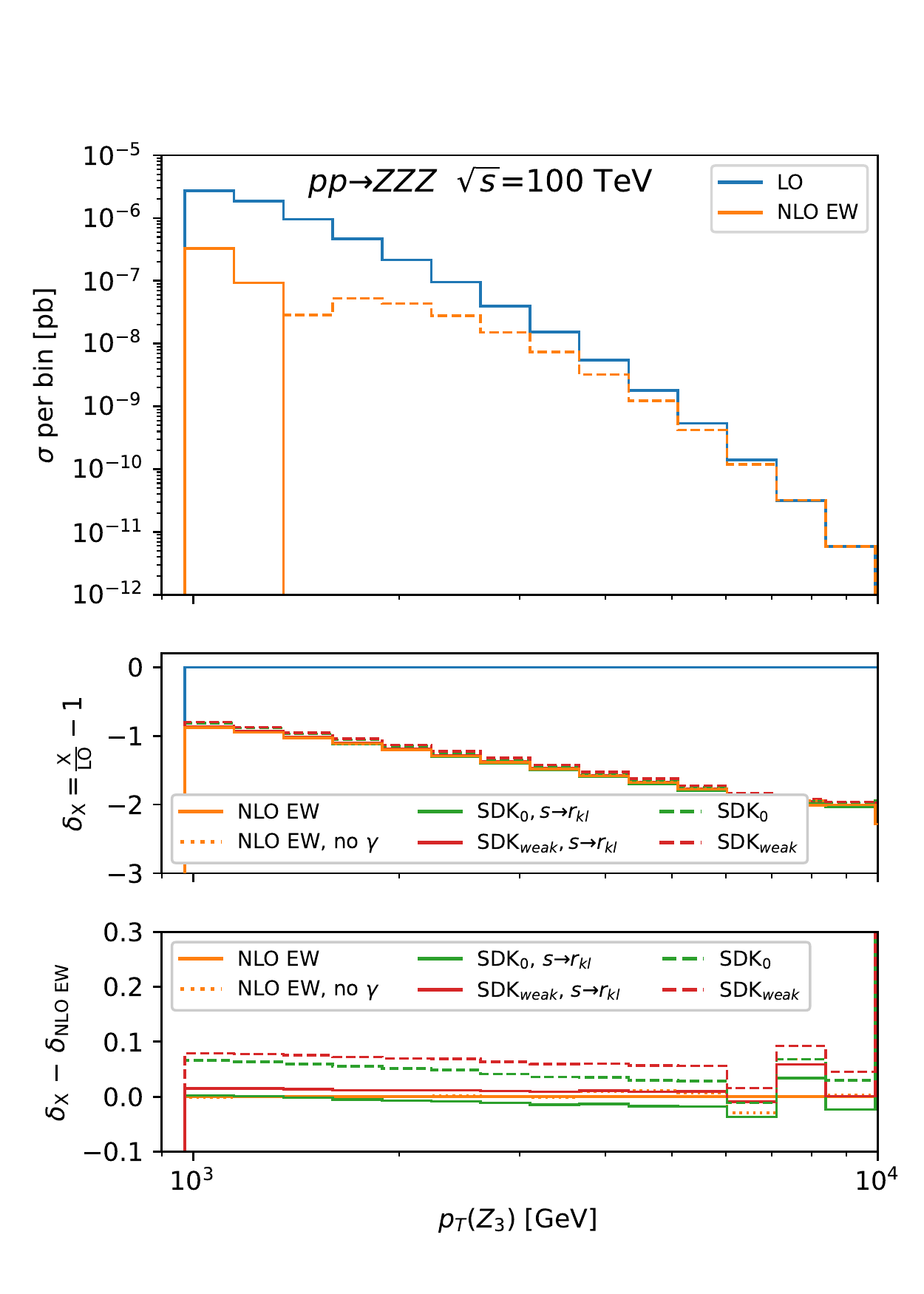}
\includegraphics[width=0.32\linewidth]{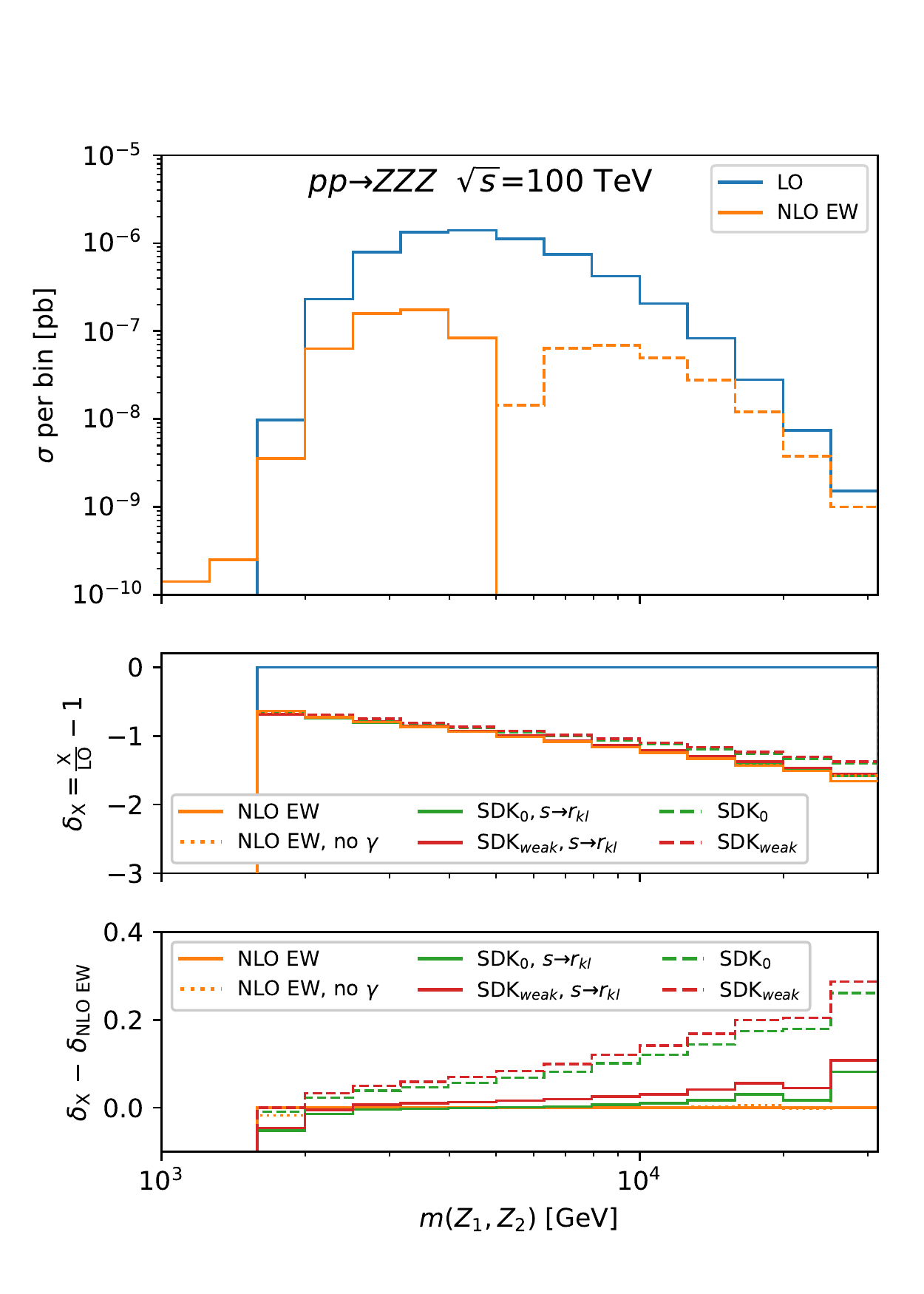}
\includegraphics[width=0.32\linewidth]{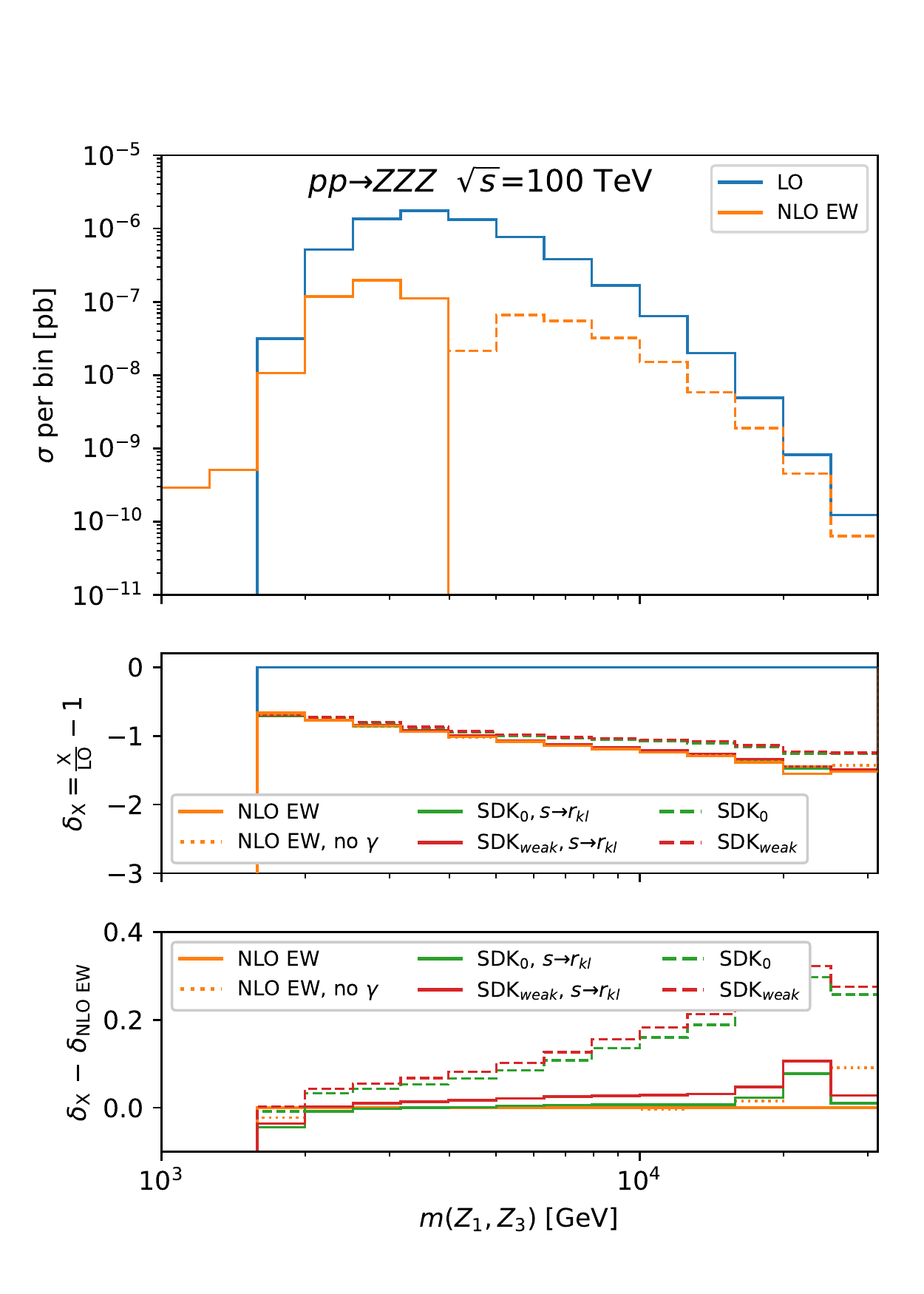}
\includegraphics[width=0.32\linewidth]{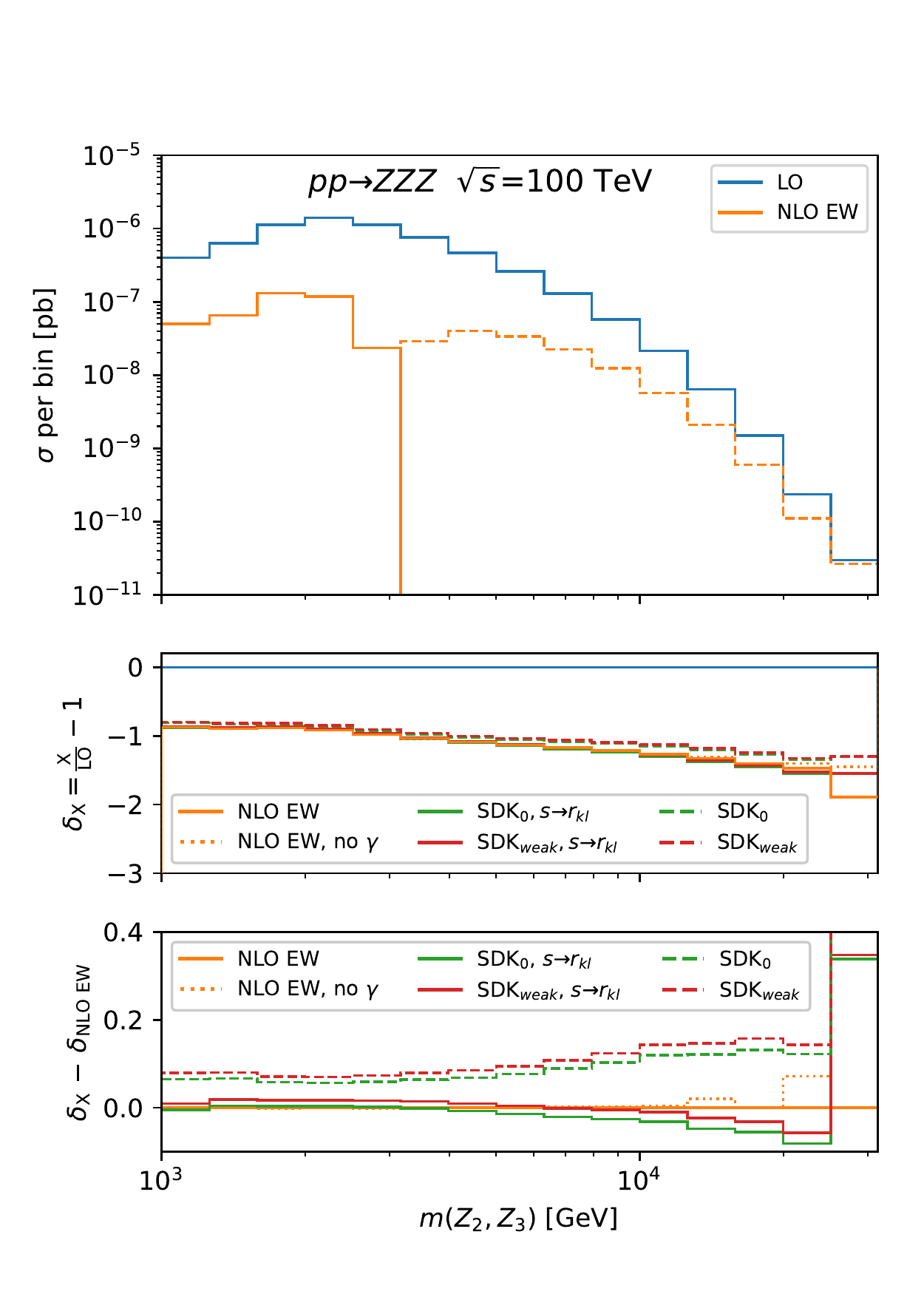}
\end{center}
\caption{Same as Fig.~\ref{fig:DY}, but for $ZZZ$ hadroproduction at 100 TeV. \label{fig:ZZZ}}
\end{figure}

\subsection{ZZZ}
\label{sec:ZZZ}

In Fig.~\ref{fig:ZZZ} we show plots, with the same layout of those in Fig.~\ref{fig:DY}, for the process $p p \TO ZZZ$.  This process has a neutral final state, so we do not expect large differences between the  $\rm SDK_{0}$ and  $\rm SDK_{weak}$ approaches. On the other hand, being a $2\TO 3$ process, the effect of the $\SS^{s\TO r_{kl}}$ terms is supposed to be more relevant. The upper plots of Fig.~\ref{fig:ZZZ} correspond to the transverse-momentum distributions of respectively the hardest $Z$-boson ($p_T(Z_1)$), the second-hardest $Z$-boson ($p_T(Z_2)$) and the softest one  ($p_T(Z_3)$). The lower plots instead correspond to the invariant masses $ m(Z_i, Z_j)$ of the three different $Z$-boson pairs.

All the results have been obtained by applying the following cuts:
\begin{equation}
p_T(Z_i)>1~\tev\,, \qquad |\eta(Z_i)|<2.5\,, \qquad m(Z_i,Z_j)>1~\tev\,, \qquad \Delta R (Z_i,Z_j)>0.5\,. 
\label{eq:ZZZcuts}
\end{equation}
Similarly to \eqref{eq:DYcuts}, these cuts resemble realistic experimental cuts for high-energy objects, but they also avoid additional logarithmic enhancements from collinear splittings appearing in the real-radiation processes or even at the Born. 

First of all, it is important to notice the size of the EW corrections. For most of the spectrum of all distributions, they are negative and larger than the LO in absolute value, reaching $\sim - 200 \%$ of it in the tail. Since they are negative, this means that fixed-order NLO EW corrections are also negative  in this regime and therefore non-physical. These distributions are a clear example of how large  Sudakov logarithms, and in turn NLO EW corrections, can be at high energy. Also they clearly point to the necessity of resumming them in order to obtain sensible predictions. Here, on the other hand, we are not providing phenomenological predictions but rather showing the accuracy of the LA and testing its implementation in {\mglong}. 

As expected, for all distributions, the difference between green and red lines ($\rm SDK_{0}$ and  $\rm SDK_{weak}$) amounts to only few percents of the LO, with no clear logarithmic enhancement in the high-energy limit. Also as expected, the impact of the $\SS^{s\TO r_{kl}}$ terms (solid versus dashed lines) is much larger for this process than for Drell-Yan production. In the upper plots of Fig.~\ref{fig:DY}, the $p_T(Z_i)$ distributions, the dashed lines are differing from the solid ones by 5-10\% of the LO for the full spectra, with the latter in turn differing only by a very few percents from the exact NLO EW prediction. The difference between dashed and solid lines is even larger in the lower plots, the $m(Z_i,Z_j)$ distributions, and especially a clear logarithmic trend can be observed. It is worth to stress that for all these distributions, with the exception of the far tail in the  $m(Z_i,Z_j)$ ones, the inclusion  of the $\SS^{s\TO r_{kl}}$ terms leads to an accuracy of very few percents for corrections spanning from $\sim$-80\% to $\sim$-200\%. This is not the case for the pure LA without  the $\SS^{s\TO r_{kl}}$ terms.

\begin{figure}[!t]
\begin{center}
\includegraphics[width=0.38\linewidth]{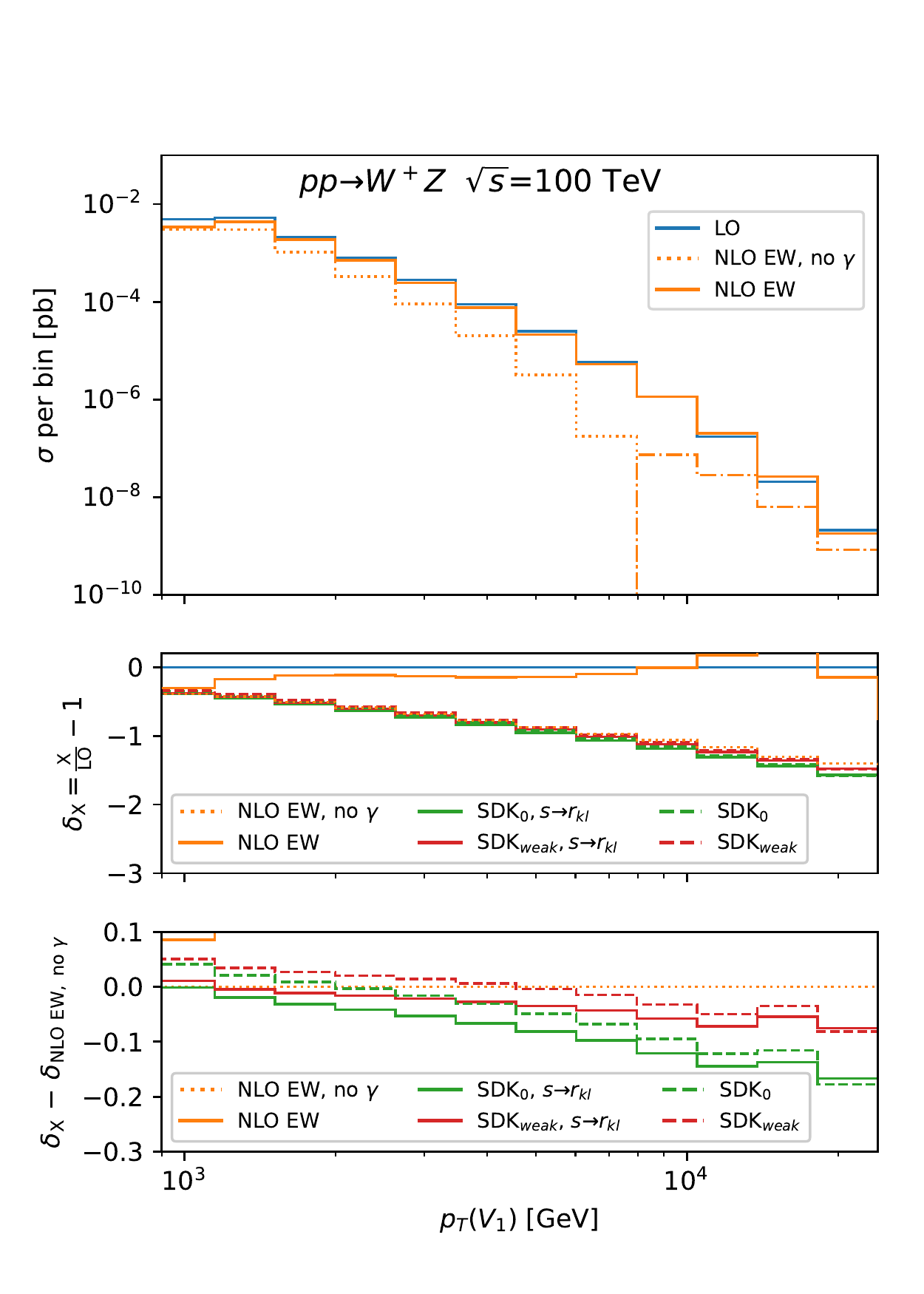}
\includegraphics[width=0.38\linewidth]{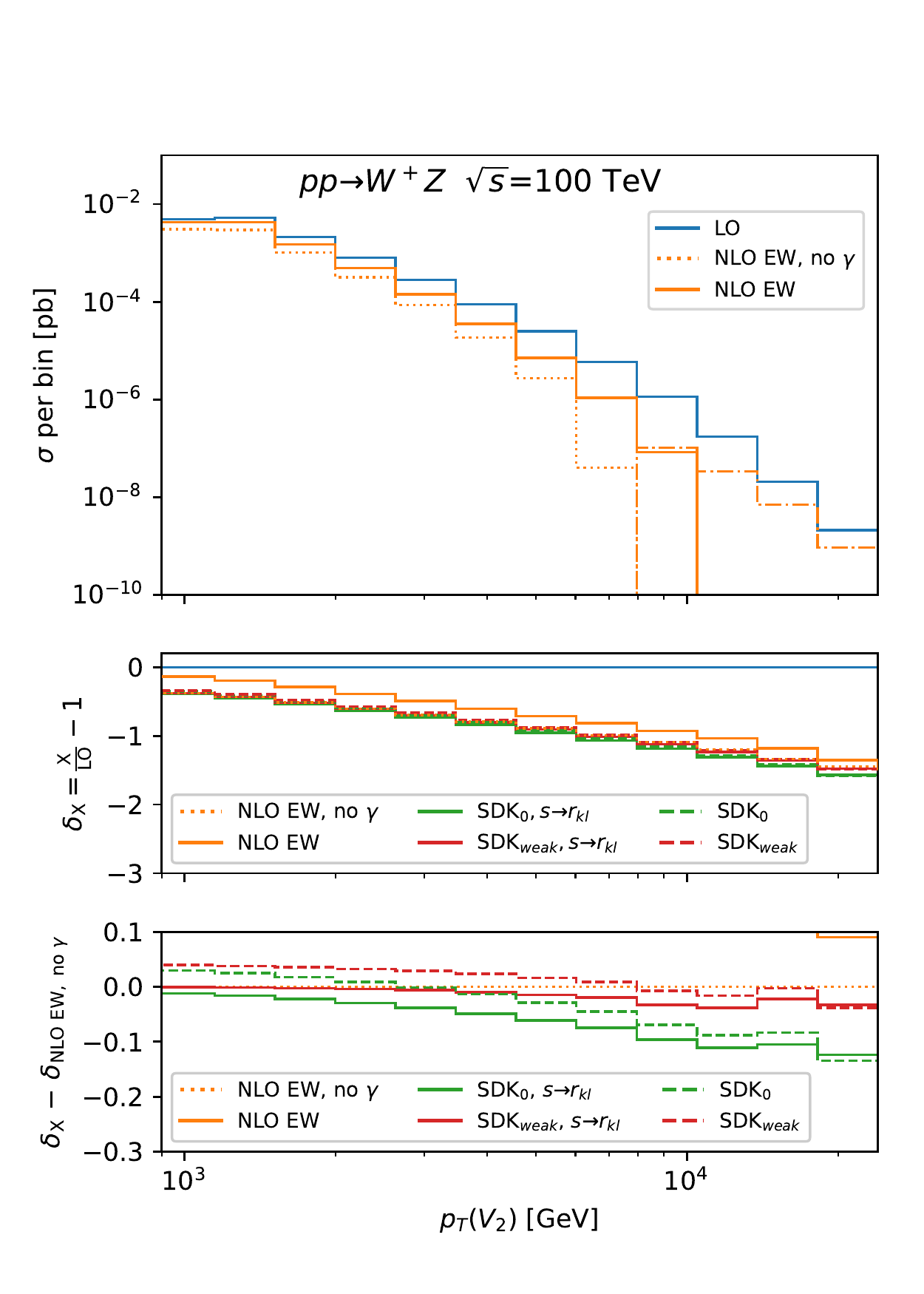}
\includegraphics[width=0.38\linewidth]{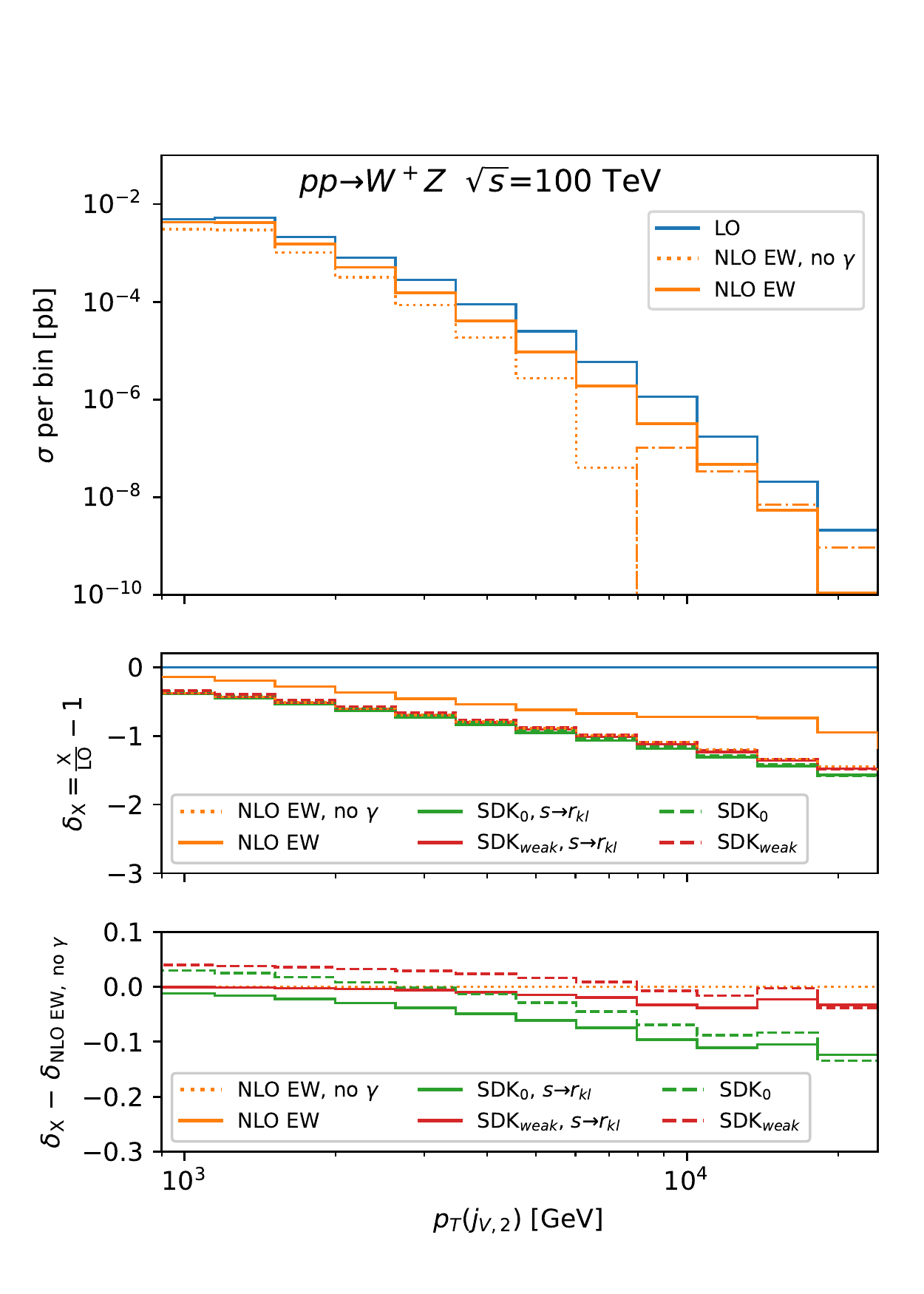}
\includegraphics[width=0.38\linewidth]{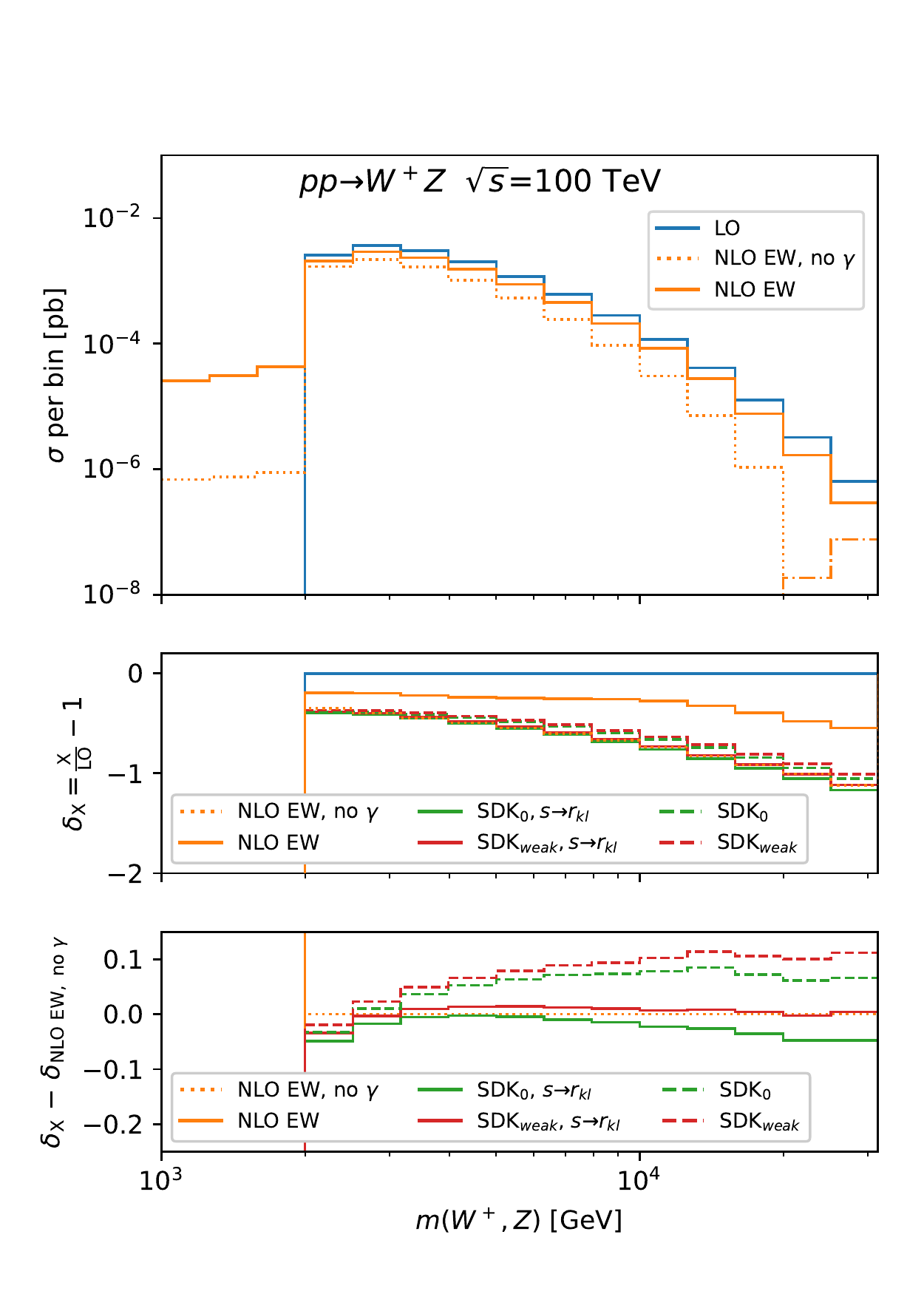}

\end{center}
\caption{Differential distributions for recombined $V$ bosons in $W^+Z$ production at 100 TeV. Comparison between exact NLO EW predictions excluding photon-initiated processes  (dotted orange) and their LA approximations in the ${\rm SDK_{weak}}$ (red) and ${\rm SDK_{0}}$ (green) approaches, including (solid) or excluding (dashed) the $\SS^{s\TO r_{kl}}$ terms.  More details are given in the text. \label{fig:WZ}}
\end{figure}

\subsection{WZ}

We now move to the case of a couple of processes where both the inclusion of the $\SS^{s\TO r_{kl}}$ terms and the use of $\rm SDK_{weak}$ is relevant. We start by showing differential distributions for the process $pp\TO W^+ Z$, where results have been obtained by using the following cuts
\begin{equation}
p_T(V_i)>1~\tev\,, \qquad |\eta(V_i)|<2.5\,, \qquad m(W^+,Z)>1~\tev\,, \qquad \Delta R (W^+,Z)>0.5\,. 
\label{eq:WZcuts}
\end{equation}
Again, these cuts resemble realistic experimental cuts for high-energy objects, but they also avoid (part of the) additional logarithmic enhancements from collinear splittings appearing in the real-radiation processes or even at the Born. 

In Fig.~\ref{fig:WZ} we show the transverse momentum of the hardest ($p_T(V_1)$) and softest ($p_T(V_2)$) recombined vector-bosons and their invariant mass ($m(W^+,Z)$). Similarly to the case of leptons \eqref{eq:recQED}, the recombination is performed by recombining any {\it charged} vector boson $V_i$ with photons that satisfy the condition $\Delta R(V_i, \gamma) < 0.4$. We also show the transverse momentum ($p_T(j_{V,2})$) of the softest jet that is tagged as a $V$-jet, namely containing one of the two vector bosons. They layout of the plots is very similar to the ones of Figs.~\ref{fig:DY} and \ref{fig:ZZZ}, but with an important difference: in the second inset we plot for each line in the first inset the difference with the {\it dotted} orange one, {\it i.e.}, the exact NLO EW prediction where the photon PDF has been set equal to zero. The reason is the following. The $WZ$ production is affected  by giant $K$-factors not only at NLO QCD \cite{Frixione:1992pj, Frixione:1993yp}, but also at NLO EW \cite{Baglio:2013toa, Mangano:2016jyj} precisely due to the opening of the $q\gamma\TO W^+ Z q'$ channel. For large value of $p_T(V_1)$ this channel induces large positive corrections of order $L(p_T^2(V_1),M^2_{V_2})$, due to configurations in which $V_1$ recoils against the emitted quark $q'$  and $V_2$ is soft and collinear to it. This effect has nothing to do with virtual NLO EW corrections and cannot be expected to be captured by the {\denpoz} algorithm, both in its $\rm SDK_{0}$ and  $\rm SDK_{weak}$ adaptations for physical observables. Therefore, in order to test the LA, we use as reference the NLO EW prediction where the photon PDF has been set equal to zero. We also plot in the main panel the NLO EW prediction without photon PDF as dotted orange line and its absolute value as dashed-dotted when it becomes negative. We remind the reader that the photon PDF is entering the process starting with NLO EW real emission of quarks, while it is not present at LO. 

Before commenting each plot of Fig.~\ref{fig:WZ}, it is again important to note how LA in general approximates very well the exact NLO EW predictions excluding photon-initiated processes, reaching almost -200\% of the LO in the tail of the distributions. This can be appreciated in the first inset of each plot. There, as in the main panel, the size of the $q\gamma\TO W^+ Z q'$ channel can also be appreciated. Its impact is maximal in the $p_T(V_1)$ distribution, due to the aforementioned $L(p_T^2(V_1),M^2_{V_2})$ enhancement, and minimal in the  $p_T(V_2)$ distribution since  large values of $p_T(V_2)$ forbid the kinematic configurations that precisely lead to the enhancement for  the $p_T(V_1)$ distribution. 

In the top-left plot ($p_T(V_1)$) we can see that none of the curves is really flat. This is not surprising, since the LA works very well when the cross section is dominated by the Born-like kinematic, which is not the case for large $p_T(V_1)$, where $V_1$ can in principle recoil against a very hard photon. Still, the solid red line  ($\rm SDK_{weak}$ with $\SS^{s\TO r_{kl}}$ terms included) leads to the best approximation and differs from the reference value, the dotted orange line, by less than 10\% of the LO also in regions where the NLO EW corrections are $\sim -200\%$. The superiority of the $\rm SDK_{weak}$ approach (red) over the $\rm SDK_{0}$ one (green) and of the inclusion of the $\SS^{s\TO r_{kl}}$ terms (solid) over their exclusion (dashed) can be better appreciated in the top-right plot ($p_T(V_2)$) and especially in the bottom-right one $m(W^+, Z)$. The difference between the dotted orange and solid red curves is only very few percents of the LO over the full range.

In the bottom-left plot, the $p_T(j_{V,2})$ distributions, all the curves are equal to those of the top-right one $p_T(V_2)$ but one: the solid orange line. In this context the only difference between a recombined $V$-boson and a $V$-tagged jet is that in the latter the $V$ boson can be recombined also with quarks. When the  $q\gamma\TO W^+ Z q'$ channel is included, soft+collinear configurations for $(V_2)$ in the final-state splitting $q^{(\prime)} \TO q' V_2$ are avoided at large $p_T(V_2)$, but the purely collinear ones still survive. The recombination of $V_2$ with quarks has therefore a sizeable impact. The bottom line is that the  $\rm SDK_{weak}$ approach takes into account only those emissions that are unavoidable from an IR-safety point of view, but additional real-emission contributions can be present and sizeable (for instance the photon-induced quark radiation), they can lead to enhancements factorising different Born matrix elements (for instance the case discussed here where $\sigma(q\gamma\TO W^+ Z q')\propto \sigma(q\gamma\TO W^+ q')\times L(p_T^2(W^+),\MZ^2)$ or $\sigma(q\gamma\TO W^+ Z q')\propto \sigma(q\gamma\TO Z q)\times L(p_T^2(Z),\MW^2)$) and their impact can depend on how the different particles are clustered among each other (for instance $V$-bosons versus $V$-tagged jet). These contributions cannot be taken into account via the {\denpoz} algorithm.  The LA and in particular  the {\denpoz} algorithm as implemented in {\mglong}  allow for a fast and, as shown in the previous examples, very reliable approximation of fixed-order exact NLO EW corrections. On the other hand, it cannot substitute the exact calculation. LA can be used as a starting point for improving the fixed-order NLO EW, by {\it e.g.} resumming the large Sudakov logarithms or alternatively for performing fast simulations with {\mglong} including the EW dominant effects at high-energy. The latter option, however, should be always cross-checked with an exact calculation before being  used for phenomenological predictions.
\begin{figure}[!t]
\begin{center}
\includegraphics[width=0.32\linewidth]{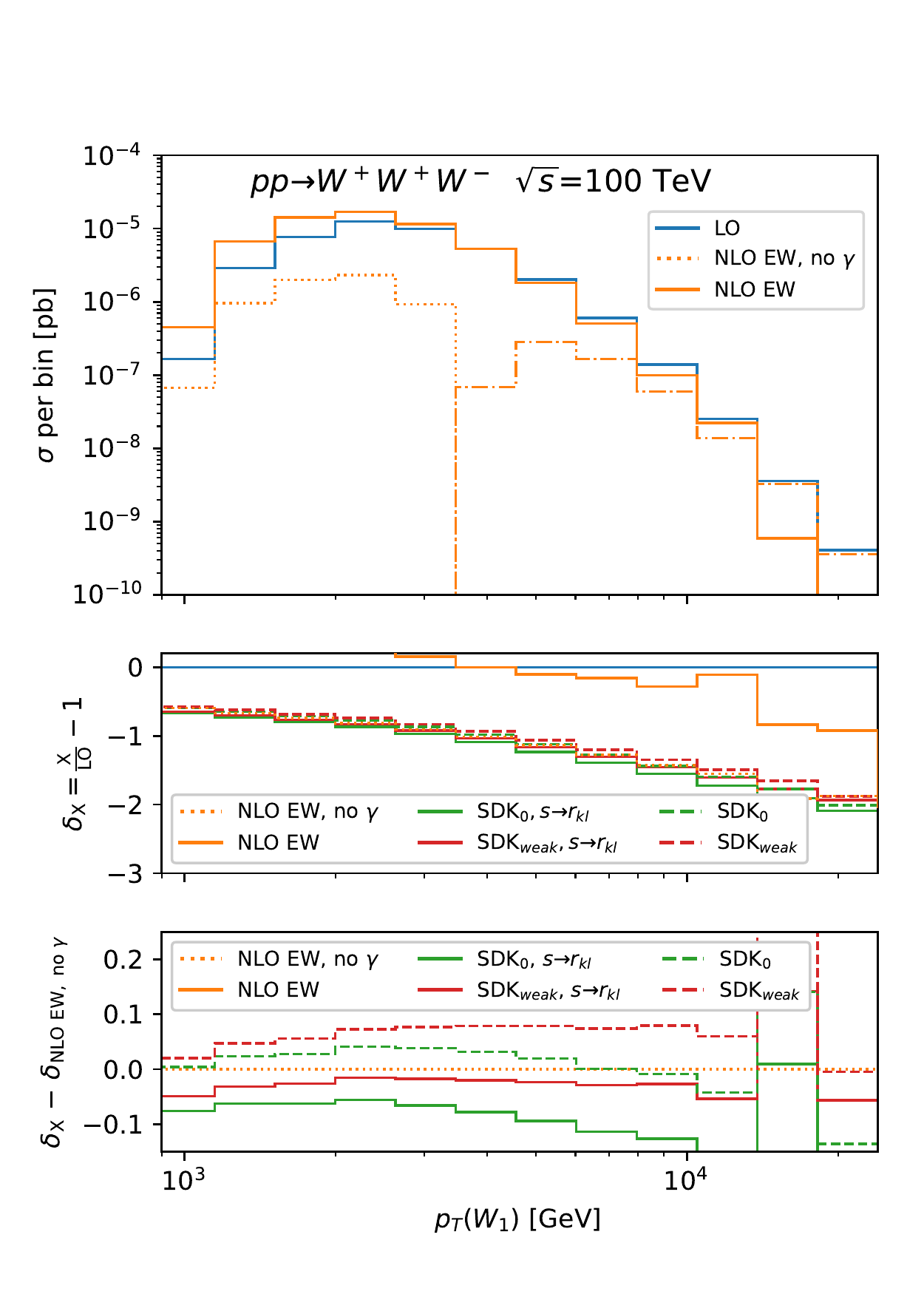}
\includegraphics[width=0.32\linewidth]{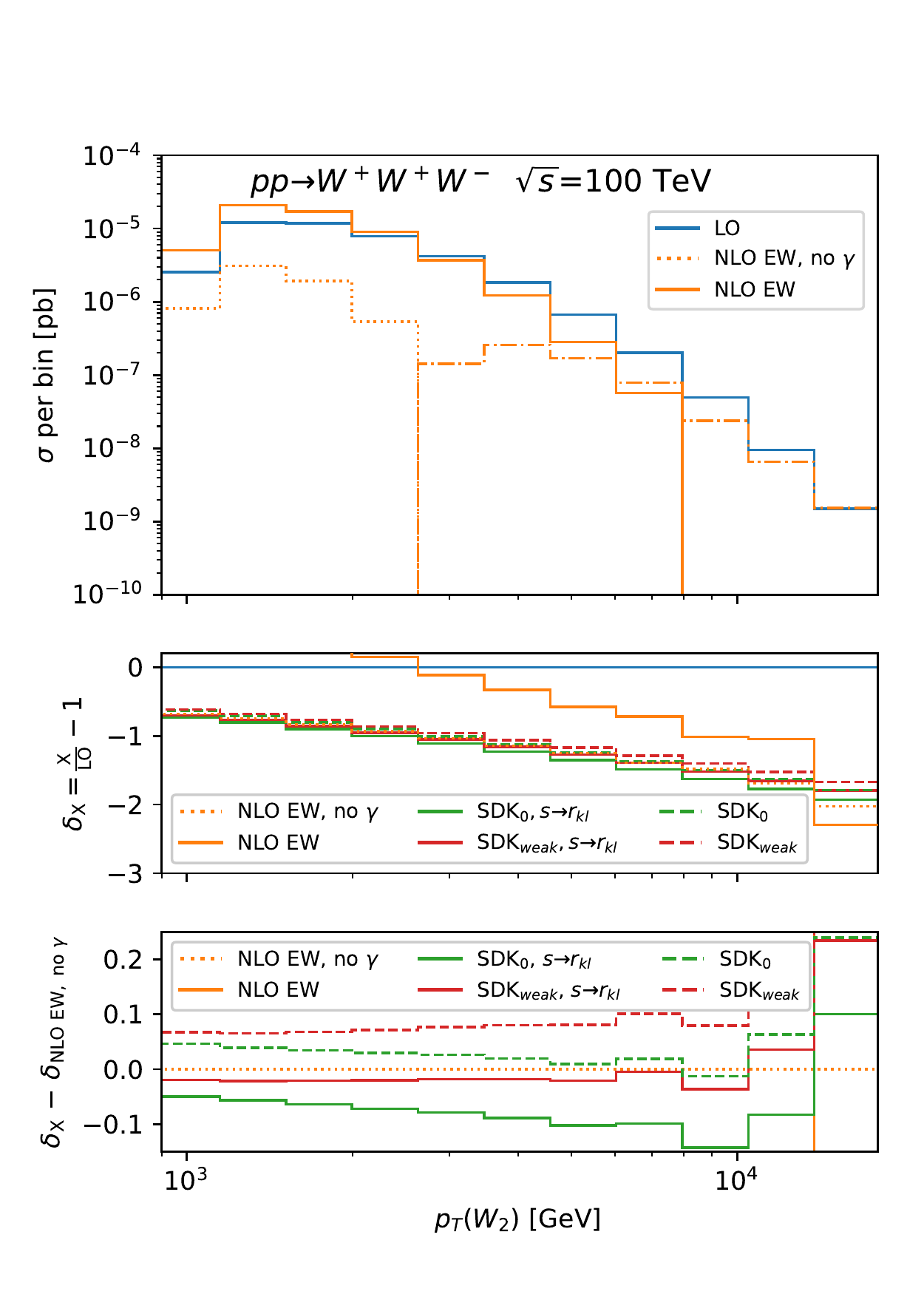}
\includegraphics[width=0.32\linewidth]{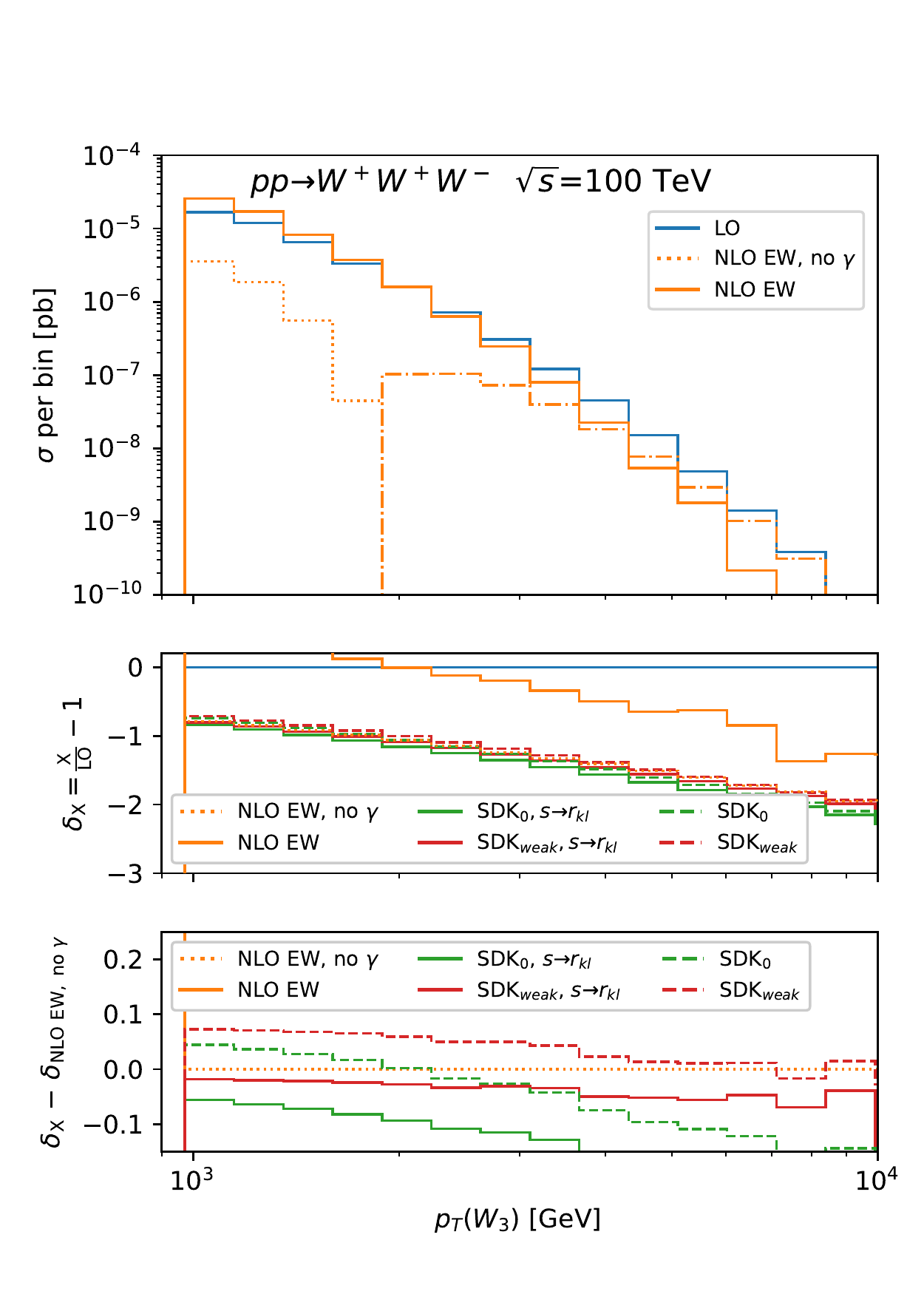}
\includegraphics[width=0.32\linewidth]{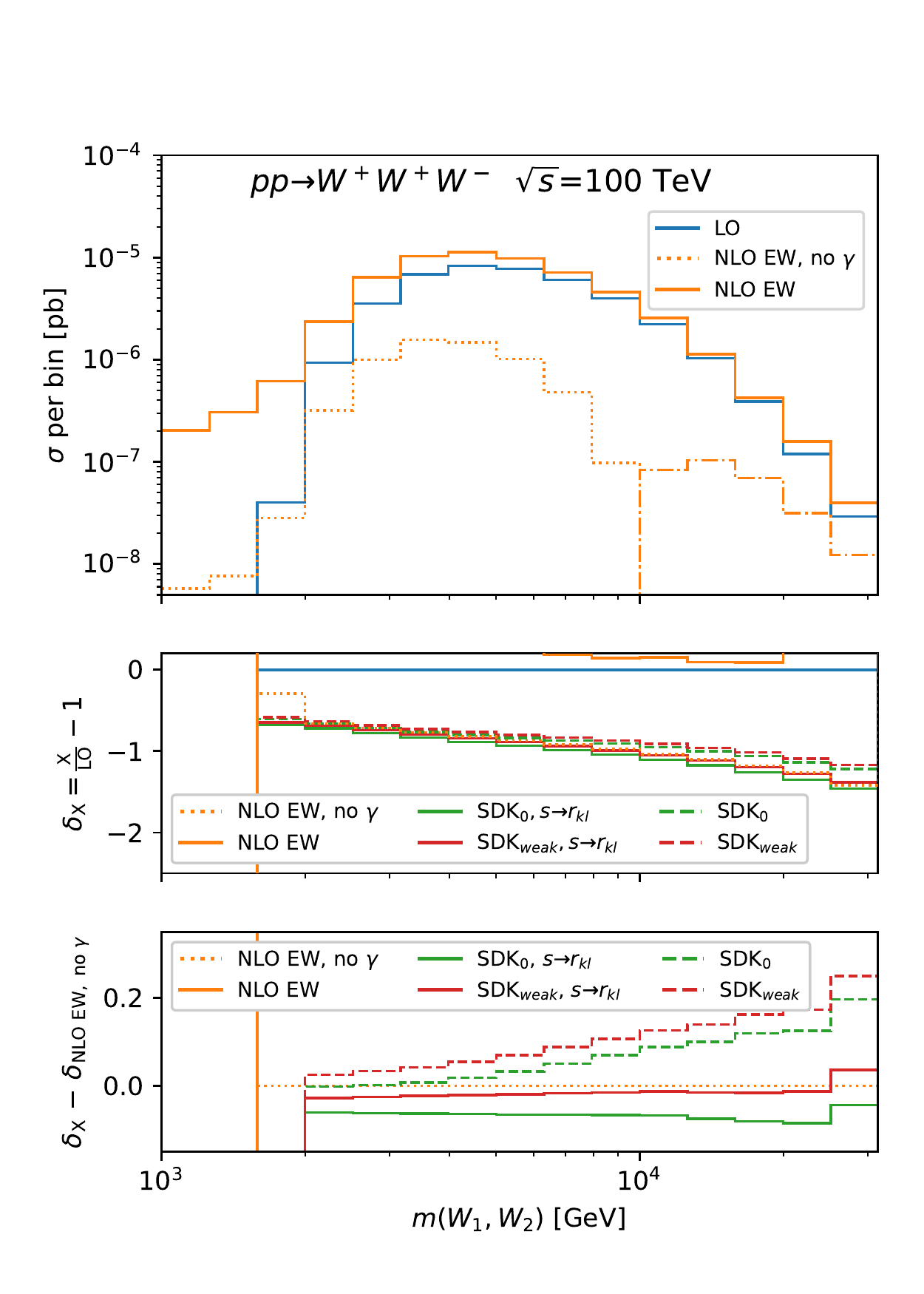}
\includegraphics[width=0.32\linewidth]{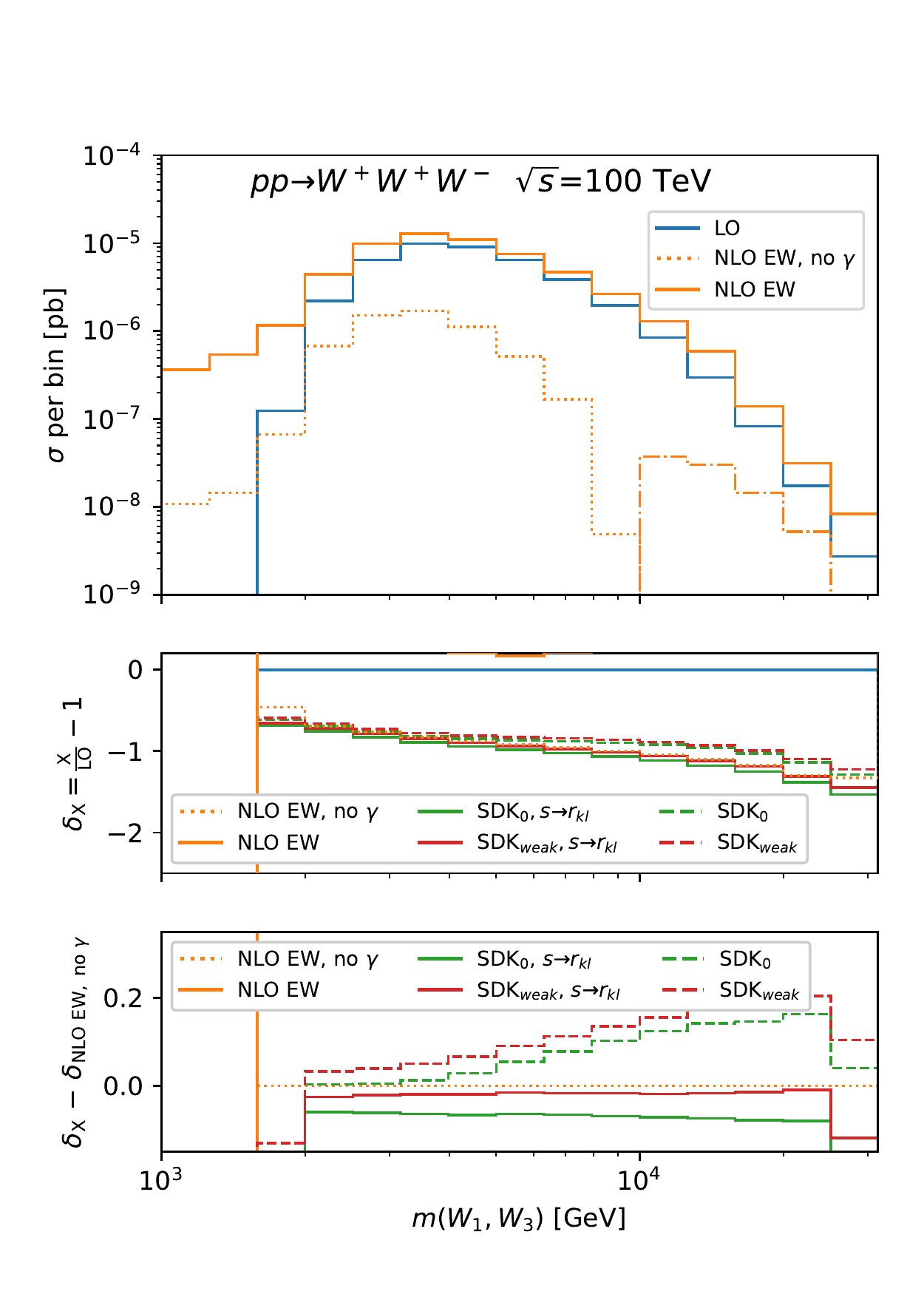}
\includegraphics[width=0.32\linewidth]{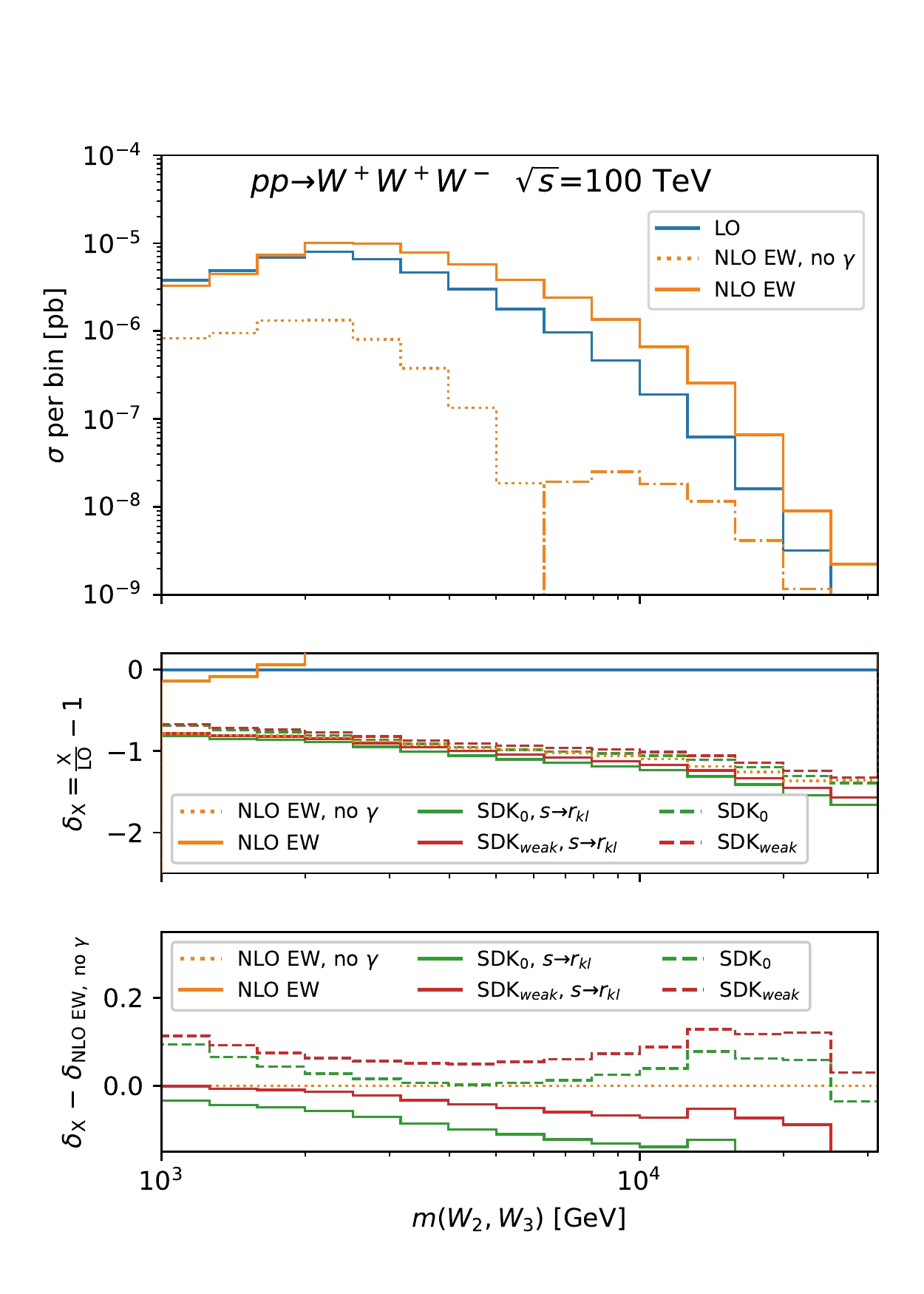}
\end{center}
\caption{Same as Fig.~\ref{fig:WZ}, but for $W^+ W^+ W^-$ hadroproduction at 100 TeV. \label{fig:WWW}}
\end{figure}
\subsection{WWW}
\label{sec:WWW}

As last example, we show in Fig.~\ref{fig:WWW} distributions for the process $p p \TO W^+ W^+ W^-$, where for the calculations the following cuts have been employed
\begin{equation}
p_T(W_i)>1~\tev\,, \qquad |\eta(W_i)|<2.5\,, \qquad m(W_i,W_j)>1~\tev\,, \qquad \Delta R (W_i,W_j)>0.5\,,
\label{eq:WWcuts}
\end{equation}
following the same rationale behind the cuts that have been used for the other processes considered in this section. In Fig.~\ref{fig:WWW} we consider the same observables already considered in Fig.~\ref{fig:ZZZ}, with $Z$ bosons replaced by $W$ bosons. The layout of the plots is the same of those in Fig.~\ref{fig:WZ} for $W^{+}Z$ production, where the reference line is the dotted orange one, the NLO EW prediction without photon-initiated processes.  Indeed, also in this case the opening of the photon-induced channels in the NLO EW corrections ($q\gamma\TO W^+ W^+W^- q'$) induces additional effects that cannot be captured by the {\denpoz} algorithm.

Similarly to the case of $W^{+}Z$ production in Fig.~\ref{fig:WZ}, we see that both the choices between $\rm SDK_{0}$ and $\rm SDK_{weak}$ approaches and between the inclusion or exclusion of $\SS^{s\TO r_{kl}}$ terms are relevant. Also, the exclusion of the photon-initiated processes is unavoidable for the comparison between LA and exact NLO EW results. As any other observable considered in this chapter, once again we see very large effects from NLO EW corrections (reaching almost $-200\%$ of the LO). Moreover, the solid red line, the LA in the $\rm SDK_{weak}$ approach including $\SS^{s\TO r_{kl}}$ terms, is able to approximate within the 10\% level  the exact NLO EW effects excluding photon-initiated processes (dotted orange line) over the full range of the observables considered the plots of  Fig.~\ref{fig:WWW}. The difference between the two aforementioned accuracies, the solid red line in the second inset, is in general a constant over the range of each observable considered. The only exception is $m(W_2, W_3)$ where a logarithmic trend is visible also for the red-solid line. A similar effect, although milder, could be observed also for the $m(Z_2, Z_3)$ distribution in $ZZZ$ production in Fig.~\ref{fig:ZZZ}. To the best of our knowledge, given the cuts in \eqref{eq:WWcuts}, this effect is not due to any possible enhancement arising from real emission contributions. However, as explained in Sec.~\ref{sec:Res_Amp}, possible additional angular-dependent effects involving ratios of invariants can be present and not captured in the {\denpoz} algorithm, regardless of the inclusion of the  $\SS^{s\TO r_{kl}}$ term.

\section{Conclusions and outlook}
\label{sec:conclusions}

In this work we have revisited the algorithm of Denner and Pozzorini \cite{Denner:2000jv} for the calculation of one-loop EW Sudakov logarithms, denoted in the previous sections as {\denpoz} algorithm. We have introduced several novelties that concern different aspects of the electroweak Sudakov approximation. Moreover, we have implemented the {\denpoz} algorithm, together with the novelties introduced, in the {\mglong} framework. Thanks to this new implementation we have provided in a completely automated approach several numerical results for different production processes, both for the virtual contribution in specific phase-space points and for physical (IR-safe) differential cross sections at a 100 TeV proton--proton collider.  All the numerical results that we have obtained corroborate the relevance of the novelties introduced in this work and also demonstrate the correctness of the implementation of the algorithm in the code.

In particular we have introduced the following novelties:  
\begin{itemize}
\item We have reframed the {\denpoz} algorithm by setting the mass of the photon and light-fermion masses exactly to zero, regularising IR divergences by mean of Dimensional Regularisation (DR), as in modern NLO EW calculations and  Monte Carlo implementations. Reframing the algorithm with the language of modern calculations, it allows for further developments and compatibility with the state-of-the-art tool implementations.
\item We have modified part of the expressions in order to take into account additional angular dependences, without assuming that all the invariants are of the same size of $s$. These modifications correspond to the terms that have been dubbed in the work as $\SS^{s\TO r_{kl}}$; they have been obtained keeping track of the dependence on any invariant $r_{kl}$ in the derivation of the subleading soft-collinear logarithms. As already mentioned in various points in the text, even with this improvement, the full control  of logarithms involving the ratios of  the different $|r_{kl}|$  and $s$ cannot be achieved via the {\denpoz} algorithm. Information on the internal structure of the diagrams is unavoidable for this purpose. On the other hand, the $\SS^{s\TO r_{kl}}$ terms  are sensitive to the presence of a hierarchy among the various invariants that characterise a specific phase-space point and substantially improve the approximation of the aforementioned class of logarithms. In order to support this statement, we have showcased for several processes and observables the superiority of the Sudakov  approximation including the $\SS^{s\TO r_{kl}}$ terms. 
\item We have identified an imaginary term that was omitted in Ref.~\cite{Denner:2000jv}, which cannot be in general neglected for $2\TO n$ processes with $n> 2$. This term is also present in the new  $\SS^{s\TO r_{kl}}$ terms that have been introduced. We have shown the relevance of this imaginary term for correctly capturing effects of order $\alpha \log(s/\MW^2)$.  
\item In this work we did not focus only on one-loop EW corrections to amplitudes, as done in the pioneering work Ref.~\cite{Denner:2000jv}, but we have considered also  the virtual NLO EW corrections to the LO cross sections. The two cases are trivially related only when the entire LO factorises a single combination of $\as$ and $\alpha$ powers. However, more in general, NLO EW corrections can also include effects due to loops of QCD origin.
 We provide the additional terms, denoted in the paper as $\deltaQCD$, that are necessary in order to take into account in Sudakov approximation also contributions from the aforementioned QCD loops.  We have proved the relevance of $\deltaQCD$ for correctly capturing effects of order $\alpha \log^n(s/\MW^2)$ with $n=1,2$, when also QCD loops contribute to virtual NLO EW corrections to the LO cross sections.
\item We describe how to modify the {\denpoz} algorithm in order to exclude the contribution of photons and also the contribution of gluons from the QCD terms mentioned in the previous bullet. We dub this approach as $\rm SDK_{weak}$. In the context of differential cross sections and IR physical observables,  we show how this approach is superior to the standard approach used in the literature, dubbed in this work as $\rm SDK_{0}$, which corresponds to simply removing the IR-divergent  logarithms involving the scale $\MW$ and the IR cut-off.  
\end{itemize}

We discuss also the technical steps of the implementation of the {\denpoz} algorithm, together with the novelties introduced in this work, in the {\mglong} framework. The choice of the {\mglong} framework, which already automates the calculation of the exact NLO EW corrections, has been crucial for the validation of our work and for demonstrating the correctness of our implementation of the Sudakov approximation, also denoted in the paper as leading approximation (LA). Systematic comparisons between exact $\ord(\alpha)$ corrections, the NLO EW, and their LA with different assumptions ($\SS^{s\TO r_{kl}}$, imaginary term, $\deltaQCD$, $\rm SDK_{weak}$, {\it etc.}) have been performed in order to check that all the logarithmic terms are correctly captured.

Besides being essential for validation, the implementation in a framework as  {\mglong} opens up for several new possibilities. First of all, the necessary ingredients for performing completely automated NLO EW corrections matched with resummed EW Sudakov logarithms are now available in a single tool. Then, a fast and stable method for approximating dominant effects from EW corrections is now available. Especially, the quality of the approximation can be checked via the NLO EW exact calculation, which however we remind the reader cannot in general  be substituted by its LA. Sudakov effects can be also more easily integrated in Monte Carlo simulations and generations of events. These are only a few of the possible outcomes of this work, which sets the basis for future works both at the phenomenological and formal level.  

Another important possible development of this work is, {\it e.g.}, the possibility of compute approximate EW corrections also in extensions of the SM. Indeed, only tree-level Feynman rules are needed. However, for a given BSM model, the associated Universal Feynrules Output (UFO) model~\cite{Degrande:2011ua}, which is the format used in {\mglong} for encoding the Feynman rules,  should provide some
extra informations that are not present at the moment: how particles are arranged in SU(2) multiplets and the values of their electroweak couplings (charge, isospin component and/or hypercharge). At the moment, this information
had to be written by hand in the code. We thus envisage an extension of the UFO format in this direction. 

As a final remark, we want to stress once again that the Sudakov approximation cannot substitute exact NLO EW calculations. As we said, this approximation can be exploited for performing fast simulations including the EW dominant effects at high energy. However, it should be always cross-checked with an exact calculation before being used for phenomenological predictions and especially comparisons with data. Moreover, as explained many times in the text, the approximation works when all the invariants $r_{kl}$ satisfy the relation $|r_{kl}|\gg \MW^2$. This implies, for instance, that the Sudakov approximation can never be directly exploited for a process involving resonances. In the SM this typically means the production and decay of the heavy particles $W,Z,H$ or $t$. On the other hand, given such a type of process, the Sudakov approximation can be applied for the calculation of the same process including only the production but not the decays of the aforementioned heavy particles, and the decay can be subsequently taken into account.

 \section*{Acknowledgements}
 
We are indebted to Ansgar Denner and Stefano Pozzorini for their valuable comments on the final version of this paper. 
We want to thank Hua-Sheng Shao for his help in the initial phase of this project and Stefano Forte for discussions.  
 We are also grateful to  the developers of {\sc MadGraph5\_aMC@NLO} for the long-standing collaboration and for discussions.    
  Most of the work of D.P.~has been supported by the Deutsche Forschungsgemeinschaft (DFG) under Germany's Excellence Strategy - EXC 2121 ``Quantum Universe'' - 390833306. M.Z.~is supported by the ``Programma per Giovani Ricercatori Rita Levi Montalcini'' granted by the Italian Ministero dell'Universit\`a e della Ricerca (MUR). 

\addcontentsline{toc}{section}{References}
\bibliographystyle{JHEP}
\bibliography{biblio}

\end{document}